\documentclass[useAMS,usenatbib]{mn2e}
\usepackage{epsfig,amsmath,amssymb,amsfonts,mathrsfs,latexsym,graphicx,deluxetable,lscape,bm,color}
\newcommand{\gsim}{\, \raisebox{-0.8ex}{$\stackrel{\textstyle >}{\sim}$ }}
\newcommand{\lsim}{\, \, \raisebox{-0.8ex}{$\stackrel{\textstyle
      <}{\sim}$ }}
\begin{document}

\title[Gravitational waves from accreting neutron stars]{Detecting
  gravitational wave emission from the known accreting neutron stars}
\author[Watts et al.]{Anna L. Watts$^{1,2}$, Badri Krishnan$^3$, Lars Bildsten$^4$
and Bernard F. Schutz$^3$
\\ $^1$ Max Planck Institut f\"ur Astrophysik,
Karl-Schwarzschild-Strasse 1, 85741 Garching, Germany \\ $^2$ Astronomical Institute ``Anton Pannekoek'', Kruislaan 403, 1098 SJ Amsterdam, The Netherlands; A.L.Watts@uva.nl \\ $^3$ Albert-Einstein-Institut, Max-Planck-Institut f\"ur Gravitationsphysik, Am
  M\"uhlenberg 1, 14476 Golm, Germany \\ $^4$ Kavli Institute for Theoretical Physics, Kohn Hall, University
  of California at Santa Barbara, CA 93106, USA}

\maketitle

\begin{abstract}
Detection of gravitational waves from accreting neutron stars (NSs) in our
galaxy, due to ellipticity or internal oscillation, would be a breakthrough in
our understanding of compact objects and explain the absence of NSs rotating
near the break-up limit. Direct detection, however, poses a formidable
challenge. Using the current data available on the properties of the accreting
NSs in Low Mass X-Ray Binaries (LMXBs), we quantify the detectability for the
known accreting NSs, considering various emission scenarios and taking into
account the negative impact of parameter uncertainty on the data analysis
process. Only a few of the persistently bright NSs accreting at rates near the
Eddington limit are detectable by Advanced LIGO if they are emitting
gravitational waves at a rate matching the torque from accretion. A larger
fraction of the known population is detectable if the spin and orbital
parameters are known in advance, especially with the narrow-band Advanced LIGO.
We identify the most promising targets, and list specific actions that would
lead to significant improvements in detection probability.  These
include astronomical observations (especially for unknown orbital periods),
improvements in data analysis algorithms and capabilities, and further detector
development.
\end{abstract}


\section{Introduction}

A number of interferometric gravitational wave observatories have been
built with the intention of opening a new observational window for
studying astrophysical objects.  These include the
LIGO\footnote{http://www.ligo.caltech.edu},
GEO\footnote{http://geo600.aei.mpg.de/},
VIRGO\footnote{http://wwwcascina.virgo.infn.it} and
TAMA\footnote{http://tamago.mtk.nao.ac.jp} detectors.  The TAMA and
LIGO Scientific Collaborations have demonstrated their ability to
reach sensitivity goals, take year-long stretches of data with good
duty cycles, and analyse the data to set, with confidence, upper
limits on the emission from a number of possible sources. At present the large LIGO and VIRGO detectors are 
performing a significant sensitivity upgrade.  This is therefore a good time to carry out a realistic study of the challenges of searching for    
gravitational waves from one class of sources that future
upgrades may render detectable: spinning neutron stars (NSs) in Low Mass X-ray Binaries (LMXBs).

For the last ten years, ever since a paper by one of us \citep{bil98} suggested
that LMXBs could be steady beacons of gravitational waves, the study of LMXBs
(and especially of the source Scorpius X-1) has been one of the scientific goals
of the development of detectors with greatly improved sensitivity. Since then,
X-ray astronomers have gathered a wealth of new data on these sources,
astrophysicists have built better models, and gravitational wave
scientists have gained considerable experience of their ability to extract weak
signals from data streams. It is timely, therefore, to revisit these estimates.

The expectation that LMXBs could be strong steady sources of gravitational waves
originates in one of the most important outstanding questions about NSs: why they all spin at frequencies much less than the break-up limit
of 1 kHz \citep{lat07}. Simple estimates \citep{cook94} of the spin-up
timescales for accreting NSs in LMXBs suggest that there should be no difficulty
in reaching at least 1 kHz. The fastest rotating accreting pulsar is at 599 Hz
\citep{gal05} and many NS rotate much more slowly \citep{cha03}. The millisecond
radio pulsars, the likely offspring of LMXBs, also spin at slower rates than
expected. The current record stands at 716 Hz \citep{hes06} and surveys in
nearby globular clusters continue to reveal a paucity of rapid rotators
\citep{ran05, hes07, freire07}. Whether the apparent spin limit is genuine
remains to be resolved. Nevertheless, given the current sample, it certainly
appears that there is some brake that prevents accreting NSs from reaching the
break-up limit.

The candidate mechanisms fall into two main camps.  In the first, accretion
torques are reduced and eventually balanced by the interaction between the
accretion disk and the NS's magnetic field \citep{gho78, whi97, ande05}.  The
second possibility, and our focus here, is the loss of angular momentum via the
emission of gravitational radiation \citep{papa78, wag84, bil98}.  There are
many conceivable ways for an accreting NS to develop a quadrupolar asymmetry
that leads to gravitational wave (GW) emission: crustal mountains \citep{bil98,
ush00, mel05, pay06, has06}, magnetic deformation \citep{cut02, has07} or
internal r-mode oscillations \citep{bil98, ande99, levi99, ande00,
  hey02, ande02, wag02, nay06, bon07}. The exciting prediction from all
GW emission scenarios is the possible direct detection of
an accreting NS by a ground-based interferometric gravitational wave detector.

Searches for periodic GWs from NSs have already been performed with 
the LIGO and GEO detectors. These include searches of known radio
pulsars \citep{abb04,abb05b,abb07b}, assuming the phase of the GW
signal to be locked to the known NS rotation, and wide parameter
surveys for hitherto unknown NSs \citep{abb05a,abb07a,abb08}.  There
have also been searches for GWs from the accreting NS Scorpius X-1.
The first \citep{abb07a} used a coherent statistic on a template grid
utilising 6 hours of data (limited by computational requirements) from
the second science run of the two LIGO 4 km interferometers, followed
by a coincidence analysis between candidates from the two detectors.
The second search \citep{abb07c} used a method of cross-correlating the
outputs of the two LIGO 4 km detectors using about 20 days of
coincident data taken during the fourth science run.  Data from resonant bar detectors have also been used in
these searches; see e.g. \citet{ast02,ast05} for a blind all-sky
search in a narrow frequency band using data from the EXPLORER
detector.

These early searches, while so far yielding only upper limits, have
served to develop and prove data analysis methods that will be used on
data from future, more sensitive searches. The VIRGO detector is
approaching its design sensitivity and the LIGO detectors have just
completed a full run at their first-stage design sensitivity. An
initial upgrade of LIGO is about to commence, followed by a full
upgrade to Advanced LIGO\footnote{http://www.ligo.caltech.edu/advLIGO}
in the next decade. VIRGO expects to parallel these developments.
These impending improvements in GW sensitivity, coupled with the
experience of performing realistic data analysis, make the time right
to assess what the relevant accreting NS properties tell us about the
ultimate detectability of NS sources.

This paper explains the prospects for detecting GWs from accreting NSs, and
identifies  the most promising targets. We start in Section \ref{overview} with the best case scenario, highlighting the data analysis challenges and the need for detailed information on the accreting NS properties, which are presented
in Section \ref{stars}. Section \ref{sec:search} explains fully
the GW search data analysis challenges and the direct consequences of parameter
uncertainty on integration times and detection statistics, allowing us to assess future detectability in
Section \ref{detect}. We close in Section \ref{disc} by highlighting where progress can be made in the short-term on NS source properties, and the implications for current and future gravitational wave searches.

\section{Overview and Best Case Estimates}
\label{overview}

The potential strength of a periodic GW signal at frequency $\nu$ from specific LMXBs
with accretion rates $\dot M$  was
estimated by \citet{bil98} (hereafter B98) in the mountain
scenario. In this model a quadrupole moment $Q$, that is stationary in the
rotating frame of the star (a `mountain'), leads to the emission of gravitational
waves with a predominant frequency $\nu =2 \nu_s$,
$\nu_s$ being the spin frequency of the NS. The assumption is that the
accretion torque on a NS of mass $M$ and radius $R$
\begin{equation}
N_a = \dot{M} \left(G M R \right)^{1/2},
\end{equation}
is balanced by the GW torque
\begin{equation}
N_{\rm gw} =  -\frac{32 G Q^2 \Omega_s^5}{5 c^5},
\label{mountaintorque}
\end{equation}
where $\Omega_s = 2\pi \nu_s$ is the angular frequency of the star.
The accretion rate is  estimated from the bolometric X-ray flux $F$ by
assuming that the luminosity is $L \approx G M\dot{M}/ R$, yielding
\begin{equation}
\dot M = \frac{4\pi R d^2 F}{GM},
\label{mdotflux}
\end{equation}
where $d$ is the distance to the source.
The predicted gravitational wave amplitude $h_0$ (as
defined in \citet{jks})\footnote{Note that flux formulae are
 often given in terms of an angular and time-averaged amplitude $h$
 rather than in terms of $h_0$.  In \citet{owe98} and \citet{ande99},
 for example, the quoted amplitude $h^2 = 2 h_0^2/5$.} can be written in terms of the
GW luminosity,  $\dot{E}_{\rm gw} = N_{\rm
  gw} \Omega_s$, as
\begin{equation}
\label{gflux}
h_0^2 = \frac{5 G}{2 \pi^2 c^3 d^2 \nu^2} \dot{E}_{\rm gw}.
\end{equation}
Under the condition of torque balance,
  \begin{equation}
h_0 = 3 \times 10^{-27}F_{-8}^{1/2}\left({R \over {\rm 10 km}}\right)^{3/4}\left({1.4 M_\odot \over M}\right)^{1/4}
\left(\frac{{\rm 1~kHz}}{\nu_s}\right)^{1/2},
\label{mount_amp}
\end{equation}
where $F_{-8}=F/10^{-8} {\rm erg \ cm^{-2}\ s^{-1}}$. Equation (\ref{mount_amp})  makes it clear that the GW signal strength depends on two observables, the flux on the sky from the LMXB and the NS spin rate \citep{wag84,bil98}.
This amplitude is then compared directly to the best case detectable
amplitude in a long search, $h_0^{\rm sens}$, which
we will derive shortly.

These amplitudes are sufficiently weak that long stretches of data must be
folded, using predicted signal templates. This is called matched filtering, and
its sensitivity improves with the square-root of the observation time, provided
that the template manages to keep phase with the real signal to within about one
radian over the entire duration. Where spin and orbital parameters are poorly
constrained, many templates must therefore be searched. Analysis can become
statistically and computationally untenable, and computational loads may limit
integration times and thereby detectability. Contrast for example the most
recent searches for radio pulsars, where a precise timing ephemeris is available
\citep{abb07b}, with searches for the closest accreting neutron star Scorpius
X-1 \citep{abb07c}.  Poor constraints on the spin and orbital parameters for Sco
X-1 necessitate multiple templates.  This reduces the feasible integration
time substantially, resulting
in upper limits that are approximately two orders of magnitude larger than those
obtained for the best radio pulsar. We begin our discussion there.

\subsection{Spin Frequencies of Accreting NSs}

The discoveries and studies with the {\it Rossi X-ray Timing Explorer}
(RXTE) over the last ten years have dramatically improved our
understanding of stellar spin rate, $\nu_s$ \citep{str06,vand06}.
Having some constraint on the spin is, as will become clear in later
Sections, the most important factor determining the feasibility of
gravitational wave
searches. For this reason we concentrate only on sources
for which there is some measurement or estimate of $\nu_s$.  We
now summarize the three relevant categories of accreting NSs:
accreting millisecond pulsars, burst
oscillation sources, and kiloHertz Quasi-Periodic Oscillation (kHz
QPO) sources.

The spin is measured directly in the accreting millisecond pulsars, where
fixed hotspots, presumably at the magnetic footpoints, are a permanent
asymmetry.  In 1998 only one such object was known; there are now
ten members of this class, three of which show only intermittent pulsations.
The other measures of spin are indirect.  Probably the most
reliable are burst oscillations, seen during Type I
X-ray bursts (when accreted material burns in an unstable
thermonuclear flash). In 1998 there were six burst oscillation
sources: there are now
twelve stars with burst oscillations seen in
multiple bursts (including three pulsars), and seven stars with tentative detections.

For the three accreting millisecond pulsars that
also show burst oscillations, the burst oscillation frequency is at or
very close to the known spin frequency \citep{cha03, str03, alt07}.  It would
therefore seem reasonable to equate the burst oscillation frequency
with the spin frequency for the non-pulsing sources (as was done in
B98).  The frequency
for a given source is highly consistent from
burst to burst, implying that there is at least a strong
dependence on stellar spin \citep{str98, mun02}.  The
detection of highly coherent oscillations lasting several hundred
seconds during a superburst adds further
support to this hypothesis \citep{str02}. There are,
however, some complicating
factors.  Firstly, burst oscillations can exhibit frequency drifts of
up to a few Hz.  Secondly, there are some differences in the
properties of the burst oscillations of the non-pulsing LMXBs as
compared to the pulsars (see for example \citet{wat06}), suggesting
that the mechanism may differ.  In models that involve global modes of the
surface layers, for example, the observed frequency would be offset from
spin frequency by several Hz \citep{hey04, pir05}.

The third class of sources to be considered are those which exhibit twin
kHz QPOs.  Early observations suggested that although the frequencies
could shift, their separation remained relatively constant, implying a
link to stellar
spin.  We now know that separation varies (often quite
substantially) as accretion rate changes, and the cause of the kHz QPOs
is still not understood.  However, most models still depend in some way on the stellar spin, either directly or via
the influence on the space-time in the inner regions of the accretion
disk (see \citet{vand06} for a recent and comprehensive review).
Observations of twin kHz QPOs in two of the accreting millisecond
pulsars have done little to
resolve the situation:  in one case separation is rather close to the
spin frequency; in the other it is slightly less than half the spin
\citep{wij03, lin05}.  Whether there is any link between kHz QPO
separation and stellar spin will doubtless emerge in due course.
However, given that most models still predict some relation, we follow
B98 and include these stars in our analysis.

\subsection{ Optimal Gravitational Wave Detection}
\label{temps}

Before being scaled down for their current sensitivity upgrades, the
first generation LIGO interferometers operated for nearly two years at
or better than their design sensitivity, by the end of which the first
generation Virgo detector was not far behind. The ``Enhanced'' LIGO
and VIRGO interferometers should begin operation in 2009, and the
Advanced LIGO interferometers are now funded and should be operational
by 2014. Advanced VIRGO is expected on the same schedule. Beyond this,
there are ambitious plans for a third generation Einstein Telescope
(ET) in Europe.

\begin{figure}
  \begin{center}
    \includegraphics[width=\columnwidth, clip]{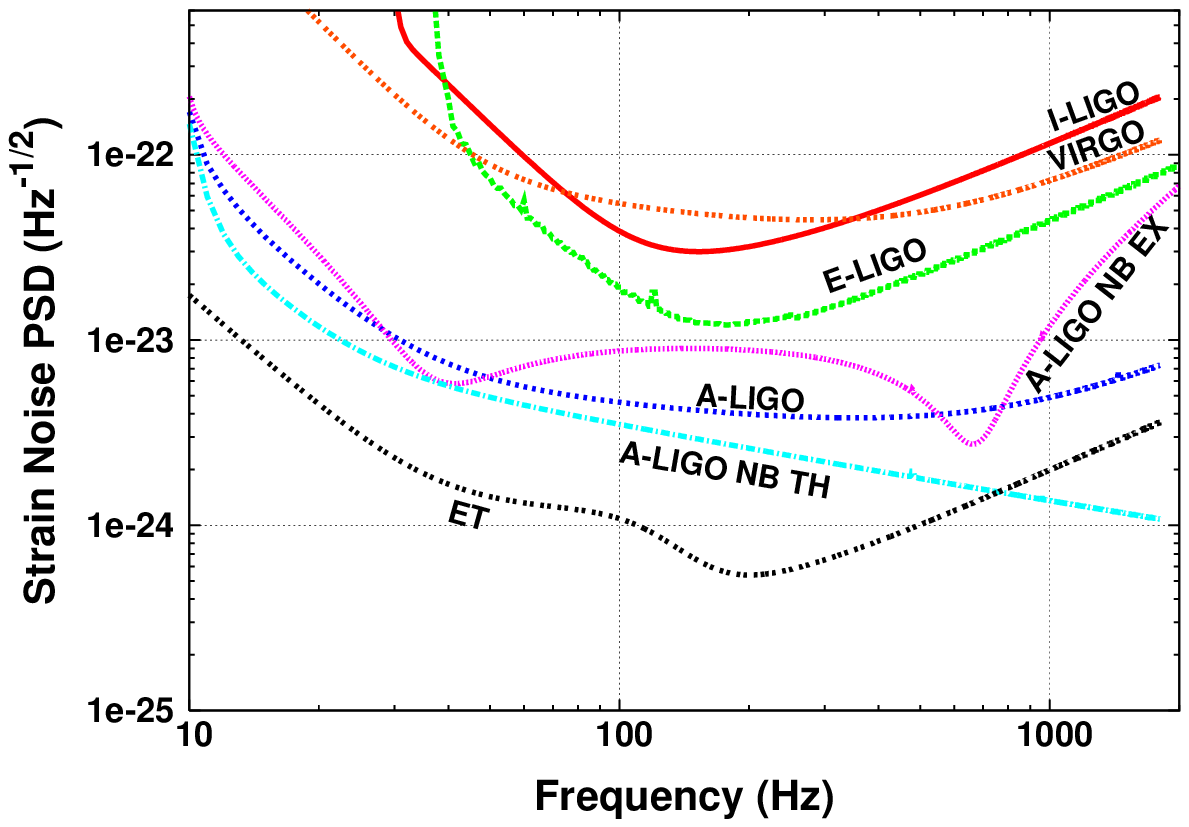}
    \includegraphics[width=\columnwidth, clip]{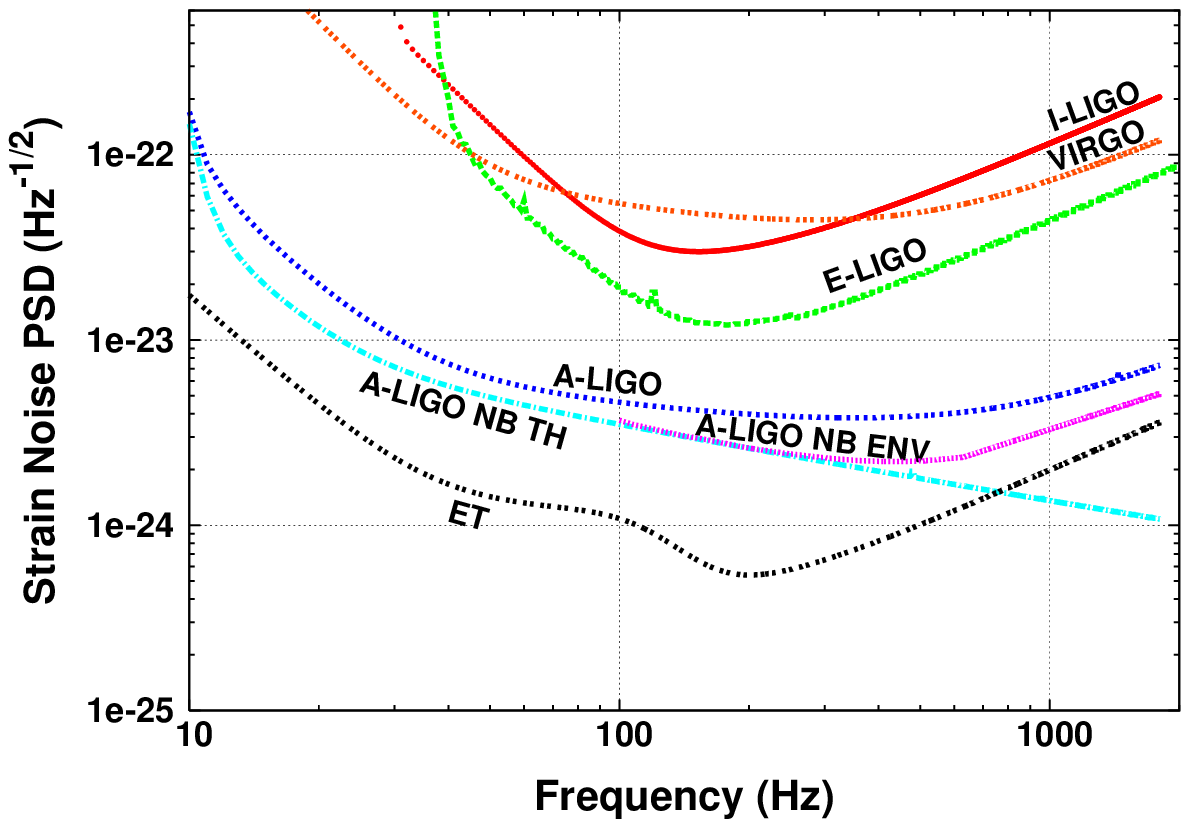}
    \caption{The noise curves for Initial LIGO, Virgo, Advanced LIGO,
      and the third generation Einstein Telescope (ET)
      interferometers.  The Initial, Enhanced and Advanced LIGO curves
      are labelled ``I-LIGO'', ``E-LIGO'' and ``A-LIGO'' respectively;
      ``ET'' is Einstein Telescope.  There are three curves for
      Advanced LIGO in each panel. The top panel shows a nominal broad
      band configuration (the so called ``zero de-tuned''
      configuration), an example of a narrow band curve (``A-LIGO
      NB EX''), and the total thermal noise (``A-LIGO TH''), i.e. the sum
      of the suspension and mirror thermal noise curves. The lower
      panel shows the lower envelope of the narrow band curves
      (``A-LIGO NB ENV'') instead of the narrow band example.  The
      thermal noise is shown because it is sometimes taken as a
      theoretical lower bound on the narrow banding for frequencies
      above, say, 100 Hz; as seen from the lower panel, this
      is not a good approximation at higher frequencies.}
    \label{fig:sh}
 \end{center}
\end{figure}

The sensitivity of a gravitational wave detector is determined by the
power spectral density of its instrumental strain noise, normalized to
an equivalent gravitational-wave amplitude $h(t)$. This is just the
Fourier power spectrum of $h(t)$ and is called $S_h(\nu)$: the noise
power per unit frequency. It is conventional to plot $h(\nu)=
[S_h(\nu)]^{1/2}$, which allows comparison of the noise to the signal's
amplitude if one knows the signal's bandwidth. The sensitivity curves
of current and future detectors are shown in Figure \ref{fig:sh}.  The
design noise curves for Initial LIGO, Virgo and ET are taken from
analytic models available, for example, in the LIGO software
repository\footnote{\texttt{http://www.lsc-group.phys.uwm.edu/daswg/projects/lal.html}}. 
The most recent science runs of the LIGO detectors have in fact
reached their design goals over a broad frequency range of interest,
above $\sim 40$ Hz.  The Enhanced LIGO noise curve in
Figure~\ref{fig:sh} is a realistic estimate of what can be achieved;
thus at low frequencies (below $\sim$ 40 Hz), the Enhanced LIGO noise
curve lies above the Initial LIGO design curve.

The Advanced LIGO detector configurations have not yet been finalized,
and we therefore need to consider different possibilities.  Advanced
LIGO can potentially be operated in a narrow band mode where
sensitivity is gained in a relatively narrow frequency range at the
expense of broad band sensitivity (see e.g. \cite{meers88,buon02}).
This could be particularly relevant for periodic signals where the
frequency is well known.  The first panel of Figure \ref{fig:sh} shows
an example of such a narrow band noise curve\footnote{The various
  Advanced LIGO noise curves shown here have been obtained using
  version 6.2 of the Matlab script ``Bench''. It is worth mentioning
  that a newer version of Bench, v7.0, is currently under development
  which includes an improvement in coating thermal noise; it is
  probably worth re-calculating the Advanced LIGO noise curves once
  this and later versions become available.}  The noise curve can be
tuned to target a broad range of frequencies by changing a number of
interferometer parameters, and for our purposes, the value of the
noise at the minimum is especially important.  Thus, the lower panel
of Fig.~\ref{fig:sh} shows the lower envelope of the narrow band noise
curves above 100 Hz.  Between 100 Hz and about 400 Hz, the narrow band
curves are limited by thermal noise, and by quantum noise at higher
frequencies\footnote{To generate the lower envelope of the narrow band
  noise curves, we have used Bench to calculate the narrow band noise
  curves for a range of choices for a few interferometer parameters.
  These parameters are the phase and transmittance of the signal
  recycling cavity, and the transmittance the input test mirror and
  the power recycling mirror.  The envelope is then obtained by
 calculating the convex hull of the minima of the various noise
  curves.}.  See Section \ref{sec:disc-da} for additional discussion.
It is important to keep in mind that the narrow-band ``envelope'' does
not represent any particular interferometer configuration but is
rather a superposition of many configurations.  It is only useful for
targeting narrow-band signals with frequency uncertainties of, say,
$\mathcal{O}(10)\,$Hz in which case we can choose the appropriate
element from the set of configurations used to produce the envelope.
Finally, note that possible designs for the third generation detectors
are still being explored, and thus the noise curve for the Einstein
Telescope is much more preliminary.

To start addressing the question of detecting GWs from LMXBs, we start
by asking how strong the signal would need to be for detection if we
knew all source parameters to sufficient accuracy that only one
template was needed (a coherent fold). Computational cost is no issue,
and we can easily integrate for long periods $T_{\rm obs}$.  Such a
fully coherent search with $D$ detectors of comparable sensitivity is
a best case and defines a signal-to-noise ratio (SNR) as ${\rm SNR}^2
= h_0^2T_{\rm obs}D/S_n$; thus, the SNR-squared builds up linearly
with the observation time (here $S_n(\nu)$ is the power-spectral
density of the detector noise).  Conversely, for a given choice of SNR
threshold for detectability, this leads to a minimum detectable signal
amplitude $h_0$ (following the notation of \citet{jks}):
\begin{equation}
  \label{eq:8}
  h_0 \approx 11.4\sqrt{\frac{S_n}{D T_{\rm obs}}}\,
\end{equation}
which is a useful indicator of the sensitivity of such a search.  The
factor of 11.4 corresponds to an SNR threshold which would lead to a
single trial false alarm rate of $1\%$ and a false dismissal rate of
$10\%$ \citep{abb07a} and a uniform averaging over all possible source
orientations and sky positons \citep{jks}.

\begin{figure*}
\begin{center}
\includegraphics[width=16cm, clip]{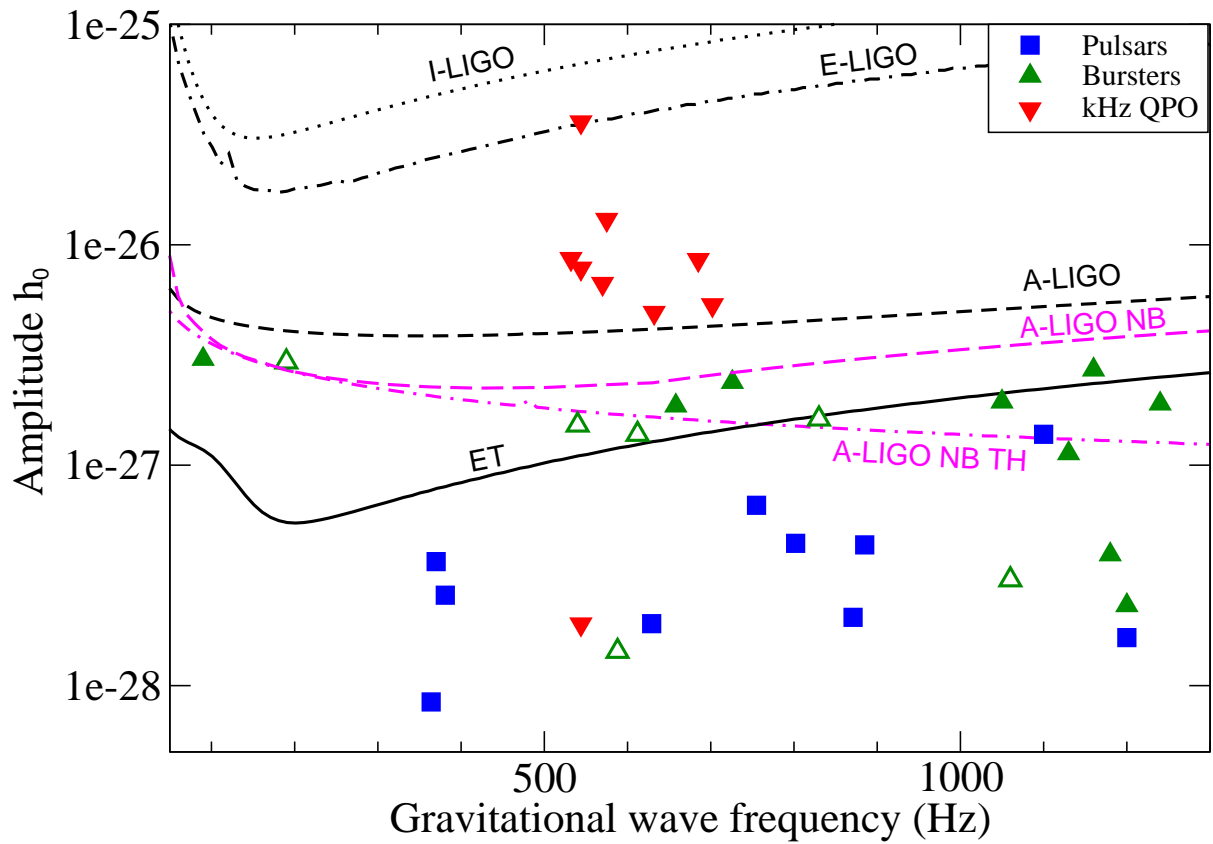}
\end{center}
\caption{Best case detectability for the mountain scenario for $T_{\rm
    obs} = 2$ years, balancing the long-term average flux, and assuming
  that all parameters are known (single template search).  The
  bursters are divided into two groups:  those for which the frequency is
  confirmed (filled) and those for which the frequency requires
  confirmation (open), see Section \ref{stars}.   The
  frequency at which the kHz QPO symbols appear is derived from the centre of the
  measured range of separations:  the predicted amplitude would be higher if the frequency were lower, and vice versa.  We show detectability thresholds for
  Initial LIGO (I-LIGO), Enhanced LIGO (E-LIGO), Advanced LIGO (A-LIGO), and the Einstein
  Telescope (ET).  We also show two detectability curves for
  Advanced LIGO Narrow Band: the expected envelope for the narrow
  band detector that includes all sources of noise (A-LIGO NB), and a curve showing
  only the thermal noise floor (A-LIGO NB TH).}
\label{mountpers}
\end{figure*}

Figure \ref{mountpers}  compares the predicted and
detectable amplitudes for this best-case scenario using the long-term flux
average derived in Section \ref{stars} and summarised in Table \ref{sdata}, and assuming that each NS
is in perfect spin balance (so that we can neglect spin derivatives), with
gravitational wave torque balancing that of accretion.
We assume that the spin frequency for the kHz
QPO sources lies in the middle of the known range of separations, and
take $T_{\rm obs}=2\,$years as a reference value.

Figure \ref{mountpers} naively implies that the kHz sources are the
most easily detected sources. However, they offer a specific
challenge, as we do not know many of their spins or orbital
parameters.  Detecting the GW signal requires a knowledge of the GW
phase evolution which depends crucially on the orbital parameters.
Ignorance or inaccurate knowledge of the orbit could then
require a search over a significant number of parameters and as we shall
see, this can have a dramatic impact on the sensitivity of the GW
search.  Hence assessing the detectability of GWs requires a more
careful description of the mechanics of the gravitational wave data
analysis process.  Most of this paper is therefore focused on clearly
assessing the data analysis challenges for present and future
generations of GW detectors that takes into account the limitations
imposed by incomplete astrophysical knowledge and finite computational
resources.

\section{Accreting Neutron Star Properties}
\label{stars}

In this Section we summarize our knowledge of the relevant search parameters for
all accreting NSs in LMXBs where there is some estimate of
the spin: the accreting millisecond pulsars, the burst oscillation
sources, and the kHz QPO sources. This sample contains both atoll and
Z sources, a classification determined by the spectral and timing properties
\citep{has89, mun02b, gie02,vand06}.   As explained in Section \ref{overview} we require flux histories, spin frequencies and orbital parameters, some of which  are measured, whilst others can only be estimated.  We also need to gauge the uncertainty on each quantity.  This sets the required search parameter space, and in
Section \ref{detect} we use this information to compute the
number of signal templates required for each source. 

This Section is rather lengthy, as we give full details of the provenance of all of the values used in our study.  There are two main reasons for this: firstly, to make clear the link between the astronomical observations and the consequences for gravitational wave searches. Secondly, many of the values that we derive involve assumptions, or draw on old or uncertain measurements: where this is the case we wanted to make it explicit, in order to drive future astrophysical modelling and observations. The information in this Section should also be a useful resource for anyone intending to carry out a search for gravitational waves from these objects, or for future detectability studies.   Readers who are not concerned with the details of the source properties and uncertainties can however skip this Section, and will find the key results summarized in Tables \ref{sdata} - \ref{khznohit}.    

\subsection{Constraining Fluxes and Accretion Rates}

The observed flux, $F$, sets the expected GW signal strength (Equation
\ref{mount_amp}).  For persistent sources we record only the long-term
average flux, $F_{\rm av}$, whereas for  transient sources we also record the outburst flux, $F_{\rm ob}$ (averaged over outbursts and quiescent periods).
Bolometric outburst fluences have been computed for the accreting millisecond pulsars,
but for the majority of sources this is not the case.  For the remaining sources we
determine $F$ using data from the RXTE All Sky Monitor (ASM),
which provides a near-continuous history of source activity from 1996 to the present
in the 2-10 keV band \citep{lev96}. Converting the 2-10 keV flux to a
bolometric flux requires detailed spectral modelling. For most
burst oscillation sources we use the results of \citet{gal07}: these authors carry out
spectral modelling using the pointed RXTE PCA data (2.5-25 keV) to estimate
bolometric flux. We compare the calculated fluxes to the ASM countrate
at the time of the observation to establish this relationship. For the transient sources
we only include in our integrated ASM histories the times when the source is in outburst  and detectable above a $3\sigma$ level. This avoids contamination from other sources in the field which would overestimate the long-term flux.  This method of estimating flux history introduces some errors, since spectral shapes (and the correction from 2-10 keV flux to bolometric flux) will change:  but it gives a reasonable estimate.  This uncertainty should however be borne in mind in Section \ref{disc} for those sources that are on the margins of detectability.

We also record position and distance.  Source position must be known
to a certain degree of precision for the long folds that
this type of analysis necessitates (Section \ref{subsec:templates}).  We have listed the most
accurate and up to date position known for the X-ray source or
its optical, infra-red or radio counterpart.  Source distance, which
is relevant for the emission modelling in Section \ref{detect}, can be
estimated in several different ways. Only for the closest source in our sample, Sco
X-1, can the distance be measured via parallax. For the
other sources different methods are used: location of
the source in a globular cluster; the presence of radius expansion
X-ray bursts (where luminosity reaches the Eddington
limit\footnote{The Eddington limit depends on the composition.
 Sometimes composition can be inferred from burst properties, but
 this is not always the case, leading to additional uncertainty.  See
 also \citet{gal08}.});
inferences about mass transfer from the long-term X-ray flux, assuming
that the binary orbit evolves due to gravitational wave emission;
absorption and spectral modelling.

We give details of pulsar frequency, burst oscillation frequency, and
twin kHz QPO separation, as measured with RXTE's Proportional Counter
Array (PCA). For sources with only a burst oscillation frequency, we
assume that $\nu_s$ lies within $\pm 5$ Hz of the burst oscillation frequency $\nu_b$.
For the kHz QPO sources we assume that the spin lies within the reported
range of kHz QPO separations.  We review both of these
assumptions in more detail in later Sections.

\subsection{Orbital Uncertainties}\label{sec:uncertainties}

 We must know (or presume) the orbital parameters in order to fold long stretches of gravitational wave data, most importantly, the orbital period $P_{\rm orb}$ and eccentricity, $e$. We also need a reference time within the orbit. Depending on the original reference we quote either $T_{90}$, the time of inferior conjunction of the companion star, or  $T_{\rm asc}$, the time of ascending node, when the Doppler shifted
frequency of the neutron star is at its lowest.  Note that $T_{90} =
T_{\rm asc} + P_{\rm orb}/4$.  The third parameter is the projected
semi-major axis of the neutron star, $a_{\rm x} \sin i$, which we
denote $a_{\rm p}$.  Depending on the measurement,
we may instead quote the amplitude of the projected orbital velocity
of the neutron star, $v_{\rm x} \sin i$ (referred to as K1
in the optical literature).  The two quantities are related by $v_{\rm
  x} \sin i = 2\pi a_{\rm x} \sin i/P_{\rm orb}$.  The orbital parameters are
measured directly for the accreting millisecond pulsars via
X-ray timing. The situation is more challenging for the
non-pulsing and intermittent sources.

Some high inclination systems show eclipses in the X-ray
lightcurve, providing both $P_{\rm orb}$ and $T_{\rm asc}$.  There are also
systems that show dips rather than full eclipses: the dips occur when the
NS is obscured by the bulge where the mass stream from the donor star joins the accretion disk \citep{whi82}. Dips certainly tell us $P_{\rm orb}$, and restrict the
inclination to lie in the range $60^\circ-75^\circ$.  What they do not
necessarily yield is $T_{\rm asc}$: in the two systems that show both dips and
eclipses, dips occur at various offsets from the eclipse
times \citep{com84, par86, mot87, sma92}.  Detailed modelling is
therefore required to determine the relationship between $T_{\rm asc}$ and the dip time
$T_{\rm dip}$. Throughout this Section we list the most
recent orbital ephemeris for each system.  In several cases the ephemerides are sufficiently out of date that we should consider $T_{\rm asc}$ to be totally
unconstrained.  However, we presume that
it would be straightforward to obtain a new ephemeris, with an error no
worse that that of the existing measurement.  We therefore use
all existing measurements in our initial assessment of detectability.

The orbital parameters can also be measured in wavebands other than
the X-ray, particularly the optical.  A number of systems show
photometric variability at the orbital period.  Maximum optical
light occurs when the NS is at inferior conjunction and we observe reprocessed
 emission from the heated face of the donor star, yielding both
$P_{\rm orb}$ and $T_{\rm asc}$, but not the projected semi-major
axis.  An alternative method that can provide all three
pieces of information is phase-resolved optical spectroscopy.  LMXBs
exhibit many emission lines, some from heated face of the donor
star \citep{ste02}, others from the accretion disk very close to the
compact object. By measuring the orbital Doppler
shifts of these lines, and using techniques such as Doppler tomography to check the
emission location, major progress has been made in computing
orbital parameters. Additional constraints on the system are possible
(assuming Roche lobe overflow and tidal locking) if one can detect
rotational broadening of absorption lines from the
donor star \citep{hor86, cas98}.

Unfortunately, for many of the systems of interest, one or all of the
orbital parameters are unknown.  We can however still place various
constraints on the systems to reduce the number of templates
required.  For those systems with measured orbital periods, for
example, we can assume that the donor star fills its Roche lobe. This fixes
$\bar{\rho}_{\rm d}$, the mean density of the donor star:
\begin{equation}
\bar{\rho}_{\rm d} \approx 110 
\left(\frac{P_\mathrm{orb}}{1 \mathrm{hr}}\right)^{-2}  {\rm ~~g~cm^{-3}}.
\label{roche}
\end{equation}
The range of donor types with this $\bar{\rho}$
limits the donor mass $M_{\rm d} = m_{\rm d} M_\odot$.  LMXB donors
include main sequence stars, evolved hydrogen-burning
stars, helium-burning stars, and brown or white dwarfs \citep{pod02}.
For $P_{\rm orb} \gsim 10$ hours, $\bar{\rho}_{\rm d} \lsim 1$
g cm$^{-3}$, the companion must be an evolved
hydrogen-burning star such as a subgiant.  For ultracompact systems
with $P_{\rm orb} \lsim 80$ minutes, the companion must be a white
dwarf, a helium star, or a highly evolved helium-rich secondary
\citep{rap84, del03, nel06}.  For intermediate orbital periods more
options are possible, maximum
donor mass being by the main sequence star - but evolved, less
massive companions or even brown dwarfs may be possible \citep{tou96,
  cha00, cha00a,bilcha}\footnote{\citet{rap84} showed that helium-rich donor
  stars would have higher masses for a given $P_{\rm orb}$ than
  expected for main sequence stars.  However, the binary evolution
  simulations of \citet{pod02} did not
  generate any such stars at intermediate orbits, so we neglect this
  possibility unless there is overwhelming evidence for the presence
  of helium-rich material (from X-ray bursts, for example).}.
Additional constraints on donor properties may come from X-ray burst
properties or spectral type:  more evolved, lower mass, donors will have later
spectral types than the main sequence star with the same
$\bar{\rho}_{\rm d}$ \citep{bara00, kol01}.  Having finally established
the range of likely donor masses, and knowing that the
NS mass  $M_{\rm x} = m_{\rm x} M_\odot$ is in the
 range $1.2-2.4 M_\odot$ \citep{lat07}, we then estimate $a_{\rm x} \sin i$ using
\begin{equation}
a_{\rm x}\sin i = 1.174  \left(\frac{P_{\rm orb}}{1{\rm
      hr}}\right)^{2/3}\frac{m_{\rm d}}{(m_{\rm d} + m_{\rm x})^{2/3}}
\sin i {\rm ~~lt-s~}.
\label{axsini}
\end{equation}
For systems that do not show dips or eclipses ($i \lsim 60^\circ$),
and where there is no other
limit on inclination, Equation (\ref{axsini})  gives only an upper limit.  For dipping
systems we assume $60^\circ \lsim i \lsim 75^\circ$, giving both
upper and lower limits.  For eclipsing systems an additional constraint comes from the half-angle of the X-ray eclipse $\theta_{\rm x}$ \citep{bra83}:
\begin{equation}
\sin i = \frac{1}{\cos\theta_{\rm x}} \left[ 1 - \left( 0.38 -
    0.2\log\frac{m_{\rm x}}{m_{\rm d}}\right)^2\right]^{1/2}.
\label{eclipse}
\end{equation}

For systems with no measured orbital period, we make
the standard LMXB assumption that $m_{\rm d}/m_{\rm x} < 0.8$.
Time-averaged accretion rate (as estimated from the X-ray
flux) can give general constraints if we assume that mass transfer is
driven by gravitational radiation:  however magnetic braking may also
play a role in mass transfer and the contribution is hard to
quantify.  The conditions required for the system to be persistent or
transient at the inferred accretion rate were also considered
\citep{kin96, dub99, int07}:
unfortunately for most of the systems in our study this added very
little in the way of tighter constraints.   For systems
where there are no better constraints on orbital period, we will assume that
$P_{\rm orb}$ lies between 10 minutes and 240 hours.  The presumed
lack of eclipses in such systems sets $i < 60^\circ$, so from Equation
(\ref{axsini}) we obtain an upper limit on $a_{\rm x} \sin i$ of 22.6 lt-s.

\subsection{Accreting millisecond pulsars}
\label{pulsar}

The ten accreting millisecond pulsars are the only systems where we clearly know the NS
spin frequency $\nu_s$. All are
transient, but their pulsation characteristics differ.  Seven of these
systems, which have short outbursts (weeks) and long periods of
quiescence (years), show persistent pulsations throughout their outburst
phases.  The other three systems are rather different.  HETE
J1900.1-2455 went into outburst in 2005 and is still active,
but only showed pulsations during the first two months of the
outburst.  The other two systems have shown pulsations only
intermittently during outburst.   We now discuss the sources in order of spin frequency, from highest to lowest.

\subsubsection{IGR J00291+5934 ($\nu_s$ = 599 Hz)}

This source was discovered in 2004 during a 13 day outburst
with total fluence (0.1-200 keV) $1.8\times 10^{-3}$ ergs cm$^{-2}$
\citep{gal05}. We assume that this is a typical outburst.
The RXTE ASM history suggests a recurrence
time of 3 years, giving $F_{\rm av} = 1.8\times 10^{-11}$
ergs cm$^{-2}$ s$^{-1}$.  The most accurate position for the source, given by the
optical/near infra-red counterpart, is RA =
$00^\mathrm{h}29^\mathrm{m}03\fs05 \pm 0\fs01$, Dec = $+59^\circ 34^\prime 18\farcs93 \pm
0\farcs05$ \citep{tor07}. This position accords with the earlier
optical position of \citep{fox04} and the Chandra X-ray
\citep{pai05}, but is offset by
$3.2\sigma$ in RA from the radio position
\citep{rup04}.  Mass transfer arguments suggest a minimum distance of
$\approx 4$ kpc \citep{gal05}, and the lack of bursts implies a
maximum distance of 6 kpc \citep{gal06a}.  There have been three
studies of the spin and orbital parameters.  \citet{gal05} used data
from the start of the outburst and the optical position of \citet{fox04}.
\citet{fal05b, bur07} used additional data from later in the outburst
and the radio position\footnote{Although an offset position can generate an apparent
  spin derivative, the offset is not sufficiently large to
account for the reported values of $\dot{\nu}_s$.}.  The studies agree on the orbital
parameters:  $P_{\rm orb} =  8844.092 \pm 0.006$ s, $T_{90} = 53345.1875164 \pm
4\times 10^{-7}$ MJD (TDB), $a_{\rm x}\sin i = 64.993 \pm 0.002$ lt-ms ($1\sigma$
uncertainties, Galloway, private communication), and $e < 2\times10^{-4}$
(3$\sigma$ upper limit).  There are however small differences in spin
parameters. In our initial assessment of detectability we use the
most recent values of \citet{bur07}:  $\nu_s
= 598.89213053 \pm 2\times 10^{-8} $ Hz,
$\dot\nu_s = (8.5 \pm 1.1) \times 10^{-13} $ Hz/s at epoch MJD 53346.184635
($1\sigma$ uncertainties).

\subsubsection{Aql X-1 (1908+005) ($\nu_s$ = 550 Hz)}

Aql X-1 is a transient atoll source with quasi-regular outbursts,
$\sim 10$ in the RXTE era.  Using the spectral modelling of
\citet{gal07} we set 5 ASM cts/s = $3.1\times 10^{-9}$ ergs cm$^{-2}$ s$^{-1}$
(bolometric).  The long-term average ASM countrate is 1.8 cts/s, while
the average countrate during the bright 45 day outburst of late 2000
was 34 cts/s.  The most recent position, from VLA observations of
the radio counterpart, is RA
= $19^\mathrm{h} 11^\mathrm{m} 16\fs01 \pm 0\fs03$, Dec = +$00^\circ 35^\prime
06\farcs7 \pm 0\farcs3$ (J2000, $1\sigma$ error bars)
\citep{rup04b}. \citet{gal07}
use PRE bursts to estimate a distance of  $3.9 \pm
0.7$ kpc or $5.0 \pm 0.9$ kpc depending on composition. \citet{case07} have reported the
detection of intermittent accretion-powered pulsations at $550.2745 \pm
0.0009$ Hz.  The orbital period,
determined from the
optical lightcurve in outburst, is $P_\mathrm{orb} = 18.9479 \pm
0.0002$ hours \citep{che98} (the quiescent period agrees at the 0.02\%
level).  The time of minimum optical light
determined by \citet{gar99} gives $T_{90} = 2450282.220 \pm 0.003$ HJD. Both time of minimum light and
orbital period were confirmed in later analysis by \citet{wel00}.
\citet{cor07} attempted to measure the projected orbital velocity of
the neutron star directly using phase-resolved optical spectroscopy:
they report a preliminary value of $v_\mathrm{x} \sin i = 68 \pm 5$
km/s.  Although the fit quality is not good and this result
requires confirmation, we will use this value in
our preliminary assessment of detectability.

\subsubsection{SAX J1748.9-2021 ($\nu_s$ = 442 Hz)}

This is a transient source located in the globular cluster NGC 6440,
with three outbursts in the RXTE era.  Using the spectral modelling of
\citet{gal07} we set 13 ASM cts/s = $5\times 10^{-9}$ ergs cm$^{-2}$ s$^{-1}$
(bolometric).  The long-term average ASM countrate is 0.2 cts/s, while
the average countrate during the bright 60 day outburst of 2005
was 10 cts/s.  The position of
the optical counterpart \citep{ver00} (which accords with the Chandra
position of \citet{poo02}) is RA = $17^\mathrm{h} 48^\mathrm{m}
52\fs14$, Dec = -$20^\circ 21^\prime
32\farcs6$ (J2000), with an error of $0\farcs5$.  \citet{kuu03}
estimate distance to the cluster as
$8.4^{+1.5}_{-1.3}$ kpc, in good agreement with the
value of $8.1 \pm 1.3$ kpc derived from PRE bursts \citep{gal07}.
\citet{alt07} discovered intermittent accretion-powered pulsations
 in two outbursts from this source (see also
\citet{gav07}). \citet{pat08} have now carried out a detailed
phase-connected timing study to determine the spin and orbital
parameters.  They find $\nu_s = 442.36108118 \pm 5\times 10^{-8}$ Hz, $P_\mathrm{orb} = 8.76525
\pm 3\times 10^{-5}$ hours, $T_{\rm asc} =
52191.507190 \pm 4\times 10^{-6}$
MJD/TDB, and $a_\mathrm{x} \sin i = 0.38760 \pm 4\times 10^{-5}$ lt-s, with $e <
1.3\times 10^{-4}$ (1$\sigma$
uncertainties and upper limits).

\subsubsection{XTE J1751-305 ($\nu_s$ = 435 Hz)}

The 2004 outburst of this source, which lasted $\approx$ 10 days, had
total fluence (2-200 keV) $(2.5\pm 0.5)\times 10^{-3}$ ergs cm$^{-2}$
\citep{mar02}. Assuming that this outburst is typical, the mean recurrence
time of 3.8 years yields $F_{\rm av} = 2 \times 10^{-11}$
ergs cm$^{-2}$ s$^{-1}$.  The most accurate position
for the source, measured with Chandra, is RA = $17^\mathrm{h} 51^\mathrm{m}13\fs49 \pm 0\fs05$, Dec = $-30^\circ 37^\prime
23\farcs4 \pm 0\farcs6$ (J2000), where the
uncertainties are the 90\% confidence limits \citep{mar02}. Mass
transfer arguments yield a lower limit on the distance of 6 kpc
\citep{mar02}.  The most recent timing study by \citet{pap08}, using
the Chandra position, gives $P_{\rm orb}  = 2545.342 \pm 0.002$ s, $T_{\rm asc} =
52368.0129023 \pm 4\times 10^{-7}$ MJD (TDB), $a_{\rm x} \sin i =
10.125 \pm 0.005 $ lt-ms,
$e < 1.3 \times 10^{-3}$ (90\% confidence level uncertainties and
upper limits). The spin
parameters are $\nu_s = 435.31799357 \pm 4\times 10^{-8} $ Hz, and the
study also suggests that $\dot{\nu}_s$ may be non-zero.

\subsubsection{SAX J1808.4-3658 ($\nu_s$ = 401 Hz)}

\citet{gal06} analyse the five known outbursts from this source, and
find a mean outburst fluence of $(6.0\pm 0.2)\times 10^{-3}$
ergs cm$^{-2}$ (0.1-200
keV).  In estimating transient gravitational wave signal we use an outburst
duration of 20 days:  we neglect the extended flaring phase often seen
at the end of outbursts in this source, since emission during this
phase is at a much lower level. Mean recurrence
time is 2.2 years, giving $F_{\rm av} = 8.6\times 10^{-11}$ ergs cm$^{-2}$ s$^{-1}$.  The most
recent and precise position, for the optical counterpart, is RA =
$18^\mathrm{h} 08^\mathrm{m} 27\fs62$, Dec =
-$36^\circ 58^\prime 43\farcs3$ (J2000), with an uncertainty of
 $0\farcs15$ \citep{har07}.  \citet{gal06} have derived a distance of
3.4-3.6 kpc using both mass transfer arguments and burst
properties.  Using the refined optical position, \citet{har07} report the
following values for the orbital parameters:  $P_{\rm orb} =
7249.156961 \pm 1.4\times 10^{-5}$ s at
time $T_{\rm asc} = 52499.9602477 \pm 1.0\times 10^{-6}$ MJD (TDB),
$\dot{P}_{\rm orb} = (3.48 \pm 0.12) \times 10^{-12}$ Hz/s,
$a_{\rm x} \sin i = 60.28132 \pm 2.4\times 10^{-4} $ lt-ms ($1\sigma$
errors).  The eccentricity $e < 0.00021$ (95\% upper limit).
The spin rate was tracked across multiple outbursts, and is given
relative to a reference frequency $\nu_0 = 400.975210$ Hz.  In the
1998 outburst, $\nu_s - \nu_0 = 0.371 \pm 0.018~\mu$Hz, with $\dot{\nu}_s$
 in the range (-7.5, 7.3)$\times 10^{-14}$ Hz/s  (95\% confidence
limits). In the 2000 outburst, $\nu_s - \nu_0  = 0.254 \pm 0.012~\mu$Hz,
with $\dot{\nu}_s$ in the range (-1.1, 4.2)$\times 10^{-14}$ Hz/s.  In
the 2002 outburst, $\nu_s - \nu_0 = 0.221 \pm 0.006~\mu$Hz, with
$\dot{\nu}_s$ in the range (-1.2, 2.5) $\times 10^{-14}$ Hz/s.  In the
2005 outburst, $\nu_s - \nu_0 = 0.190 \pm 0.015~\mu$Hz with $\dot{\nu}_s$
in the range (-0.5, 2.3)$\times 10^{-14}$ Hz/s.  Fitting the frequency
evolution across all four outbursts gives $\dot{\nu}_s = (5.6
\pm 2.0)\times 10^{-15}$ Hz/s.  In our initial assessment of
detectability, we use the spin solution for the 2002 outburst.

\subsubsection{HETE J1900.1-2455 ($\nu_s$ = 377 Hz)}

This source was first detected in June 2005, and has
remained in outburst ever since \citep{deg07}.  \citet{gal06a} and \citet{gal07a}
  report an average (bolometric) outburst flux of $\approx 9\times
  10^{-10}$ ergs cm$^{-2}$ s$^{-1}$.  We assume an outburst duration of 2 years,
  and (given that no previous outbursts are known) a recurrence time
  of at least 10 years \citep{kaa06}, yielding
  $F_{\rm av} \approx
  2\times 10^{-10}$ ergs cm$^{-2}$ s$^{-1}$.  The position of the optical counterpart,
is RA = $19^\mathrm{h} 00^\mathrm{m} 08\fs65$, Dec =
-$24^\circ 55^\prime 13\farcs7$
(J2000), with an estimated uncertainty of $0\farcs2$ \citep{fox05}.  The
distance estimated by \citet{gal07} using RXTE observations of
radius expansion bursts, is $4.7 \pm 0.6$ kpc.  This accords
with the earlier estimate of 5 kpc made by \citet{kaw05} using HETE
burst data.  Timing analysis by \citet{kaa06}, using RXTE data from
June 16 - July 7 2006, resulted in the following orbital parameters:  $P_{\rm
  orb} = 4995.258 \pm 0.005$ s, $T_{90} = 53549.145385 \pm 7\times 10^{-6}$ MJD
(TT), $a_{\rm x} \sin i = 18.41 \pm 0.01$ lt-ms, $e < 0.002$.  The spin was
$\nu_s = 377.296171971 \pm 5\times 10^{-9}$ Hz.  All errors and upper
limits are $1\sigma$
uncertainties.   On July 8th there was an apparent jump in spin rate
of $\Delta \nu_s/\nu_s \sim
  6\times 10^{-7}$,
  to $\nu_s = 377.291596 \pm 1.6\times 10^{-5}$ Hz.  Thereafter pulsations ceased
  and spin has not been
tracked since despite the fact that the source has remained in
outburst \citep{gal07a}. In our initial assessment of detectability we
do not take into account the apparent jump in spin.

\subsubsection{XTE J1814-338 ($\nu_s$ = 314 Hz)}

This pulsar has had only one outburst in the RXTE era, lasting
$\approx 50$ days.  \citet{gal06a} estimate a bolometric outburst
fluence of $(3.0 \pm 0.1) \times 10^{-3}$ ergs cm$^{-2}$ s$^{-1}$. Given a
recurrence time of at least 7.5 years, this yields an upper limit on
$F_{\rm av}$ of $1.3\times 10^{-11}$ ergs cm$^{-2}$ s$^{-1}$.  The position
derived from X-ray and optical
spectroscopy is RA =  $18^\mathrm{h} 13^\mathrm{m} 39\fs04$,
Dec = -$33^\circ 46^\prime 22\farcs3$ (J2000), 90 \%
confidence error circle of $0\farcs2$ \citep{kra05}.  These
authors use X-ray spectroscopy to infer a minimum distance of 3.8 kpc.
The upper limit on the distance, derived from burst properties,
is $8.0 \pm 1.6$ kpc \citep{str03}.  Timing analysis by \citet{pap07},
using data from the whole 2003
outburst, leads to the following orbital parameters: $P_{\rm orb}  =
15388.7229 \pm 0.0002$ s, $T_{\rm asc} = 52797.8101698 \pm 9\times
10^{-7}  $ MJD (TDB),
$a_{\rm x} \sin i = 390.633 \pm 0.009$ lt-ms, $e < 2.4\times 10^{-5}$ ($3\sigma$ upper
limit).  The associated spin parameters are $\nu_s  = 314.35610879 \pm
1\times 10^{-8}$ Hz, $\dot{\nu}_s = (-6.7 \pm 0.7)\times 10^{-14}$ Hz/s.
Uncertainties are
at the 90\% confidence level.

\subsubsection{XTE J1807-294 ($\nu_s$ = 191 Hz)}

There has been only one recorded outburst from this source, for which
\citet{gal06a} computes a bolometric
fluence of $(3.1\pm 0.2)\times 10^{-3}$ ergs cm$^{-2}$. For a recurrence time of at least 7.1 years,
$F_{\rm av}$ is at most $1.4\times 10^{-11}$ ergs cm$^{-2}$ s$^{-1}$.  In
estimating transient gravitational wave signal we use an outburst
duration of 50 days, neglecting the prolonged low flux tail at the
end of the outburst.  The most
accurate position,
measured with Chandra, is RA =
$18^\mathrm{h} 06^\mathrm{m} 59\fs8$,
Dec = -$29^\circ 24^\prime 30^{\prime\prime}$ (J2000), with an
uncertainty due to systematic errors of $1^{\prime\prime}$
\citep{mar03c}. Using mass transfer estimates,
\citet{gal06a} derives a lower limit on
the distance of 4.7 kpc.  Determination of spin and orbital parameters
in this source is
complicated by extreme variations in pulse profile.  A recent study by
\citet{rig07} finds $P_{\rm
  orb} = 2404.41665 \pm 0.00040$s, $a_{\rm x} \sin i =
4.819\pm 0.004$ lt-ms, $T_{\rm asc} = 52720.675603 \pm 6\times 10^{-6}$
MJD/TDB and $\nu_s = 190.62350694 \pm 8\times 10^{8}$ Hz (1$\sigma$ uncertainties). The eccentricity $e
< 0.0036$ ($2\sigma$ upper limit).  \citet{cho07} report similar
values apart from for $T_{\rm asc}$, where the values found by the two
studies differ by more than the quoted uncertainties.  Both
\citet{cho07} and \citet{rig08} suggest a non-zero $\dot{\nu}_s \sim
10^{-14} - 10^{-13}$ Hz/s in outburst.  In our initial assessment of
detectability we use the values and uncertainties of \citet{rig07},
although clearly the `true' uncertainty on $T_{\rm asc}$ is larger.

\subsubsection{XTE J0929-314 ($\nu_s$ = 185 Hz)}

This source has had one outburst in the RXTE era, for which
\citet{gal06a} estimates a bolometric fluence of $(5.4 \pm 0.3)\times
10^{-3}$ ergs cm$^{-2}$.  Outburst duration, which we assume to be
typical, was $\approx 60$ days.  Given a recurrence time of at least
6.3 years, $F_{\rm av}$ is at most $2.7\times 10^{-11}$ ergs cm$^{-2}$ s$^{-1}$.
The most accurate position, given by the optical
counterpart, is RA = $9^\mathrm{h} 29^\mathrm{m} 20\fs19$, Dec =
-$31^\circ 23^\prime 03\farcs2$ (J2000) with error circle
$0\farcs1$ \citep{gil05}.  \citet{gal06a} uses mass transfer
and recurrence time estimates to infer a lower limit to the distance
of 3.6 kpc (revising an earlier value of 5 kpc in \citet{gal02}).
Timing analysis by \citet{gal02}, using an earlier optical
position from \citet{gil02}, leads to the following orbital parameters:
$P_{\rm orb} = 2614.746 \pm 0.003$ s, $T_{90} = 52405.49434 \pm
1\times 10^{-5}$
 MJD (TDB), $a_{\rm x} \sin i =  6.290 \pm 0.009$ lt-ms.  The
 associated spin parameters are $\nu_s =
185.105254297 \pm 9\times 10^{-9}$ Hz, $\dot{\nu}_s = (-9.2 \pm 0.4)
\times 10^{-14}$ Hz/s.  All errors are $1\sigma$ uncertainties
(Galloway, private communication).   The eccentricity $e < 0.007$
 ($2\sigma$ limit).

\subsubsection{SWIFT J1756.9-2508 ($\nu_s$ = 182 Hz)}

This source has had one outburst in 2006 lasting 13 days, with total
fluence (1-10000 keV) $(4.5 \pm 0.8) \times 10^{-4}$ ergs cm$^{-2}$.  No
previous outbursts are known (although there are gaps in coverage),
but this suggests a recurrence time of at least 10 years. The best position for
the source, from SWIFT, is RA = $17^\mathrm{h} 56^\mathrm{m} 57\fs35$, Dec =
-$25^\circ 06^\prime 27\farcs8$ (J2000), with uncertainty $3\farcs5$.
The distance is not well constrained, but is thought to be $\approx 8$ kpc.
The orbital parameters are $P_{\rm orb} = 3282.104 \pm 0.083$ s,
$T_{\rm asc} = 54265.28707 \pm 6\times 10^{-5}$ MJD (TDB), $a_{\rm x}
\sin i = 5.942 \pm 0.027$ lt-ms, $e < 0.026$ (95\% upper limit).   The
spin rate is $\nu_s = 182.065804253~\pm~7.2\times 10^{-8}$ Hz, with
$\dot{\nu}_s < 1\times 10^{-12}$ Hz/s.  All errors are 90\% confidence,
upper limits are 95\% confidence.  All information on this source is
taken from \citet{krim07}.

\subsection{Burst oscillation sources}

We list all sources for which burst oscillations have been reported,
in order of decreasing burst oscillation frequency $\nu_b$.  For some sources,
oscillations have been detected at the same frequency in
multiple bursts: these results can be regarded as secure.  In some
cases, however, oscillations have only been seen in a single burst.
Given the number of NSs whose bursts have
now been searched for oscillations (a factor not included in quoted
statistical significances), these results should be regarded as
tentative until confirmed in a second burst.  We include them in our
survey (marked with an asterisk) since if they turn out to be promising gravitational wave
sources this would provide added impetus to confirm or alternatively
rule out the candidate burst
oscillation detection.  

\subsubsection{XTE J1739-285 ($\nu_b$ = 1122 Hz)*}

This transient atoll source has had four outbursts in the RXTE era.
Using the spectral modelling of \citet{kaa07} we set 10 ASM cts/s =
$1.4\times 10^{-9}$ (2-20 keV) and apply a bolometric correction
factor of 1.34 (the average factor found by \citet{gal07} for burst
sources).  The long-term average ASM countrate is 0.3 cts/s, and the
average countrate during the bright 200 day outburst of 2005 was 4
cts/s.   The most precise source
position, measured by Chandra, is RA =
$17^\mathrm{h} 39^\mathrm{m} 53\fs95$, Dec = -$28^\circ 29^\prime
46\farcs8$ (J2000),
with a 90\% error radius of $0\farcs6$ \citep{kra06}. The absence of
an optical/IR counterpart sets an upper limit on distance of 12 kpc
\citep{tor06}.  The absence of PRE bursts sets a more
stringent upper limit of 10.6 kpc \citep{kaa07}.  A candidate 1122 Hz burst oscillation
was detected
in part of one burst recorded by RXTE
during the 2005
outburst.  The orbital parameters of the source are unknown, and there
are no measured constraints on the properties of the companion.

\subsubsection{4U 1608-522 ($\nu_b$ = 620 Hz)}

This transient atoll source has had several outbursts during RXTE's
lifetime, and seems to be active at a low level (ASM countrate $> 3
\sigma$) even when not in outburst.  Using the spectral modelling of
\citet{gal07} we set 70 ASM cts/s = $4.8\times 10^{-8}$ ergs cm$^{-2}$ s$^{-1}$
(bolometric).  The long-term average ASM countrate is 3.6 cts/s, while
the average countrate during the bright 100 day outburst in 2005 was
30 cts/s.  The best position, from the optical
counterpart, is RA = $16^\mathrm{h} 08^\mathrm{m} 52\fs2$, Dec =
-$52^\circ 17^\prime 43^{\prime\prime}$ (B1950), with
errors of $\pm 0\farcs5$ \citep{gri78}.  No more up
to date position is available.  Source distance, derived under the
assumption that radius expansion bursts reach the Eddington limit for
pure helium, is $4.1 \pm 0.4$ kpc \citep{gal07}.  Burst oscillations
at $\approx 620$ Hz have
been detected in multiple
bursts \citep{har03, gal07}.  The last (unsuccessful) search for persistent
pulsations from this source, using data
from 1989-1991, searched only up to 512 Hz \citep{vau94}. The orbital
parameters are
not known, although a number of tentative periodicities have been
reported \citep{loc94, wac02}.   The spectral type of the companion is that of a late F/early G type
main sequence companion star, but would also match that of a more
evolved K/M type star \citep{wac02}. The mean density range for donors
of this type suggests that the
orbital period lies in the range 10 - 125 hours.  Given that the system is non-eclipsing we assume $i <
60^\circ$ and hence obtain, for the assumed orbital periods, $a_{\rm
  x} \sin i < 12.3$ lt-s.

\subsubsection{SAX J1750.8-2900 ($\nu_b$ = 601 Hz)}

This weak transient atoll source has had two outbursts in the RXTE
era.  Following the spectral modelling of \citet{gal07} we set 8 ASM
cts/s = $3.4\times 10^{-9}$ ergs cm$^{-2}$ s$^{-1}$ (bolometric).  The long-term average ASM countrate is 0.08 cts/s, while the
average countrate during the brighter 110 day outburst in 1997 was 2.8
cts/s.  The most precise
position, from BeppoSAX,
is  RA = $17^\mathrm{h} 50^\mathrm{m} 24^\mathrm{s}$, Dec = -$29^\circ
02^\prime 18^{\prime\prime}$ (J2000), with a 99\% error radius of
$1^\prime$ \citep{nat99}.  The distance, estimated from radius
expansion bursts, is $6.79 \pm 0.14$ kpc \citep{gal07}. Burst
oscillations at 601 Hz have been detected in multiple bursts
\citep{kaa02, gal07}.  The orbital parameters are not
known, and there are no observations of the companion star.  

\subsubsection{GRS 1741.9-2853 (AX J1745.0-2855) ($\nu_b$ = 589 Hz)}

This is a transient source in the crowded Galactic
Centre, with three outbursts in the RXTE era. Unfortunately the fact
that the field is crowded means that
there is no reliable ASM flux history for this source.  The most
precise position for the source, derived by Chandra, is RA =
$17\mathrm{h} 45^\mathrm{m} 2\fs33$, Dec =
-$28^\circ 54^\prime 49\farcs7$ (J2000), with an uncertainty
of $0\farcs7$ \citep{mun03}.  Analysis of
radius expansion bursts suggests a distance $6.0 \pm 1.6$ kpc
(Eddington limit assuming cosmic abundances) or $8
\pm 2$ kpc (Eddington limit for pure He) \citep{gal07}. Burst oscillations at
589 Hz have been detected in 2 bursts
\citep{str97}\footnote{The bursts were originally attributed to a different
source, MXB 1743-29, due to source confusion in the crowded field}.
The orbital parameters are not known, and there is no information on
the properties of the companion star.

\subsubsection{4U 1636-536 ($\nu_b$ = 581 Hz)}

This is a persistent atoll source.  Using the spectral modelling of
\citet{gal07} we set 18 ASM cts/s = $8.4\times 10^{-9}$
ergs cm$^{-2}$ s$^{-1}$ (bolometric).  The long-term average ASM countrate is 10 cts/s. The best position for the
source, from the optical counterpart V801 Ara, is  RA =
$16^\mathrm{h} 36^\mathrm{m} 56\fs41$, Dec = -$53^\circ 39^\prime 18\farcs1$ (B1950),
with an error circle of less than $ 1^{\prime\prime}$ \citep{bra83}.
No more up to date position is available.   The distance, estimated
from a large
sample of radius expansion bursts, is $6.0 \pm 0.5$ kpc
\citep{gal06b}. Burst oscillations at $\approx 581$ Hz
have been seen in multiple bursts and a superburst \citep{str98, str02, gal07}.  The most recent
unsuccessful search
for persistent pulsations, using data from 1987, searched only up to
512 Hz \citep{vau94}.  The latest ephemeris, derived from
phase-resolved optical spectroscopy, gives
$P_{\rm orb} = 0.15804693 \pm
1.6\times 10^{-7}$ days and $T_{90} = 2452813.531 \pm 0.002 $ HJD \citep{cas06}.
\citet{aug98} set an upper limit on the orbital period derivative of
$|P_{\rm orb}/\dot{P}_{\rm orb}|\le 3\times 10^5$ years.  No tighter
limit has yet been reported.  Orbital Doppler shifts on burst
oscillations during a superburst lead to limits $90 < v_{\rm x} \sin i < 113$ km/s,
\citep{str02, cas06}.   The quoted range corresponds to varying the
reference phase of the ephemeris across the $\pm 1 \sigma$ range.

\subsubsection{X 1658-298 (MXB 1659-29) ($\nu_b$ = 567 Hz)}

This transient source, which has eclipses and dips, was active from
1976-1979, and then again from 1999-2001.  Using the spectral
modelling of \citet{gal07} we set 2.4 ASM cts/s = $6.7\times 10^{-9}$
ergs cm$^{-2}$ s$^{-1}$ (bolometric).   Over the most recent outburst, which lasted 870 days,
the average ASM countrate was 2.7 cts/s.  Assuming a recurrence time
of $\approx 23$ years, the long-term average ASM countrate is 0.3
cts/s.  The best position
for the source, from the optical counterpart
V2134 Oph, is RA =
$17^\mathrm{h} 02^\mathrm{m} 06\fs42$, Dec =
-$29^\circ 56^\prime 44\farcs33$ (J2000), with an uncertainty of
$0\farcs1$ \citep{wac98}. Assuming that the bright radius expansion bursts
reach the Eddington limit for pure
He, \citet{gal07} derive a distance of $12 \pm 3$ kpc. Burst oscillations at
$\approx 567$ Hz have been detected in
multiple bursts \citep{wij01b}.  The most up to date X-ray
ephemeris, by
\citet{oos01}, gives $P_{\rm orb} = 0.2965045746 \pm 3.4\times
10^{-9}$ days, and $T_{90} =
2443059.225826 \pm 0.000093$ JD/TDB.  The projected semi-major axis
has not been measured, but we can set bounds on it using the
constraints outlined at the start of this section.  For the above
orbital period, \citet{wac00} find X-ray eclipse
duration half-angles $6.34^\circ \pm 0.01^\circ$ (slightly lower than
the value reported by \citet{com84}). We can therefore use Equation
(\ref{eclipse}) to restrict $\sin i$.  If we assume Roche lobe
overflow, then from Equation (\ref{roche}) the mean density of the
donor star is 2.2 g/cm$^3$, suggesting a main sequence or evolved
companion.  Maximum donor mass occurs if the donor is
on the main sequence.  In this case,  maximum $M_d = 0.78
M_\odot$ and would be of spectral type K0 \citep{wac98}.  Whilst
the data are consistent with this spectral type, there is some
indication that the spectral type is later, suggesting a more evolved
(lower mass) companion with spectral type perhaps as late as M2
\citep{wac00}.  We
will therefore adopt a minimum companion mass of $0.1M_\odot$
\citep{bara00, pod02}.  For the neutron star we will consider masses
in the range $1.2 - 2.4 M_\odot$ \citep{lat07}.  These assumptions
suggest that $a_\mathrm{x} \sin i$ lies in the range  0.24--2.03 lt-s.

\subsubsection{A 1744-361 ($\nu_b$ = 530 Hz)*}

This transient dipping source has irregular outbursts, three in RXTE's
lifetime.  Based on the spectral modelling of \citet{bha06} we set 1.2
ASM cts/s = $3\times 10^{-10}$ ergs cm$^{-2}$ s$^{-1}$ (3 - 14 keV) flux and
apply a bolometric correction factor of 1.34 (the mean
correction factor for converting 2.5 - 25 keV flux to bolometric flux
found by \citet{gal07} for other burst sources).  The long-term average
ASM countrate is 0.16 cts/s, and during the bright 100 day outburst in
2003 the average countrate was 6 cts/s.  The
best position, for the radio counterpart, is RA = $17^\mathrm{h}
48^\mathrm{m} 13\fs148 \pm 0\fs014$, Dec =
-$36^\circ 07^\prime 57\farcs02 \pm 0\farcs3$ (J2000) \citep{rup03}.  This is within
the error circle of the Chandra position \citep{tor04}.  The lack of radius
expansion in the one burst detected by RXTE sets an upper limit on the
distance of $\approx 9$ kpc
\citep{bha06}.  These authors reported a burst oscillation at $\approx
530$ Hz during the rising
phase of the one burst detected by RXTE.  They also reported a
possible orbital
period of $97 \pm 22$ minutes, traced by dips in the X-ray lightcurve,
\citep{bha06}.  In our initial assessment we will assume that this is
indeed the orbital period.  In this case the assumption of Roche lobe
overflow (Equation \ref{roche}) gives a mean donor density of $28-70$
g/cm$^3$, consistent with a main sequence star or slightly evolved
donor.   The maximum donor mass (for a star on the main sequence) is
0.22 $M_\odot$.  For a dipping source we can assume an inclination in
the range $60-75^\circ$.  For neutron star masses in the range
$1.2-2.4 M_\odot$ we can therefore set an upper limit on $a_{\rm x}
\sin i$ of 0.2 lt-s.  The minimum donor mass, of $0.07 M_\odot$, is set by the most
evolved hydrogen burning star possible.  Together with the inclination constraint, this gives
a lower limit on $a_{\rm x} \sin i$ of 0.05 lt-s.

\subsubsection{KS 1731-260 ($\nu_b$ = 524 Hz)}

This transient atoll source was in outburst from 1988 (and possibly
earlier) until
2001. Based on the spectral modelling of \citet{gal07} we set 18 ASM
cts/s = $9.7\times 10^{-9}$ ergs cm$^{-2}$ s$^{-1}$ (bolometric).  The average
countrate while the source was still in outburst was 9 cts/s:
the long-term average countrate over the RXTE lifetime is 4 cts/s.  The most precise position, for the optical counterpart, is RA =$17^\mathrm{h} 34^\mathrm{m} 13\fs47$, Dec =
-$26^\circ 05^\prime 18\farcs8$, with an error of $0\farcs4$ \citep{wij01}.  This is
within the error circle of the Chandra position \citep{rev02}.  Assuming
that the radius expansion bursts reach the limit for pure He,
\citet{gal07} derive a distance of $7.2 \pm 1.0$
kpc.  Burst oscillations at $\approx
524$ Hz have been detected in multiple bursts \citep{smi97, mun00, gal07}.
The orbital parameters have not been measured.  A study of the scatter of
asymptotic burst oscillation frequencies by \citet{mun00} suggested
that the  $v_{\rm x}\sin i$ might be as high as $340
\pm 100$ km/s, but this has not been revisited.  Identification of the
counterpart was hampered by high reddening along
the galactic plane, but it has now been detected \citep{mign02b}.   If
the companion
is on the main sequence, it has to be of spectral type later than F:
if it has evolved off the main sequence then it is not a red giant.
For this to be the case in a Roche lobe overflowing system we require
$P_{\rm orb} > 2$ hours.

\subsubsection{4U 0614+09 ($\nu_b$ = 415 Hz)*}

This is a bursting atoll source, persistent but highly variable. Based
on spectral modelling by \citet{for00}, we set 9 ASM cts/s =
$3.3\times 10^{-9}$ ergs cm$^{-2}$ s$^{-1}$ (bolometric).  The long-term average
ASM countrate is 3.3 cts/s.  The
best and most recent position, from a Spitzer observation of the IR
counterpart, is RA =
$6^\mathrm{h} 17^\mathrm{m} 07\fs35 \pm 0\fs 03$, Dec = +$09^\circ 08^\prime
13\farcs60 \pm 0\farcs05$ (J2000) \citep{mig06}\footnote{This is consistent
with the older position for the optical counterpart V1055 Ori
\citep{bra83}.}.  \citet{bran92} infer an upper limit on the distance of
3 kpc from an X-ray burst. \citet{str08} detected burst oscillations
at 415 Hz in one burst recorded by the SWIFT Burst Alert Telescope.
The orbital parameters are unknown, but this is a
  candidate ultra-compact binary \citep{jue01}.  \citet{nel04} have
  shown that the companion is most likely a C/O white dwarf.  Using
  the white dwarf models of \citet{del03} this would imply $P_{\rm
    orb}$ = 15-20 minutes and hence (assuming that $i< 60^\circ$ 
  due to the lack of dips and eclipses), $a_{\rm x} \sin
  i \le$ 0.014 lt-s. 

\subsubsection{4U 1728-34 (GX 354+00) ($\nu_b$ = 363 Hz)}

This is a persistent atoll source.  Based on the spectral modelling of
\citet{gal07} we set 4 ASM cts/s = $1.2\times 10^{-9}$ ergs cm$^{-2}$ s$^{-1}$
(bolometric).  The long-term average ASM countrate is 7.3 cts/s.  The most precise position for this
source is that of the radio counterpart, RA = $17^\mathrm{h}
31^\mathrm{m} 57\fs73 \pm 0\fs02$, Dec =
-$33^\circ 50^\prime 02\farcs5 \pm
1\farcs1$ (J2000), where the errors are $1\sigma$
uncertainties \citep{mart98}.  Assuming that the bright PRE bursts
reach the He limit, \citet{gal07}
infer a distance of $5.2 \pm 0.5$ kpc.  Burst oscillations have been detected in multiple bursts at $\approx
363$ Hz \citep{str96, gal07}.  An unsuccessful search for persistent
pulsations using Ginga data was carried out by \citet{vau94}.  The
orbital parameters are not
known, although \citet{str98b} infer  $v_{\rm x} \sin i < 20.7$ km/s from the
scatter of asymptotic frequencies of burst oscillations from a series
of bursts from 1996-7.  We will use this constraint in our initial
assessment and use it to estimate $P_\mathrm{orb}$.  If we assume
Roche lobe overflow, then we find that a main sequence or evolved star cannot
satisfy the various relations.  The companion must be either a white
dwarf or a helium star.  This is consistent with the properties of the
X-ray bursts from this source, which suggest a hydrogen poor donor
\citep{gal07}. We will therefore assume that $P_{\rm orb} < 10$ hours.

\subsubsection{4U 1702-429 ($\nu_b$ = 329 Hz)}

This is a persistent atoll source.  Based on the spectral modelling of
\citet{gal07}, we set 2 ASM cts/s = $7.8\times 10^{-10}$
ergs cm$^{-2}$ s$^{-1}$ (bolometric).  The long-term average ASM countrate is
3.2 cts/s.  The best position, measured by
Chandra, is RA = $17^\mathrm{h}
06^\mathrm{m} 15\fs314$, Dec = -$43^\circ 02^\prime
08\farcs60$ (J2000), with an uncertainty of
$0\farcs6$ \citep{wac05}.  Assuming that the PRE bursts reach the
limit for pure He,
\citet{gal07} derive a distance of $5.5 \pm 0.2$ kpc.  Burst
oscillations are detected in multiple bursts at $\approx$ 329 Hz
\citep{mar99,gal07}.  The orbital parameters are not known but there
are no dips or eclipses.

\subsubsection{MXB 1730-335 (Rapid Burster) ($\nu_b$ = 306 Hz)*}

The Rapid Burster is a transient globular cluster source with regular
outbursts that occur around every 200 days.   It is
unusual in being the only system to show both Type I and Type II X-ray
bursts, the latter being driven by spasmodic accretion.  Using the spectral
modelling of \citet{gal07} we set 10 ASM cts/s = $6.7\times 10{-9}$
ergs cm$^{-2}$ s$^{-1}$ (bolometric).  The long-term average ASM countrate is 1
ct/s, and during a typical 25 day outburst the average countrate is
8.8 cts/s.  The most
accurate position, given by the radio counterpart, is RA = $17^\mathrm{h} 33^\mathrm{m} 24\fs61$, Dec = -$33^\circ 23^\prime
19\farcs8$ (J2000), with an error of $0\farcs1$ \citep{moo00}.
This is within the error circle of Chandra
observations \citep{home01}.  The distance to the host globular
cluster, Liller 1, is $8.8^{+3.3}_{-2.4}$ kpc \citep{kuu03}.
Averaging the burst rise phase of 31 X-ray bursts recorded by RXTE
revealed a weak candidate
burst oscillation frequency of $\approx 306$ Hz \citep{fox01}.  The
orbital parameters are not known, and no optical counterpart has been
detected because of crowding in the host globular cluster.  

\subsubsection{IGR J17191-2821 ($\nu_b$ = 294 Hz)*}

This transient X-ray source was discovered only recently, with one
recorded outburst.  The average ASM countrate during the 11 day
outburst was 3.7 cts/s, giving a long-term average countrate during
the RXTE era of 0.01 cts/s.  Based on \citet{kle07a} we assume 3.5 ASM
cts/s = $1.2\times 10^{-9}$ ergs cm$^{-2}$ s$^{-1}$ (2-10 keV), and apply a
bolometric correction factor of 2.  The most
precise position for the source, from SWIFT, is RA =
259.81306$^\circ$, Dec = -$28.29919^\circ$ (J2000), with an accuracy of
$4^{\prime\prime}$ \citep{kle07b}. Burst oscillations at $\approx 294$ Hz
have been detected in one
burst \citep{mar07}. The peak flux of the X-ray bursts sets
an upper limit to the distance of $\sim 11$ kpc \citep{mar07}.
Orbital parameters are not yet known, and there is no information on
the companion.

\subsubsection{4U 1916-053 ($\nu_b$ = 270 Hz)*}

This is a persistent source in an ultracompact binary with an H-poor
donor star. Following the spectral modelling of \citet{gal07} we set
0.5 ASM cts/s = $2.7\times 10^{-10}$ ergs cm$^{-2}$ s$^{-1}$ (bolometric).  The
long-term average ASM countrate is 1.3 cts/s.  The most recent position, given by Chandra, is RA =
$19^\mathrm{h} 18^\mathrm{m} 47\fs871$, Dec = -$05^\circ
14^\prime 17\farcs09$ (J2000), with an error $0\farcs6$
\citep{iar06}\footnote{This position has a substantial offset in
  declination from the optical counterpart (V1405 Aql) reported by
  \citep{got91},
with RA =
$19^\mathrm{h} 18^\mathrm{m} 47\fs91$, Dec = -$05^\circ
14^\prime 08\farcs7$ (J2000).}.  \citet{gal07} derive a distance of $8.9 \pm 1.3$ kpc or $6.8
\pm 1.0$ kpc, depending on composition, from PRE bursts.   Burst oscillations at $\approx 270$ Hz have been
detected in one X-ray burst \citep{gal01}.  The orbital period has
been the subject of much debate in the literature,
due in part to differences between the X-ray and orbital periods
determined by dipping and photometry respectively.  The most recent
papers on this topic seem to resolve the issue \citep{cho01, ret02} by
determining that the system displays superhumps rather than being a
hierarchical triple. \citet{cho01} use X-ray dip times to derive $P_{\rm
  orb} = 3000.6508 \pm 0.0009$ s (in agreement with the value of $P_{\rm orb} =  3000.6452 \pm 0.0043$ s
reported by \citet{wen06} using the RXTE All Sky Monitor), with
$\dot{P}_{\rm orb} < 2.06 \times
10^{-11}$ ($2\sigma$ upper limit).  The X-ray dip ephemeris is
$T_\mathrm{dip} = 50123.00944 \pm 1.4\times 10^{-4}$ MJD.  Whilst it
is not entirely clear how the dip time relates to $T_{90}$ we will
assume in our initial analysis that this could be determined, and use
the uncertainty in $T_\mathrm{dip}$ as an estimate of the uncertainty
in $T_{90}$.  The projected semi-major axis has not been measured, but we can impose
some constraints.  For the known orbital period, the assumption of
Roche lobe overflow (Equation \ref{roche}) gives the mean mass of the
donor as  158
g/cm$^3$, implying a dwarf companion.  X-ray burst properties imply a
helium rich donor.  Optical spectroscopy shows large amounts of N,
suggesting that the companion is a helium white dwarf rather than a helium
star or an evolved secondary \citep{nel06}. Using the models of
\citet{del03} this implies a companion mass in the range 0.008 - 0.03
$M_\odot$ (depending on core temperature).  For dips we expect an
inclination in the range $60-75^\circ$.  If we assume that the neutron
star mass is in the range $1.2 - 2.4 M_\odot$, Equation (\ref{axsini})
implies that $a_\mathrm{x} \sin i$ must lie in the range 4--25 lt-ms.

\subsubsection{XB 1254-690 ($\nu_b$ = 95 Hz)*}

This source is persistent, at a steady level, with dips. \citet{gal07}
carry out spectral modelling and find a persistent flux level of
$9\times 10^{-10} {\rm erg \ cm^{-2} \ s^{-1}}$ (bolometric).  The Chandra position
reported by \citet{iar07} is RA =
$12^\mathrm{h} 57^\mathrm{m} 37\fs153$, Dec = -$69^\circ
17^\prime 18\farcs98$, with a 90\% uncertainty radius of
$0\farcs6$\footnote{This position is $2^{\prime\prime}$ away from the
  older optical counterpart position reported by \citet{bra83}.}.  Radius expansion in the precursor to a superburst leads
to a distance estimate of $13 \pm 3$ kpc \citep{int03}.  \citet{bha07}
reported tentative evidence of a burst oscillation at
95 Hz in one burst from this source. \citet{mot87} used optical
photometry to find  $P_{\rm orb} = 3.9334
\pm 0.0002$ hours   This accords with the period of $3.88 \pm 0.15$ hours
derived from the X-ray lightcurve \citep{cou86}, although the X-ray
dips are not always present \citep{sma99}.  \citet{barn07} have recently
updated the ephemeris derived by \citet{mot87} and report an X-ray dip
time $T_{\rm dip} = 2453151.647 \pm 0.003$.  \citet{mot87} showed that
the X-ray dips in this source occurred at phase 0.84 (with zero at
optical minimum), so they therefore derive $T_0$ = JD 2453151.509 $\pm
0.003$.  \citet{barn07} have used phase-resolved spectroscopy of the He II
$\lambda 4686$ emission line
(thought to be emitted in the inner accretion disk, close to the
compact object) to estimate the velocity of the compact object, and find a
velocity semi-amplitude $v_\mathrm{x} \sin i = 130 \pm 16$ km/s.  Only
the lower portion of this range is consistent with a main sequence or
undermassive companion star:  however we will use this value in our
initial assessment of detectability.

\subsubsection{EXO 0748-676 ($\nu_b$ = 45 Hz)}

This system, which shows both dips and eclipses, has been persistently
active since 1985.  Based on the spectral modelling of \citet{gal07},
we set 0.6 ASM cts/s = $3.6\times 10^{-10}$ ergs cm$^{-2}$ s$^{-1}$ (bolometric).  The
long-term average countrate is 0.76 cts/s.  The best position,
for the optical counterpart UY Volantis, is RA = $7^\mathrm{h}
48^\mathrm{m} 25\fs0 \pm 0\fs1$, Dec = -$67^\circ 37^\prime 31\farcs7 \pm
0\farcs7$ (B1950) \citep{wad85}.  No more up to date
position is available. The detection of PRE bursts that seem to be He
rich implies a distance
of $7.4 \pm 0.9$ kpc \citep{gal07}.  \citet{vil04} discovered burst
oscillations at 45 Hz by averaging
together spectra from 38 separate bursts detected between 1996 and
early 2003. Timing of the eclipses constrains the orbital parameters.
Attempts to compute an orbital ephemeris, however, have been complicated.  The
most recent study by
\citet{wol02} finds a large apparent period change of 8 ms over the
period 1985 to 2000 - much larger than expected from orbital models -
and intrinsic jitter that cannot be explained by any simple
ephemeris. The reason for this variability has yet to be
resolved, and it is not clear whether this represents genuine evolution
in the binary period or not.  \citet{wol02} consider various models in
their analysis, with orbital periods $P_{\rm orb} = 0.1593378 \pm 1\times
10^{-7}$ days, mid-eclipse times $T_{90} = 46111.0739 \pm 0.0013$
MJD/TDB, and $|\dot{P}_{\rm orb}| \lesssim 10^{-11}$.  More recent
analysis by \citet{wen06}, using ASM data, finds $P_{\rm orb} =
0.1593375 \pm 6\times 10^{-7}$ days.  The projected semi-major axis
has not been measured directly, but can
be constrained.  \citet{wol02} find eclipse durations $497.5 \pm 6$
s, which gives a eclipse half-angle $\theta_x = (6.543 \pm
0.015)^\circ$.  This constrains the inclination via Equation
(\ref{eclipse}).  Roche lobe overflow (Equation \ref{roche}) implies a mean density for the
donor star of 7.5 g/cm$^3$, suggesting a main sequence or evolved
companion.  Maximum donor mass corresponds to a main sequence donor,
with $M_d = 0.42 M_\odot$.  Donor mass can be lower if the companion
is evolved, so following \citet{hyn06} we take a minimum plausible
companion mass of 0.07 $M_\odot$.  For the NS we consider
masses in the range $1.2 - 2.4 M_\odot$.  This implies that
$a_\mathrm{x} \sin i$ lies in the range 0.11--0.84 lt-s.

\subsection{Kilohertz QPO sources}

\begin{figure}
\begin{center}
\includegraphics[width=8cm, clip]{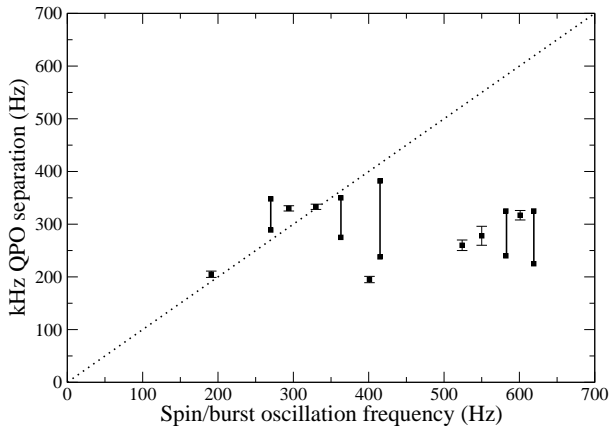}
\end{center}
\caption{A comparison of twin kHz QPO separation and spin frequency or
burst oscillation frequency for those sources that show both
phenomena.  The dotted line indicates equality of the two
measures. For some objects kHz QPO separation is consistent with
being constant:  these are shown as single points with error bars.
Note however that this may be due to poor sampling of source states.
For five objects kHz QPO separation varies:  these are shown
as two points with a line indicating the range.}
\label{khzs}
\end{figure}

These remain the most difficult sources, because of the uncertainty in
the precise relationship between kHz QPO separation (which varies) and
spin frequency. Sources where a wide range of accretion rates have been
sampled show variation, those where only a few accretion rates have been
sampled (including the pulsars that have kHz QPOs) do not.  To gauge the uncertainty, consider the kHz QPO
separations recorded for those sources where we have either a spin
frequency or a burst oscillation frequency, illustrated in Figure \ref{khzs}:

 \begin{itemize}
\item{{\bf Aql X-1}: 550 Hz intermittent pulsar, twin kHz QPO
    separation $278 \pm 18$ Hz \citep{bar08}.}
 \item{{\bf SAX J1808.4-3658}: 401 Hz pulsar.  Twin kHz QPOs
    observed once during the 2002 outburst. \citet{wij03}
    report a separation of $195 \pm 6$ Hz, later analysis by
    \citet{van05} using a different fitting technique gives a
    separation of $182 \pm 8$ Hz.}
 \item{{\bf XTE J1807-294}: 191 Hz pulsar.  Twin kHz QPOs detected on
   several occasions during the 2003 outburst. The measured separations
   were consistent with the spin frequency, although the
   weighted average separation of $205 \pm 6$ Hz exceeds the
   spin at the 2.3$\sigma$ level \citep{lin05}.}
 \item{{\bf 4U 1608-522}: 619 Hz burst oscillations. Twin kHz QPOs
   observed in the 1996 and 1998 outbursts, separations in
   the range $225 - 325$ Hz  \citep{men98a, men98b}.}
 \item{{\bf SAX J1750.8-2980}:  601 Hz burst oscillations.  Tentative
   detection of twin kHz QPOs with a separation of $317 \pm 9$ Hz
   \citep{kaa02}.}
 \item{{\bf 4U 1636-536}:  582 Hz burst oscillations. Twin kHz QPOs
   detected on multiple occasions, with separations varying from
   $240 - 325$ Hz \citep{wij97a, men98c, jon02, dis03, bar05}.}
 \item{{\bf KS 1731-260}: 524 Hz burst oscillations.  Twin kHz QPOs
   with separation $260.3 \pm 9.6$ Hz seen in one observation
   \citep{wij97}.}
\item{{\bf 4U 0614+09}: 415 Hz burst oscillations.  Twin kHz QPOs have
    been seen on several occasions \citep{for97, van00,
  van02}.  Separations vary from $238 \pm 7$ Hz up to $382 \pm 7$ Hz
  (perhaps even as low as $213 \pm 9$ Hz although this figure is
  tentative). }
 \item{{\bf 4U 1728-34}: 363 Hz burst oscillations.  Twin kHz QPOs
   detected on multiple occasions, with separations in the range
   275--350 Hz \citep{str96, men99, dis01, mig03}.}
 \item{{\bf 4U 1702-429}: 330 Hz burst oscillations. Twin kHz QPOs
   detected once in 1997, with a separation of $333
   \pm 5$ Hz \citep{mar99}.}
 \item{{\bf IGR J17191-2821}: 294 Hz burst oscillations.  Twin kHz QPOs
      with separation 330 Hz \citep{kle07}.}
 \item{{\bf 4U 1916-053}: 270 Hz burst oscillations.  Twin kHz QPOs
   detected several times in 1996. Separation was consistent
   with being constant at $348 \pm 12$ Hz on four occasions; on the
   fifth it was $289 \pm 5$ Hz \citep{boi00}.}
 \end{itemize}
It is far from clear that there is a direct (or indeed any) relationship between kHz
QPO separation and spin, particularly for the high frequency
sources.  See \citet{yin07} and \citet{men07} for an extended
discussion of this issue.

We now summarize the properties for the kHz QPO sources.  Note that we exclude
from our data set the peculiar X-ray binary Cir
X-1 (1516-559), since it is not clear that our accretion torque model
applies.  Although this source has twin kHz QPOs \citep{bou06}, it is
thought to be a high mass X-ray binary \citep{jon07} with a highly
eccentric orbit \citep{mur80, oos95} where accretion disk
formation is only sporadic \citep{joh99, joh01}.

\subsubsection{Cyg X-2 (2142+380)}

This is a persistent Z source that has X-ray bursts.  \citet{gal07}
carried out spectral modelling and found that the long-term average flux
is $1.1\times 10^{-8}$ ergs cm$^{-2}$ s$^{-1}$.  The best
position for this source, from the optical counterpart, is RA =
$21^\mathrm{h} 42^\mathrm{m}36\fs91$, Dec =
+$38^\circ 05^\prime  27\farcs9$, with an accuracy of $0\farcs5$
(B1950)\citep{gia67, bra83}. No more up to date high precision position is
available. \citet{gal07} use the PRE bursts to estimate a
distance of $10 \pm 2$ kpc or $14 \pm 3$ kpc depending on composition.
However optical observations suggest a distance of only $7.2 \pm 1.1$
kpc \citep{oro99}.  \citet{wij98b} detected twin kHz QPOs in RXTE data from a few hours on
July 2 1997,
when the source was on the horizontal branch.  Peak separation was
$346 \pm 29$ Hz.  \citet{kuz02} re-analysed the same data and found a
separation of $366 \pm 18$ Hz.  Unsuccessful searches for persistent
pulsations were carried out using Ginga data from
1987-1989 \citep{woo91, vau94}. \citet{kuu95} set
upper limits on the presence of burst oscillations for frequencies
below 256 Hz using EXOSAT data from
1983-1985, and
\citet{sma98} set upper limits on burst oscillations in the 200-600 Hz
range for one burst
observed with RXTE.  The binary orbit can be constrained by optical observations of the
companion star V1341 Cyg.  The most recent spectroscopic
observations, by \citet{cas98} give
$P_{\rm orb} = 9.8444 \pm 0.0003$ days,  $T_{90} = 2449339.50 \pm
0.03$ HJD, and $e = 0.024 \pm 0.015$ (consistent at the $2\sigma$
level with being zero, which we assume in our initial analysis).
Error bars are $1\sigma$
uncertainties. By measuring the projected velocity of the secondary
star and its rotational broadening, then assuming tidal locking and
Roche lobe overflow, they infer $v_{\rm x} \sin i = 29.9
\pm 3.6$ km/s.  We use this value in our initial assessment of
detectability.

\subsubsection{GX 340+0 (1642-455)}

This is a persistent Z source that has not shown X-ray bursts. Based
on spectral modelling by \citet{for00} we set 25 ASM cts/s =
$2.3\times 10^{-8}$ ergs cm$^{-2}$ s$^{-1}$ (bolometric).  The long-term average
countrate is 30 cts/s.  The best
position, given by the radio counterpart, is RA =
$16^\mathrm{h} 45^\mathrm{m} 44\fs60 \pm 0\fs02$,
Dec = -$45^\circ 37^\prime 53\farcs6 \pm 0\farcs3$ (J2000)
\citep{pen93}.  \citet{chr97} use Einstein observations to establish
an upper limit on the distance of $11 \pm 3.3$
kpc\footnote{\citet{for00} quote a lower distance for this source, but
  the value given does not tally with that in the original reference
  that they cite.}.  Twin kHz QPOs have been detected in RXTE
data from 1997-8, with
separations in the range  $275 \pm 24$ Hz to $413 \pm 21$ Hz
\citep{jon98, jon00}.  However, for the sample as a whole, the
separation is
formally consistent with being constant
at $339 \pm 8$ Hz.   Unsuccessful searches for persistent pulsations
have been carried out using Ginga data \citep{woo91, vau94}. The
orbital parameters are not known.

\subsubsection{4U 1735-44}

This is a bright persistent atoll source that has both bursts and
superbursts.  Using the spectral modelling of \citet{gal07} we set 18
ASM cts/s = $9.1\times 10^{-9}$ ergs cm$^{-2}$ s$^{-1}$ (bolometric).  The
long-term average countrate is 14 cts/s. The best position, for the optical counterpart V926 Sco,
is RA = $17^\mathrm{h} 35^\mathrm{m}19\fs28$, Dec = -$44^\circ
25^\prime 20\farcs3$, uncertainty $<
1^{\prime\prime}$\citep{bra83}. No more up to date position is
available.  The distance inferred from the radius expansion bursts
(assuming He limit) is
$8.5 \pm 1.3$ kpc \citep{gal07}.  Twin kHz QPOs were seen in RXTE observations on May 30-31 1998.  Peak
separation varied from $296 \pm 12$ Hz up to $341 \pm 7$ Hz,
inconsistent at the $3.1\sigma$ level with being constant
\citep{for98}. \citet{jong96} placed upper limits on the presence of
burst oscillations (up to 256 Hz) using EXOSAT data: no figures have
been reported for the RXTE burst sample. The orbital parameters can be
constrained by optical observations.
\citet{cas06} report a recent spectroscopic ephemeris, with
$P_{\rm orb} = 0.19383351 \pm 3.2\times 10^{-7}$ days, in agreement with the most
recent photometric ephemeris \citep{aug98}.  The time of inferior
conjunction of the donor star $T_{90} = 2452813.495 \pm 0.003$ HJD.
Further measurements and source modelling suggest that velocity
semi-amplitude of the donor star is in the range 215-381 km/s, with a
mass ratio $m_{\rm d}/m_{\rm x}$ in the range 0.05-0.41. This would suggest $v_{\rm x} \sin
i < 156$ km/s and we use this limit in our initial estimate of
detectability.

\subsubsection{GX 5-1 (1758-250)}

This is a persistent non-bursting Z source, located in the highly
absorbed galactic bulge region.  Based on spectral modelling by \citet{for00}, we set 50 ASM cts/s =
$3.8\times 10^{-8}$ ergs cm$^{-2}$ s$^{-1}$ (bolometric). The long-term average
ASM countrate is 70 cts/s.  The best position, for the radio
counterpart, is RA = $18^\mathrm{h} 01^\mathrm{m} 08\fs233$,
Dec = $-25^\circ 04^\prime 42\farcs044$ (J2000), positional
uncertainty $0\farcs040$ \citep{ber00}.  Distance to this source is
poorly constrained, although \citet{chr97} give an upper limit of $9
\pm 2.7$ kpc\footnote{The distance of $6.4\pm 0.6$ kpc derived by
  \citet{pen89},
  cited incorrectly as $9.2\pm 0.7$ kpc by \citet{fen00, mig06a} is
  highly model-dependent.}.  Twin kHz QPOs have been reported in RXTE
observations from 1996 -
2000 \citep{wij98c, jon02b}.  Peak separation is not constant, but
varies from $232 \pm 13$ Hz up to $344 \pm 12$ Hz.  Unsuccessful
searches for persistent pulsations were carried out using Ginga data
from 1987 \citep{woo91, vau94}.  The orbital parameters are unknown.

\subsubsection{4U 1820-30}

This is a persistent bursting atoll source with regular dipping
cycles.  Using the spectral modelling of \citet{gal07} we set 32 ASM
cts/s = $2.1\times 10^{-8}$ ergs cm$^{-2}$ s$^{-1}$ (bolometric).  The long-term
average ASM countrate is 20.6 cts/s.  The best
position, from Hubble Space Telescope observations of the optical counterpart, is  RA =
$18^\mathrm{h} 23^\mathrm{m} 40\fs453 \pm
0\fs012$, Dec = -$30^\circ 21^\prime 40\farcs08
\pm 0\farcs15$ (J2000, $1\sigma$
errors), \citep{sos95}\footnote{Obtaining a more accurate radio position is
complicated by the presence of a nearby pulsar \citep{mig04}.}.  The
source is located in the globular cluster NGC 6624, and optical
observations imply a cluster distance of $7.6 \pm
0.4$ kpc \citep{hea00}.  The distance estimated from X-ray bursts is
$\approx 6.6$ kpc \citep{vac86, kuu03}.  Twin kHz QPOs have been
detected by RXTE on several occasions
\citep{sma97, zha98b, alt05}.  Measured separations vary in the
range 220-350 Hz, although the uncertainties are such that the
separation is consistent with
being constant at $\approx 275$ Hz. \citet{dib05} have placed
extremely stringent limits on the presence of persistent pulsations up
to 2000 Hz using archival RXTE data, for all $a_{\rm x} \sin i
< 16.8$ lt-ms (superceding earlier work by \citet{woo91} and
\citet{vau94}).  \citet{jong96} placed upper limits on the
presence of burst oscillations up to 256 Hz for EXOSAT bursts.
The most up to date X-ray ephemeris, using all data from Ariel 5,
Ginga and RXTE, gives $P_{\rm orb} = 685.0119 \pm 1.02\times 10^{-4}$
s, $\dot{P}_{\rm orb} = (-7.54 \pm 3.21)\times 10^{-13}$ s/s, and the
time of maximum X-ray light is $2442803.63564 \pm 2.2\times 10^{-4}$
HJD \citep{cho01b}.  Noting that an ephemeris of similar accuracy can
also be derived
from UV data \citep{and97}, we will assume that the reference time
could if required be related to a known phase in the orbit.  There is also a super-orbital periodicity of
$\approx 176$ days,
most likely due to perturbations of the orbital eccentricity (up to $e
= 0.004$ by a third member of the system \citep{cho01b, zdz07}, but we
neglect this in our initial assessment.
The projected semi-major axis has not been measured, but can be
constrained.  The assumption of Roche lobe overflow (Equation
\ref{roche}) gives a mean donor density of 3300 g/cm$^3$, implying a
white dwarf companion. The donor must also be helium-rich to explain
the X-ray burst properties.  Helium white dwarf models of
\citet{del03} suggest a donor mass in the range $0.07 - 0.08
M_\odot$ (depending on temperature).  \citet{and97}
inferred an inclination in the range $35-50^\circ$ from
observations of the UV counterpart.  \citet{bal04}, analysing
superburst data, inferred a slightly lower value, so we will consider
a minimum inclination of $30^\circ$.  Then for neutron star masses in
the range $1.2-2.4 M_\odot$ we predict $a_\mathrm{x} \sin i$ in the
range 7--20 lt-ms.  Recent modelling by \citet{zdz07} suggests that
 the super-orbital variability may eventually
pose even tighter constraints on the system.

\subsubsection{Sco X-1 (1617-155)}

This the closest accreting neutron star in our study, and is a persistent Z source.
Using the spectral modelling of \citet{for00} we set 920 ASM cts/s =
$4\times 10^{-7}$ ergs cm$^{-2}$ s$^{-1}$ (bolometric).  The long-term average
ASM countrate is 892 cts/s.  The most accurate position
for the source, given by VLBA measurements, is RA = $16^\mathrm{h} 19^\mathrm{m}55\fs0850$, Dec = -$15^\circ
38^\prime 24\farcs9$, with an uncertainty of $0\farcs5$
\citep{brad99}.  The source has also has a measurable proper motion,
which we neglect in this initial analysis \citep{brad99}. The distance, measured by
parallax, is $2.8 \pm 0.3$ kpc
\citep{brad99}.  Twin kHz QPOs are observed, with separations in the
range 240--310 Hz
\citep{vand96, vand97, men00}.  There have been unsuccessful searches
for persistent pulsations up to 256 Hz using EXOSAT \citep{mid86} and
up to 512 Hz using Ginga data \citep{woo91, her92, vau94}.
Photometric observations of the optical counterpart V818 Sco imply an
orbital period of $68023.84 \pm 0.08$ s \citep{gotl75}, although
analysis of RXTE ASM data by
\citet{vande03} suggests that the true period could in fact be slightly longer,
at 68170 s. Analysis by \citet{ste02} indicates $v_{\rm x} \sin i = 40 \pm 5$ km/s.  Assuming the
orbital period of \citet{gotl75}, \citet{ste02} derive an ephemeris
with  $T_{90}  =  2451358.568 \pm 0.003$ HJD\footnote{\citet{abb07a}
  use a slightly larger uncertainty on $T_{90}$ in their searches for
  gravitational waves from Sco X-1, to
    account for the time elapsed since the \citet{ste02} measurement.
  For consistency with the rest of our analysis we use the smaller
  uncertainty, assuming that the measurement could be re-done today to
the same level of accuracy.}.

\subsubsection{GX 17+2 (1813-140)}

This is a persistent Z source with X-ray bursts. Using the spectral
modelling of \citet{gal07} we set 40 ASM cts/s = $1.6\times 10^{-8}$
ergs cm$^{-2}$ s$^{-1}$ (bolometric).  The long-term average ASM countrate is 45
cts/s.  The most accurate
position, from VLA observations of the radio counterpart, is  RA = $18^\mathrm{h} 16^\mathrm{m}
1\fs389 \pm 0\farcs06$, Dec = -$14^\circ 02^\prime 10\farcs62 \pm
0\farcs04$ (J2000),
$1\sigma$ uncertainties in position \citep{deu99}.  This is within the
Chandra error circle for the X-ray position \citep{cal02}.  Analysis
of X-ray bursts suggests a distance of either  $9.8 \pm 0.4$ kpc or
$12.8 \pm 0.6$ kpc \citep{gal07}. However, there
are questions over how to  correct for the super-Eddington persistent
flux, and the true distance could be lower \citep{kuu02}.  Twin kHz QPOs have been observed on multiple occasions \citep{wij97b,
  hom02}.  Separation, which varies from $239 \pm 17$ Hz up to $308
  \pm 14$ Hz, is not constant at the 97\% confidence level.  Upper
  limits on the presence of persistent pulsations in Ginga data were
  reported by \citet{woo91} and \citet{vau94}.  Upper limits on the presence of
  burst oscillations in EXOSAT and RXTE data have been reported by
  \citet{kuu97} and \citet{kuu02}.  The orbital parameters are not
  known.

\subsubsection{XTE J2123-058}

This is a transient bursting atoll source in the Galactic Halo, with
one recorded outburst in 1998.  Using the spectral
modelling of \citet{gal07} we set 6.4 ASM cts/s = $2.1\times 10^{-9}$
ergs cm$^{-2}$ s$^{-1}$ (bolometric).  The average ASM countrate over the 50 day
outburst was 2.7 cts/s, yielding a long-term average countrate over
RXTE's lifetime of 0.03 cts/s.  The most accurate
position, as measured by Chandra, is RA = $21^\mathrm{h} 23^\mathrm{m}
14\fs54$, Dec = -$05^\circ 47^\prime 53\farcs2$ (J2000, uncertainty
$0\farcs6$) \citep{tom04}. \citet{tom01} infer a distance of $8.5 \pm
2.5$ kpc, consistent with the distance of $9.6 \pm 1.3$ kpc inferred
by \citet{cas02}.  Twin kHz QPOs were detected during one observation, with separations
in the range $255 \pm 14$ to $276 \pm 9$ Hz \citep{hom99, tom99}.
\citet{cas02} use spectroscopic and photometric measurements to derive
$P_{\rm orb} = 21447.6 \pm 0.2$ s ($1\sigma$ errors).  \citet{tom02},
however, derive $P_{\rm orb} = 21442.3 \pm 1.8$ ($1\sigma$
errors). This discrepancy, and results from earlier photometric
measurements by \citet{tom99, zur00}, have yet to be resolved. The
most recent time of minimum optical light $T_{90} = 2451779.652 \pm
0.001$ HJD.  The projected velocity $v_{\rm x} \sin i$ is also constrained.
\citet{cas02} attempt to measure this directly, and find $v_{\rm x}
\sin i = 140 \pm 27$ km/s.  \citet{tom01, tom02} measure the projected
orbital velocity and rotational velocity of the companion.  Assuming a
Roche lobe filling and tidally locked companion they infer the mass
ratio, and hence a projected orbital velocity for
the neutron star of $v_{\rm x}
\sin i = 110^{+54}_{-36}$ km/s.  \citet{sha03} use the
projected orbital velocity of the companion measured by \citet{cas02}
and use more sophisticated models of the system to give the mass
ratio.  The resulting projected orbital velocity is $v_{\rm x}
\sin i = 103^{+46}_{-7}$ km/s (90\% confidence).  In our initial
analysis, however, we use the direct measurement by \citet{cas02}.

\subsubsection{GX 349+2 (1702-363, Sco X-2)}

This persistent Z source does not show X-ray bursts.  Using the
modelling of \citet{zha98c}, we set 43 ASM cts/s = $1.4\times 10^{-8}$
ergs cm$^{-2}$ s$^{-1}$ (2-10 keV).  There is no detailed spectral modelling
available for this source, so we adopt a bolometric correction factor of 2.  The
long-term average ASM countrate is 50.2 cts/s. The best position, from VLA measurements of
the radio
counterpart, is RA =
$17^\mathrm{h} 02^\mathrm{m}22\fs93$, Dec = -$36^\circ
21^\prime 20\farcs3$ (B1950,
accuracy $0\farcs5$)
\citep{coo91}.  \citet{iar04} infer a distance of 3.6-4.4 kpc from
BeppoSAX observations.  Twin kHz QPOs were detected by RXTE in January 1998, with a separation
of $266 \pm 13$ Hz \citep{zha98c}, confirmed by \citet{one02}.  The
source has however rarely been observed in the state where kHz QPOs
are prevalent.   A search for persistent pulsations in Ginga data from 1989 was
unsuccessful  \citep{vau94}.  The binary period, measured using optical photometry and spectroscopy, has been the subject of
  some debate \citep{sou96,
  wac96, barz97}, but is now established as $P_{\rm orb} = 22.5 \pm 0.1$ hours ($1\sigma$ error) \citep{wac97}.
  The other binary parameters have not been measured, although we can
  constrain $a_{\rm x} \sin i$.  The assumption of Roche lobe overflow
  (Equation \ref{roche}) gives a mean donor density 0.2 g/cm$^3$,
  which requires a donor that has evolved off the main sequence.  In
  the absence of better constraints we will assume a mass ratio 
  $m_{\rm d}/m_{\rm x} < 0.8$.  For the range of neutron star masses considered ($1.2 -
  2.4 M_\odot$) this implies $a_{\rm x} \sin i < 7$ lt-s.

 \section{Searching with Parameter Uncertainties}
\label{sec:search}

Let us summarize briefly the gravitational wave emission from a
neutron star in a binary system. We assume that the centre of mass of
the binary is not accelerating in the solar system barycentre (SSB)
frame.  The timing model for the arrival times of the wave-fronts of
the GW is taken to be the usual one \citep{tay89}.  Let $T$ be the
arrival time of the wave at the SSB, $\tau$ the proper time of
emission in the rest frame of the neutron star, and $t$ the time in
the rest frame of the gravitational wave detector.  The quantities $T$
and $\tau$ are related by
\begin{equation}
  \label{eq:1}
  T - T_0 = \tau + \Delta_R + \Delta_{E} + \Delta_{S}
\end{equation}
where $\Delta_R$ is the Roemer time delay accounting for the light
travel time across the binary, $\Delta_{E}$ and $\Delta_{S}$ are
respectively the orbital Einstein and Shapiro time delays in the
binary, and $T_0$ is a reference time.  There are no additional timing
delays due to dispersion.  It turns out that, for our purposes, the
Roemer delay is the most significant contribution.  If $\mathbf{r}$ is
the vector joining the centre of mass of the binary system with the
neutron star, and $\mathbf{n}$ is the unit vector pointing from the
SSB to the source, then
\begin{equation}
  \label{eq:2}
  \Delta_R = -\frac{\mathbf{r}\cdot\mathbf{n}}{c}\,.
\end{equation}
There is then a similar relation between $T$ and the arrival time $t$ at the
earth based detector, and we assume that this can be corrected for
since the sky-position is known.

In the models that we are considering, the intrinsic gravitational wave
frequency $\nu$ depends on the spin frequency $\nu_s$.  The phase of the GW at the SSB is
\begin{equation}
  \label{eq:3}
  \phi(t) = \Phi_0 + \Phi(t)
\end{equation}
where
\begin{equation}
  \label{eq:4}
  \Phi(T) = 2\pi \nu \left(T -
    \frac{\mathbf{n}\cdot\mathbf{r}(T)}{c}\right) \,.
\end{equation}
Inclusion of frequency derivatives in this phase model is straightforward,
and we do not write it down explicitly.  Since the gravitational wave
amplitudes are expected to be very weak and the output of the GW
detectors dominated by noise, knowledge of the waveform, especially
its phase, is crucial for detection.

The phase $\Phi(t)$ depends on the orbital parameters introduced in Section
\ref{sec:uncertainties}: $P_{\rm orb}$, $\dot{P}_{\rm orb}$, $T_{\rm asc}$,
$a_{\rm x}\sin i$, and $e$.   In addition, there are 2 parameters specifying the
orientation of the orbital plane:  the inclination angle $\iota$ and the
argument of periapsis $\omega$.  Of these 7 parameters, only 6 are required to
define the phase model because of the projection along the line of sight
$\mathbf{n}$; see \citet{dhu01} for further details.  Taking the spin frequency
$\nu_s$ and its time derivative $\dot{\nu}_s$ into account, we therefore have a
total of 8 parameters which determine the frequency evolution of the signal:
$(\nu_s,\,\dot{\nu}_s,\,a_{\rm x}\sin i,\,e,\,P_{\rm orb},\,\dot{P}_{\rm
orb},\,T_{\rm asc},\,\omega)$.

This is clearly a very large parameter space, and a search over all
these parameters using a sufficiently large data volume will be a big
data analysis challenge.  Let us therefore make some simplifying
assumptions: $\dot{\nu}_s=0$, $\dot{P}_{\rm orb} = 0$, and $e=0$; i.e.
we assume a neutron star spin perfectly balanced between accretion and
gravitational radiation, and a circular orbit which does not decay
appreciably over the course of the observation time. These assumptions
may not hold for the sources and for the large observation
times that we are considering, and an actual search might very well
have to take some or all of these effects into account. However, for
assessing the detection prospects as we want to do here, this
simplification is useful, since adding the extra parameters will further increase the number of templates.  For some 
of the more promising sources at or near the detection threshold, these
assumptions will need to be revisited in greater detail. Some of these extra
parameters may need to included, and the resulting search might
again become computationally difficult; this will be studied in
greater detail in future work.

In the case when the orbit is circular ($e=0$), which we shall assume
in the rest of this paper, the argument of periapsis and the initial
orbital phase combine additively into a single parameter so that we
are left with only 4 search parameters: $\bm\lambda = (\nu_s,\,a_{\rm
  x}\sin i,\,P_{\rm orb},\,T_{\rm asc})$; we shall denote the
components of $\bm \lambda$ by $\lambda^i$ with $i=0\ldots 3$.

\subsection{Template counting}
\label{subsec:templates}

To determine the computational cost involved in searching the parameter space
described above, we need to calculate the number of templates required.  A
calculation of the required number of templates to search a portion of the
parameter space is based on demanding a certain maximum mismatch between the
templates at neighbouring points in parameter space. This also guarantees that the
true signal will not have more that the given mismatch to at least one of
the search templates. The mismatch between waveform templates is measured simply
as the fractional loss in the signal-to-noise ratio (SNR) when one waveform is
filtered (folded) by the other. This fractional loss can be regarded as a
distance measure between points in parameter space, and this leads naturally to
the definition of a parameter-space metric $g_{ij}$
\citep{sath91,sath94,owen96,prix07b}. Using the metric, we write the proper distance
squared (the ``mismatch'') between two infinitesimally separated parameter space
points as
\begin{equation}
  \label{eq:17}
  m = g_{ij}d\lambda^id\lambda^j
\end{equation}
The size of the parameter space is then given by the volume measure determined
by the metric in the usual way.  Then, assuming that we cover this parameter
space by a lattice grid, the number of templates is the total parameter space
volume divided by the volume of each unit cell which makes up the lattice.

The optimal choice of the lattice is determined by a solution to the so called
sphere-covering problem \citep{prix07}. For our purposes, we shall use a simple
cubic grid, and there are two reasons why it is acceptable to use this
approximation. First, the dimensionality of the reduced parameter space that we are
looking at is low enough that the improvement in the template placement
efficiency is not more than a factor of about 2 or 3 \citep{prix07}.
Furthermore, this improvement in the efficiency does not actually lead to a
corresponding factor of 2-3 improvement in the sensitivity; the gain in the
coherent integration time afforded by this improvement is much smaller because
the computational cost typically scales as a large power of the coherent
integration time. Finally, the size of each unit cell is chosen based on the
fractional loss in SNR, i.e. the mismatch $m$, that we are willing to tolerate;
we shall use a reference value of $m=30\%$ in this paper.

The first detailed study of the parameter space metric for a neutron
star in a binary orbit was carried out in \citet{dhu01} for a coherent
matched filter search.  The search for gravitational radiation from
Sco X-1 reported in \citet{abb07a} was the first and so far, only
application of this study.  The aim of this section is mainly to
collect some template counting equations for later use.  These
equations can all be derived in a more or less straightforward manner
from the results of \citet{dhu01}.  The main difference is that
\citet{dhu01} use notation and variables targeted towards
gravitational wave data analysis, while here we choose to use notation
more familiar to an astronomy/astrophysics audience.

The first issue is the number of parameters which must be searched.
Let us denote by $\Delta\lambda^i = \lambda_{\rm max}^i - \lambda_{\rm
  min}^i$ the uncertainty in $\lambda^i$ from astronomical
observations; we assume the region to be rectangular.  Since the
proper length of the line in the $\lambda^i$ direction is
$\int_{\lambda_{\rm min}^i}^{\lambda_{\rm
    max}^i}\sqrt{g_{ii}}d\lambda^i$, a useful upper bound on this
proper length is to use the maximum value of $g_{ii}$, i.e.
$\left(g_{ii}^{\rm max}\right)^{1/2}\Delta\lambda^i$.  This proper
distance can then be compared with our reference mismatch $m$, and we
 get a measure of the number of templates required in the
$\lambda^i$ direction:

\begin{equation}
  \label{eq:18}
  N_{\lambda^i} = \Delta\lambda^i \sqrt{\frac{g_{ii}^{\rm max}}{m}}\,.
\end{equation}
If $N_{\lambda^i}< 1$, it indicates the $\lambda^i$ direction can be
covered by just a single template, and the effective dimensionality of
our parameter space is reduced by 1.  The number of templates for the
frequency are:
\begin{equation}
  \label{eq:27}
  N_\nu = \Delta \nu \frac{\pi T_{\rm obs}}{ \sqrt{3m}}\,.
\end{equation}
The uncertainty $\Delta \nu$ relates directly to the uncertainty in
spin $\Delta \nu_s$. We shall ignore the correlations of $\nu$ with the other
parameters; this approximation will suffice for our purposes.  For the
other directions, the expressions for $N_{\lambda^i}$ have simple
expressions in two regimes: $T_{\rm obs} \ll P_{\rm orb}$ and $T_{\rm
  obs}\gg P_{\rm orb}$.

In the limit of large observation times, $T_{\rm obs} \gg P_{\rm orb}$,
we have:
\begin{eqnarray}
  \label{eq:19}
  N_{a_{\rm p}} &=& \Delta a_{\rm p}  \frac{\sqrt{2} \pi \nu_{\rm max}}{\sqrt{m}}\,,\\
  N_{T_{\rm asc}} &=& \Delta T_{\rm asc}\frac{\sqrt{8}\pi^2  a_{\rm
      p}^{\rm max} \nu_{\rm max}}{  P_{\rm orb}^{\rm min}\sqrt{m}} \,,\\
  N_{P_{\rm orb}} &=& \Delta P_{\rm orb} \frac{\sqrt{8}\pi^2 a_{\rm
      p}^{\rm max} \nu_{\rm max} T_{\rm obs} }{ \sqrt{3m}(P_{\rm orb}^{\rm min})^2}\,.
\end{eqnarray}
In the other limiting case $T_{\rm obs} \ll P_{\rm orb}$, we get
\begin{eqnarray}
  \label{eq:20}
  N_{a_{\rm p}} &=& \Delta a_{\rm p}\frac{2 \pi^3  \nu_{\rm max}
    T_{\rm obs}^2 }{ \sqrt{45m} (P_{\rm orb}^{\rm min})^2}\,,\\
  N_{T_{\rm asc}} &=& \Delta T_{\rm asc}  \frac{8\pi^5 a_{\rm p} \nu_{\rm max}
    T_{\rm obs}^3 }{ \sqrt{175m} ( P_{\rm orb}^{\rm min})^4}\,,\\
  N_{P_{\rm orb}} &=& \Delta P_{\rm orb}  \frac{4\pi^3 a_{\rm p} \nu_{\rm max}
  T_{\rm obs}^2 }{ \sqrt{45m} (P_{\rm orb}^{\rm min})^3}\,.
\end{eqnarray}
For a search over all three parameters at once, the total number of templates in
$(a_{\rm p}, P_{\rm orb}, T_{\rm asc})$ space is, for $T_{\rm
  obs} \gg P_{\rm orb}$:
\begin{equation}
  \label{eq:22}
  N_{\rm a_{\rm p}P_{\rm orb}T_{\rm asc}} = \frac{\pi^5 \nu_{\rm max}^3
    T_{\rm obs}}{\sqrt{8m^3}}\Delta[a_{\rm p}^3] \Delta[P_{\rm
    orb}^{-2}]\Delta[T_{\rm asc}]\,,
\end{equation}
while for $T_{\rm obs} \ll P_{\rm orb}$ it is:
\begin{equation}
  \label{eq:23}
    N_{\rm a_{\rm p}P_{\rm orb}T_{\rm asc}} =
    \frac{\sqrt{48}\pi^{13}\nu_{\rm max}^3  T_{\rm
        obs}^9}{\sqrt{7027611500m^3}}\Delta[a_{\rm p}^3]\Delta[P_{\rm
      orb}^{-10}] \Delta[T_{\rm asc}] \,.
\end{equation}
Note that $N_{\rm a_{\rm p}}N_{\rm P_{\rm orb}}N_{\rm T_{\rm asc}} \neq
N_{a_{\rm p}P_{\rm orb}T_{\rm asc}}$.  This happens because the correlations
between the different parameters (i.e. the off-diagonal terms in the metric) can
be very important, especially for short observation times.

Similarly, we shall require the equations in the case when one of the
coordinates can be ignored and the search can be performed in a 2-dimensional
subspace.  The equations for $T_{\rm obs} \gg P_{\rm orb}$ are:
\begin{eqnarray}
  \label{eq:24}
  N_{a_{\rm p}T_{\rm asc}} &=& \frac{\pi^3 \nu_{\rm max}^2}{P_{\rm orb}m}
  \Delta[a_{\rm p}^2]\Delta[T_{\rm asc}]\,,\\
  N_{P_{\rm orb}T_{\rm asc}} &=& \frac{\pi^4  (a_{\rm p}^{\rm max})^2
    \nu_{\rm max}^2 T_{\rm obs}}{m\sqrt{3}}  \Delta[P_{\rm
    orb}^{-2}]\Delta[T_{\rm asc}]  \,,\\
  N_{a_{\rm p}P_{\rm orb}} &=& \frac{\pi^3 \nu_{\rm max}^2T_{\rm obs}}{m\sqrt{3}}
  {\Delta[a_{\rm p}^2] \Delta[P_{\rm orb}^{-1}]}\,.
\end{eqnarray}
Finally, for $T_{\rm obs} \ll P_{\rm orb}$ we get:
\begin{eqnarray}
  \label{eq:25}
  N_{a_{\rm p}T_{\rm asc}} &=& \frac{2\pi^8\nu_{\rm max}^2T_{\rm
      obs}^5}{45\sqrt{35}mP_{\rm orb}^6}   \Delta[a_{\rm
    p}^2]\Delta[T_{\rm asc}]\,,\\
  N_{P_{\rm orb}T_{\rm asc}} &=& \frac{4\pi^8 \nu_{\rm max}^2T_{\rm
      obs}^5 (a_{\rm p}^{\rm max})^2}{135\sqrt{35}m} \Delta[P_{\rm
    orb}^{-6}]\Delta[T_{\rm asc}]\,,\\
  N_{a_{\rm p}P_{\rm orb}} &=& \frac{8\pi^8\nu_{\rm max}^2T_{\rm
      obs}^6}{2835\sqrt{5}m} \Delta[P_{\rm orb}^{-6}]\Delta[a_{\rm
    p}^2]\,.
\end{eqnarray}
We take the total number of templates to be $N_\nu$ (if it exceeds
unity) multiplied by the number of templates in $(a_{\rm p}, P_{\rm
  orb}, T_{\rm asc})$ space.

While these equations might not seem very illuminating, two important features
are worth remembering. First, and probably most importantly, the scaling of the
number of templates with $T_{\rm obs}$ is very different in the two regimes
$T_{\rm obs} \ll P_{\rm orb}$ and $T_{\rm obs} \gg P_{\rm orb}$.  For example,
in Equations (\ref{eq:22}) and (\ref{eq:23}), the scaling is $\mathcal{O}(T_{\rm
obs})$ when $T_{\rm obs} \gg P_{\rm orb}$, while it is $\mathcal{O}(T_{\rm
obs}^9)$ for small $T_{\rm obs}$.  This will have important consequences for GW
data analysis, as we shall see later.  Second, computational cost issues become
more important at higher frequencies and for tighter orbits, because the number
of templates typically increases when $\nu_{\rm max}$ increases or $P_{\rm
orb}$ decreases.

Finally, we say a few words about the positional accuracy required for
the GW searches.  We do not wish to consider searches over
sky-position, and thus it is important to know the position
sufficiently accurately beforehand.  The periodic wave searches get
their sky-position information from the Doppler pattern of the
frequency evolution.  The sky-position accuracy $\Delta\theta$ depends
strongly on the coherent observation time $T_{\rm coh}$.  For short
observation times ($\ll 1\,$yr) $\Delta\theta$ increases as roughly
$\mathcal{O}(T_{\rm obs}^2)$ or $\mathcal{O}(T_{\rm obs}^3)$
\citep{brad98,prix07b}.  This increase in the sky-resolution
eventually saturates when $T_{\rm coh}$ becomes comparable to a year.
The ultimate limit on $\Delta\theta$ is the diffraction limit, with
Earth's orbit being the aperture size.  Thus, the smallest error box
for position will be $\Delta\theta \sim \lambda_{\rm gw}/1\,{\rm AU}$
where $\lambda_{\rm gw}$ is the wavelength of the GW.  This
corresponds to about $8^{\prime\prime}$ at $50\,$Hz and it is
inversely proportional to frequency; this requirement is easily met
for all the sources we are considering.  It might in fact be possible
to use a sky-position mismatch as a veto to rule out potential
candidates. For a candidate with given values of the frequency and
orbital parameters, we could calculate the detection statistic at
mismatched sky-positions and verify that the SNR does decrease as
expected.  The work of \citep{prix05} (see also Appendix A of
\cite{kri04}) which studies correlations in frequency and sky-position
mismatch might be useful for this purpose.

\section{Future Detectability}
\label{detect}

\subsection{Best case detectability for various emission models}

We will start by looking at the best case scenarios for detection, and
ask what would happen if we knew all of the parameters to sufficient
accuracy that we only had to search one template for each source,
using the flux information collected in Section \ref{stars} and
summarized in Table \ref{sdata}.  In
this case computational cost is not an issue, and we can integrate for
long periods. We will assume that the system is in perfect
spin balance (so that we can neglect spin derivatives), and that
gravitational wave torques are the only negative torques operating in
the system.  This means that we neglect any possible spin-down effects
due to the interaction of the neutron star magnetosphere with the
accretion disk, or any magnetic dipole spin-down.

\begin{figure}
\begin{center}
\includegraphics[width=8cm, clip]{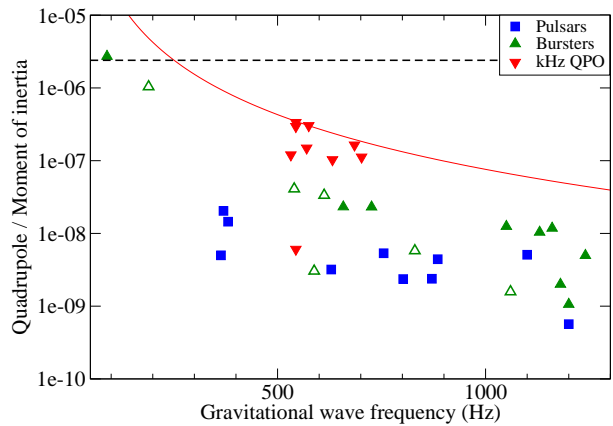}
\end{center}
\caption{The quadrupole $Q$ required for spin balance in the
  mountain scenario given the long-term average flux.  $Q$ is scaled
  by $10^{45}$ g cm$^2$ (the
  approximate moment of inertia of a neutron star). The
  frequency at which the kHz QPO symbols appear is the centre of the
  measured range of separations. The solid line illustrates how the requisite
  $Q$ would vary for Sco X-1 if this is not the spin frequency. The dashed line
  shows the maximum feasible non-magnetic quadrupole calculated by
  \citet{has06}. The uncertainty in the inferred quadrupole is not
  shown, but can be substantial since it depends on the distance to
  the star (see Table \ref{sdata}). }
\label{infquad}
\end{figure}

In Section \ref{overview} we gave an overview of the spin balance model and
calculated the best case detectability assuming gravitational wave
emission due to a `mountain', balancing the long-term average flux.  The results were shown in Figure 
\ref{mountpers}.  We also however need to consider whether the quadrupole $Q$ required for spin
balance is feasible.

\begin{equation}
\label{Qreq}
Q = 7.4\times 10^{35} {\rm g~cm^2} \left(\frac{R_{10}^3}{M_{1.4}}\right)^{1/4}
d_{\rm kpc} F_{-8}^{1/2}
\left(\frac{{\rm 1~kHz}}{\nu_s}\right)^{5/2}
\end{equation}
where $d_{\rm kpc} = d$/(1 kpc).  Figure \ref{infquad} shows the
required quadrupole (scaled by  $10^{45}$ g cm$^2$, the approximate
moment of inertia of a neutron star), assuming the long-term average
flux.   We also show the maximum
sustainable quadrupole for an accreted
crust computed by \citet{has06} (note that these authors compute
$Q_{22} = (15/8\pi)^{1/2} Q$).  The values required for the slower
spinning bursters are at the upper boundary of what is thought to
be feasible, and if the spin of the kHz QPO sources is slower than the
measured separations, the required quadrupole may exceed this value.
However, magnetic confinement could support larger quadrupoles
\citep{mel05, pay06}.

It is clear from Figure \ref{mountpers} that if the kHz QPO sources do have spins in
the range inferred from the kHz QPO separation (or higher) then Sco
X-1 is the only source that is in principle marginally detectable by Enhanced
LIGO.  Predicted amplitude would of course rise if the spin frequency
were substantially lower, and if this were the case both Sco X-1 and
GX 5-1 could be within detectable range for Enhanced LIGO (although
the inferred quadrupole would be large, approaching the maximum
thought possible).  

For Advanced LIGO,
several of the kHz QPO sources are in principle detectable for a
single template search in the broad band configuration.  Two of the
burst oscillation sources (XB 1254-690 and 4U 1728-34) are also
marginally detectable within the given narrow band envelope.  Several
other burst oscillation sources might be detectable if it were
possible to push the narrow band envelope further down towards the
thermal noise floor.  If the
proposed Einstein Telescope reaches its design specification, several
more of the burst and kHz QPO sources might be reachable, although the
pulsars remain undetectable.

\begin{figure*}
\begin{center}
\includegraphics[width=16cm, clip]{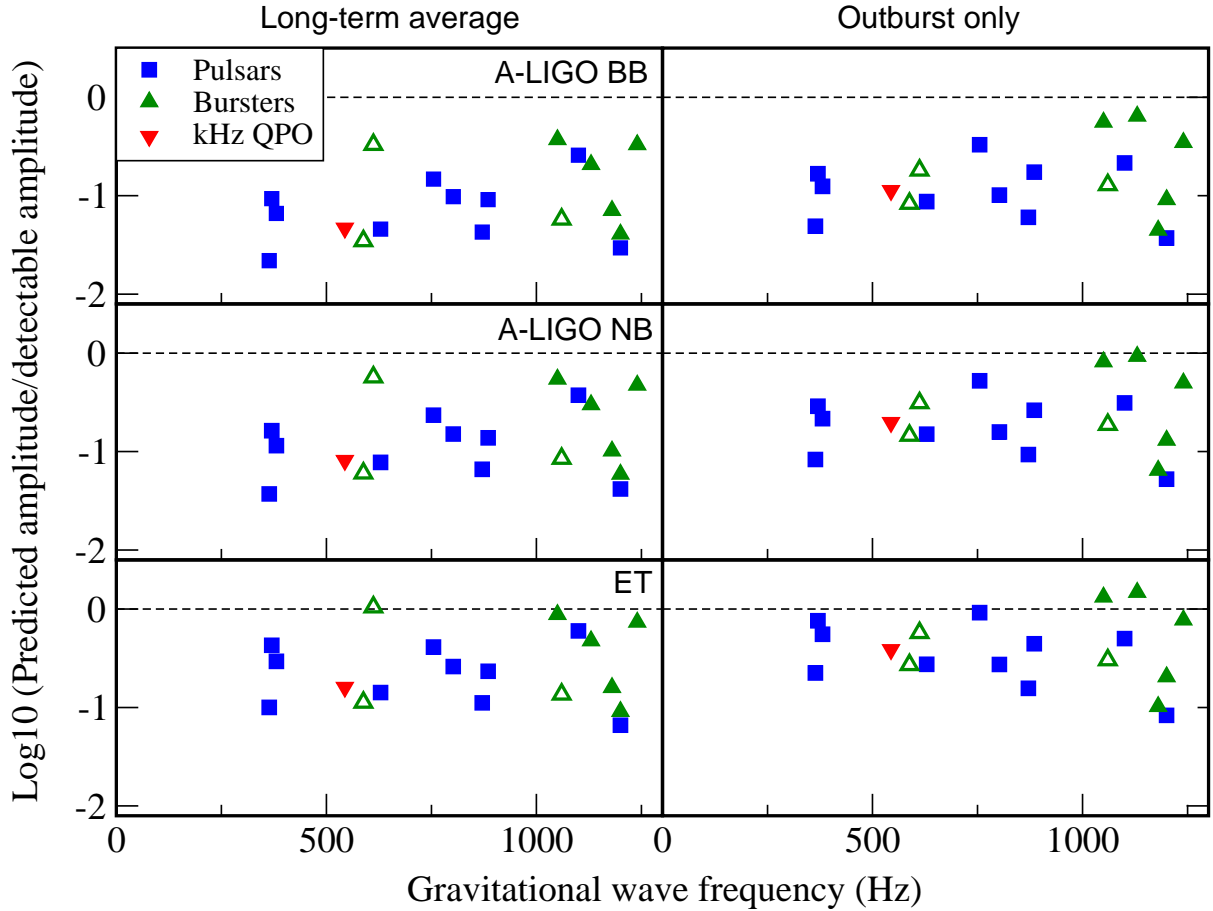}
\end{center}
\caption{Best case detectability (single template) for the transients, in the mountain scenario. The left-hand panels show the detectability if gravitational wave emission balances the long-term average accretion rate, for $T_{\rm obs}$ = 2 years.  The right-hand panels show the detectability if we if we consider the quadrupole necessary to balance the accretion torque during outburst.  $T_{\rm obs}$ for each source is now either   
  the outburst duration or the maximum integration time (2 years),
  whichever is longer.  In order to compare sources with different
  integration times, we show $\log_{10}$ of the ratio of the predicted
to the detectable amplitude. The three rows are for different detector configurations: Top - Advanced LIGO broad band; Middle - Advanced LIGO Narrow Band Envelope; Bottom - Einstein Telescope.  }
\label{mounttrans}
\end{figure*}

One of the major uncertainties in our modelling is the response to
variations in accretion rate.  In order to gauge this effect we therefore
consider an alternative model for the transients, one in which the 
gravitational wave torque balances the accretion torque in outburst.
This assumes that the gravitational wave emission mechanism responds
very rapidly to the accretion. Although response timescales are not
well studied, this scenario is not unreasonable - an
accretion-induced mountain or unstable oscillation, for example, may
well grow during outburst and
decay during quiescence.  Figure \ref{mounttrans} compares the best case detectability for the transients if we balance the outburst flux rather than the long-term average flux (Table \ref{sdata}).  The
integration time in outburst is taken to be either the typical outburst duration, or the
maximum integration time, whichever is larger.  The required
quadrupoles are all below the \citet{has06} limit.  Many of the
sources show an improvement:  the reduction in $T_{\rm obs}$ is more
than compensated for by the increase in flux.  The burst oscillation
sources X 1658-298 and KS 1731-260, for example, which are undetectable if we consider
time-average flux, lie within this scenario on or just above the Advanced LIGO
Narrow Band envelope.  The intermittent pulsar HETE J1900.1-2455 also becomes a more promising source for the Eintein Telescope.  Further theoretical consideration should clearly be given to the issue of
torque response timescales.  

\begin{figure*}
\begin{center}
\includegraphics[width=16cm, clip]{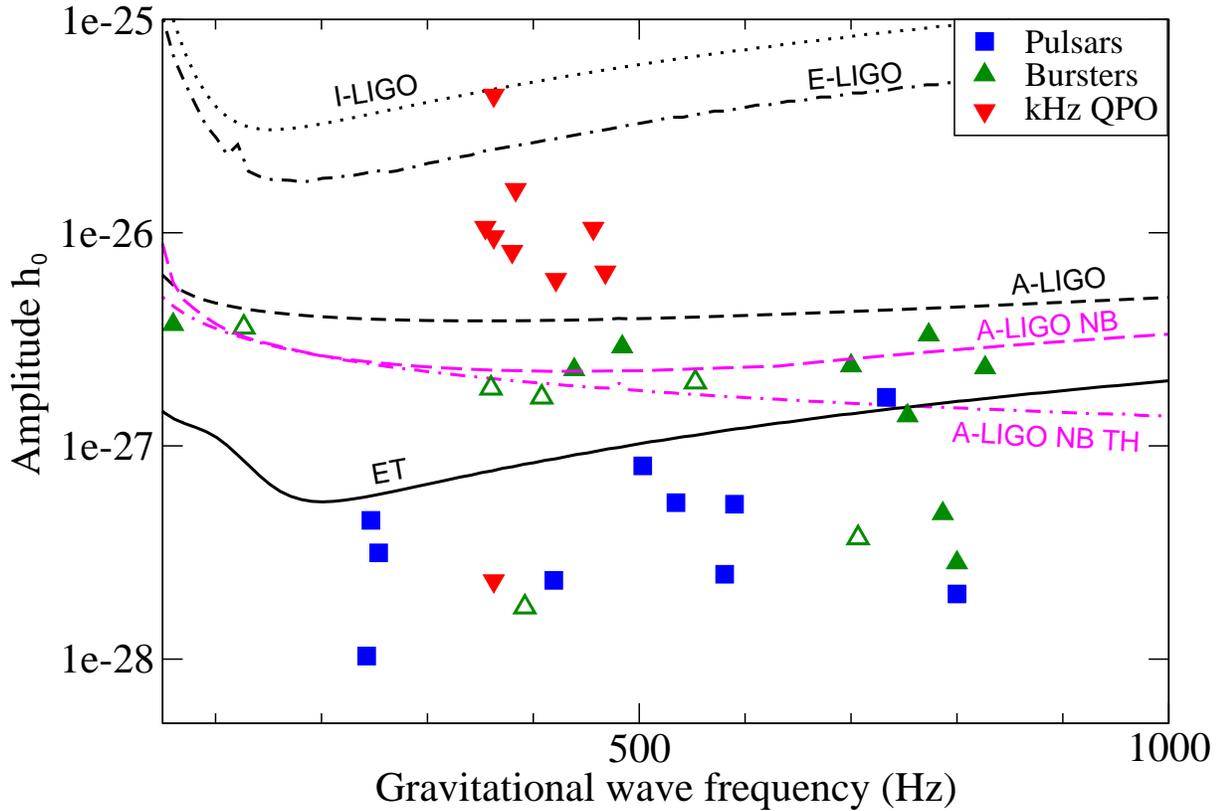}
\end{center}
\caption{Best case detectability (single template search) using the long-term average flux,
  for the r-mode scenario.  Symbols and lines as for Figure \ref{mountpers}.  }  
\label{rmodepers}
\end{figure*}

We also consider the situation
that would result if the accretion torque is balanced
by spin-down due to an internal r-mode rather than a mountain.  We
first make the assumption that the star has a spin rate and
temperature such that it can sustain an unstable r-mode
(so that it lies in the `r-mode instability window', see
\citealt{ande01}).  This assumption may not be warranted, particularly for
some of the more slowly rotating stars.  We also assume that
the r-mode amplitude is steady: r-mode unstable stars may well
experience duty cycles with short-lived periods of strong spin-down
\citep{levi99, ande00, hey02, kin03}, but there are some scenarios in which
 r-mode emission may be persistent \citep{ande02, wag02, nay06, bon07}. The
 gravitational wave torque associated with the dominant $l=m=2$ r-mode
 ($l$ and $m$ being the
standard angular quantum numbers) is

\begin{equation}
N_\mathrm{gw} =  - 3 \tilde{J} \Omega_s \alpha^2 M R^2/\tau_g
\label{rmodetorque}
\end{equation}
\citep{ande01}.  The frequency of the emitted gravitational radiation is not a
harmonic of the spin frequency:  for the $l=m=2$ r-mode it is 4/3 times the spin
frequency\footnote{We assume that this calculation of the
  relationship between mode
  frequency and spin is exact.  In practice there is some theoretical
  uncertainty, which would further increase the number of templates to
  be searched.}. The quantity $\alpha$ is the mode amplitude.  For an $n=1$
 polytrope model of the neutron star, $\tilde{J}
= 1.635\times 10^{-2}$ and the radiation
reaction timescale $\tau_g$ is given by

\begin{equation}
\tau_g = 47 M_{1.4}^{-1} R_{10}^{-4} \left(\frac{{\rm
      1~kHz}}{\nu_s}\right)^6 {\rm ~s}
\end{equation}
\citep{ande01}.  The associated luminosity is $\dot{E}_{\rm gw} = N_{\rm gw}
\omega_m /
m$, where $\omega_m = \omega_{\rm gw}$ is the angular frequency of the
mode.  Assuming spin balance, $N_a = N_{\rm gw}$, the resulting
gravitational wave amplitude is

\begin{equation}
h_0 = 3.7 \times 10^{-27} \left(\frac{R_{10}^3}{M_{1.4}}\right)^{1/4}
F_{-8}^{1/2}
\left(\frac{{\rm 1~kHz}}{\nu_s}\right)^{1/2}
\label{rmode_amp}
\end{equation}
In Figure \ref{rmodepers} we compare the predicted and
detectable amplitudes for the r-mode scenario for $T_{\rm obs} = 2$ years.

\begin{figure}
\begin{center}
\includegraphics[width=8cm, clip]{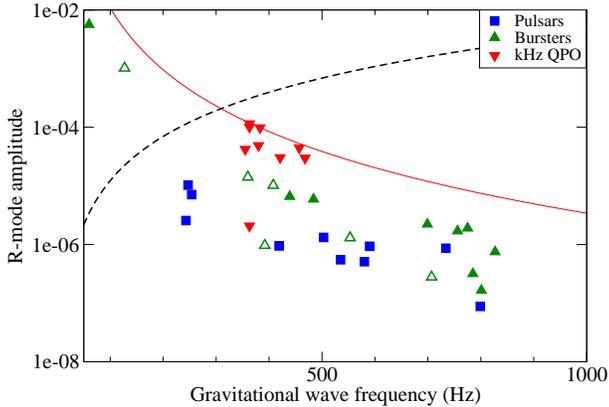}
\end{center}
\caption{The r-mode amplitude $\alpha$ required for spin balance in the
 r-mode scenario, using the long-term average
  flux for all sources (persistent and transient).  The dashed line
  shows the maximum (saturation) amplitude calculated by \citet{arr03}.
  The uncertainty in the inferred amplitude is not shown, but can be
  substantial since it depends on the distance to the star. The solid line indicates the r-mode amplitude that would be required for Sco X-1 if the spin frequency differs from that inferred from kHz QPO separation. }
\label{infamp}
\end{figure}

We now consider whether the required r-mode
amplitude is feasible.  Assuming spin balance, the mode amplitude is

\begin{equation}
\alpha = 6.9\times 10^{-8} \left(M_{1.4}^5
  R_{10}^9\right)^{-1/4} d_{\rm kpc} F_{-8}^{1/2}
\left(\frac{{\rm 1~kHz}}{\nu_s}\right)^{7/2}
\end{equation}
Figure \ref{infamp} shows the amplitude required for spin balance
for each source for the long-term average flux.   The required amplitudes should be
compared to the maximum saturation amplitude $\alpha_s$
computed by \citet{arr03}:

\begin{equation}
\alpha_s \approx 8\times 10^{-3} \left(\frac{\nu_s}{{\rm 1~kHz}}\right)^{5/2}
\end{equation}

The values required for the two slowest
spinning bursters exceed the amplitude limit, as would the kHz QPO
sources if the spin is much slower than the
measured separations.
However for all other sources the
required amplitude is below the \citet{arr03} limit.

The gravitational wave frequency is lower within the r-mode scenario than in the
mountain scenario. Spin balance therefore requires a higher GW amplitude, making
the r-mode emission model rather more optimistic. For Advanced LIGO
many of the kHz QPO sources lie above the broad band noise curve. Two
of the burst oscillation sources with reasonable values of $\alpha$, 4U 1636-536 and 4U 1728-34, sit well above the Advanced LIGO narrow band
envelope, while 4U 1702-429 and KS 1731-260 lie on the
curve (XB 1254-690, which would also be marginal for the narrow band configuration, has an unfeasibly high $\alpha$ ).  For the Einstein
Telescope, however, these sources are well above the noise curve, and the
intermittent pulsar Aql X-1 is also marginally detectable.  

If we consider the transient scenario (balancing outburst torque
rather than time-average torque) the r-mode model also makes rather
more optimistic predictions than the mountain model.   Three burst oscillation
sources that are marginally detectable at best when balancing
long-term flux now come within
range of Advanced LIGO's Narrow Band configuration:  XB 1658-298, KS
1731-260 and 4U 1608-522.  By the time we reach the sensitivity of the
Einstein Telescope the intermittent pulsar HETE J1900.1-2455 may also be
detectable in outburst.

\subsection{Data analysis techniques}
\label{subsec:damethods}

We consider two kinds of searches over our parameter space (either of
these methods can also be used as parts of an optimized multi-stage
hierarchical scheme \citep{brad00,cgk05,meins07}):
\begin{description}
\item[(i)] A full coherent matched-filter search using all available
  data, possibly from multiple detectors.  This involves a demodulation
  of the data for a given parameter space point using the so called
  $\mathcal{F}$-statistic; details can be found in \citep{jks,cs05}.
  Such a coherent demodulation statistic can be augmented by the
  technique of combining sidebands as described in \citet{mes07}
  (adapted from similar techniques used in radio pulsar searches, \citet{ran03}).
\item[(ii)] A power folding method where the available data is broken
  up into smaller segments (often called ``stacks'' in the GW
  literature); the duration of each stack is $T_{\rm coh}$ and $N_{\rm
    stacks}$ denotes the number of stacks. If there were no gaps in
  the data, we would have $T_{\rm obs} = N_{\rm stacks}T_{\rm coh}$,
  but duty factors of $\sim 80 \%$ are more likely.  Each segment is
  coherently demodulated using (i) and excess power from each segment
  is combined without maintaining phase coherence.  Examples of such
  power folding methods are: a) stack-slide \citep{brad00} where the
  relevant statistic is simply a weighted sum of
  $\mathcal{F}$-statistic values (at the appropriate frequency bins)
  \begin{equation}
    \label{eq:29}
    \rho = \sum_{k=1}^{N_{\rm stacks}} w_k \mathcal{F}_{(k)}\,,
  \end{equation}
  or b) the Hough transform method \citep{kri04} where one adds
  weighted binary number counts
  \begin{equation}
    \label{eq:30}
    \rho = \sum_{k=1}^{N_{\rm stacks}} w_k n_{(k)}
  \end{equation}
  where $n_{(k)}$ is either 0 or 1 depending on whether the $\mathcal{F}$
  exceeds a certain threshold.  In each case, the weights $w_k$ are
  chosen to optimize the sensitivity.
\end{description}
For both (i) and (ii), the computational cost is proportional to the
total number of templates $N_{\rm temp}$ and to the amount of data
available.  Thus, for (i), it is approximately $A T_{\rm obs}N_{\rm
  coh} $ for some constant $A$, while for (ii) it is proportional to $
N_{\rm stacks} T_{\rm coh}N_{\rm temp}$.  The precise value of the sensitivity of
these searches clearly depends on the details of the analysis method
and software, and the quality of data, such as the duty cycle of the
detector (which might reduce the amount of data actually available),
the presence of noise artifacts such as spectral disturbances,
uncertainties in the calibration of the detector, and so on.  It is thus not
possible to estimate the sensitivity to better than, say, $\sim
5-10\%$ without actually carrying out the search, and it is in fact
even pointless to try and do so in this paper.  It is of course
possible to get semi-realistic estimates and this is what we shall do,
but these uncertainties should always be kept in mind.

Starting with the statistical factor, note that if the threshold corresponding
to a single trial false alarm rate is $\alpha$, the probability that the
threshold is crossed at least once in $N_{\rm trials}$ independent trials is $FA
= 1-(1-\alpha)^{N_{\rm trials}} \approx \alpha N_{\rm trials}$ when $\alpha
N_{\rm trials} \ll 1$ \citep{jks}. So we choose $\alpha = FA/N_{\rm trials}$ to
ensure that the total false alarm probability is $FA$.  We equate the number of
trials with the total number of templates that must be searched. Strictly
speaking, this is not true because the different templates are not completely
independent.  This is usually not a significant effect as long as the mismatch
$m$ used to construct the template bank is not too small; we shall use a
reference value of $m=0.3$.  To make matters more complicated, while there are
reliable calculations for the template counting for (i) as presented in
Sec.~\ref{subsec:templates}, there are as yet no reliable estimates for (ii). We
shall therefore consider only the number of templates in each coherent segment.
Fortunately, the dependence of the statistical factor is not very steep (in fact
slower than logarithmic) with the number of trials, so this does not make a
significant difference to our results.

\begin{figure}
\begin{center}
\includegraphics[width=8cm, clip]{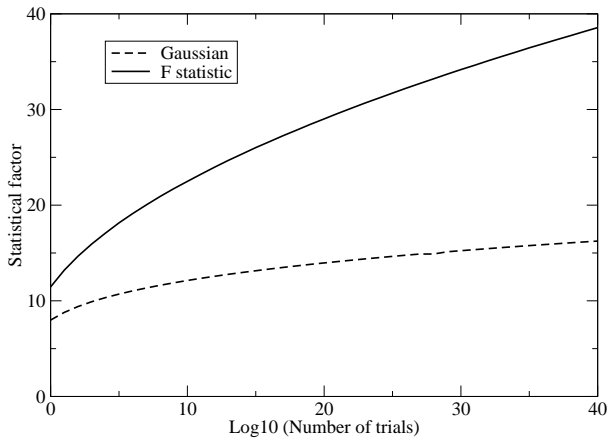}
\end{center}
\caption{Dependence on the number of independent trials (templates) of
  the statistical factor $F_{\rm stat}$ that appears in the
  detectability equation (\ref{eq:9}). Note that $F_{\rm stat} = 11.4$
  when $N_{\rm trials} = 1$ which reproduces equation (\ref{eq:8}). }
\label{StatFactor}
\end{figure}

Some details of the statistical calculation are in order.  The choice
of $\alpha$ determines the thresholds for both the coherent
$\mathcal{F}$-statistic search and the semi-coherent search.  The
exact relation is however different.  The $\mathcal{F}$-statistic
follows a $\chi^2$-distribution with 4 degrees of freedom (it is
actually $2\mathcal{F}$ which is $\chi^2$ distributed).  For the
semi-coherent statistic $\rho$ is, to a reasonable approximation,
simply Gaussian (assuming $N_{\rm stacks}$ to be sufficiently large);
it is actually $\chi^2$ with $4N_{\rm stacks}$ degrees of freedom.
The assumption of gaussianity may be questionable here because we are
after all dealing with the tail of the distribution, and the central
limit theorem may not be reliable especially when $N_{\rm stacks}$ is
not particularly large. A further complication arises when $\rho$ is
not simply a sum of the $\mathcal{F}$-statistic values but is perhaps
a Hough statistic \citep{kri04} when the distribution of $\rho$ is
closer to a binomial (which can also be approximated by a Gaussian).
We shall nevertheless assume gaussianity for our purposes and leave a
more detailed study for future work.  The relations between $\alpha$
and the corresponding threshold $\mathcal{F}_{\rm th}$ and $\rho_{\rm
  th}$ are then different in the two cases:

\begin{equation}
  \label{eq:26}
  \alpha = (1+\mathcal{F}_{\rm th})e^{-\mathcal{F}_{\rm th}} \,,\qquad
  \alpha = \frac{1}{2}\textrm{erfc}\left(\frac{\rho_{\rm
        th}}{\sqrt{2}\sigma} \right)
\end{equation}
where $\textrm{erfc}$ is the complementary error function, and
$\sigma$ is the standard deviation of $\rho$. In the presence of a
signal with amplitude $h_0$, the mean values of $\mathcal{F}$ and
$\rho$ are increased by an amount proportional to the SNR$^2$, and thus to
$h_0^2$.  Taking a fixed value of the false dismissal rate $\beta$
determines the minimum value of $h_0$ needed to exceed the thresholds
$\rho_{\rm th}$ and $\mathcal{F}_{\rm th}$; we shall always take $\beta=0.1$ as
the reference value.  Folding in a uniform averaging over all possible
pulsar orientations along with the statistical factor in a single
parameter $F_{\rm stat}$, we get the following expression for the
sensitivity
\begin{equation}
  \label{eq:9}
  h_0^{\rm sens} \approx \frac{F_{\rm stat}}{N_{\rm stacks}^{1/4}}\sqrt{\frac{S_h}{D T_{\rm coh}}}\,.
\end{equation}
where $F_{\rm stat}$ varies with the number of trials required, as
shown in Figure~\ref{StatFactor}.  This factor was also discussed in
\citep{abb07a} and \citep{abb05a} which are examples of (i) and (ii)
respectively.  In each case, $F_{\rm stat}$ increases slower than
logarithmically with $N_{\rm trials}$, and the statistical factor is
worse for the $\chi^2$ distribution as expected because it is not as
sharply peaked as the Gaussian. Finally, it is perhaps worth
mentioning that averaging over the pulsar orientation $\cos\iota$ may
not always be appropriate, especially if there happen to be
independent estimates of the neutron star orientation.  If the value
of $\cos\iota$ happens to be anywhere in the top $1-p$ percentile of a
uniform distribution (i.e. if we average over all $|\cos\iota| \geq
p$), then
\begin{equation}
  \label{eq:28}
  F_{\rm stat} \rightarrow \frac{F_{\rm stat}}{\sqrt{1 + 11(p+p^2)/16 + (p^3+p^4)/16}}\,.
\end{equation}
So for example, if $p = 0.9$, we get $F_{\rm stat} \rightarrow F_{\rm
  stat}/1.5$.  This could be important for sources which happen to be
near the detection threshold.

So far we have incorporated the statistical and geometrical factors in
Equation (\ref{eq:9}).  For template bank based searches, the sensitivity is
further degraded because of the discreteness of the template
grid.  For a template bank created with a maximum mismatch of $m$,
assuming a cubic grid, the average degradation in the SNR is
$(1-m/3)$.  This corresponds to a degradation of $\sqrt{1-m/3}$ in
$h_0$ which is $\approx 0.95$ for $m=0.3$.

The next question we need to address is the computational cost and
whether it is necessary or worthwhile to do the semi-coherent search
(ii).  Clearly, for a given amount of data while we ideally want to
make the coherent integration time $T_{\rm coh}$ as large as possible,
and the number of segments $N_{\rm stacks}$ correspondingly small, the
choice of these parameters is dictated by the computational cost.
This computational burden is in fact, by far, what limits the search
sensitivity (As we just saw, $F_{\rm stat}$ has a very weak dependence
on the number of templates; we need to change the number of templates
by orders of magnitude before it has an appreciable effect).  The
bigger effect of reducing the number of templates by better
astrophysical modelling or observations is that it allows us to have a
larger coherent integration time $T_{\rm coh}$.

For a given source, we would first like to know if a semi-coherent
search would be useful.  The obvious answer is: whenever computational
cost is an issue, but a more quantitative answer is also easy to get.
Let us compare the sensitivities of (i) and (ii) for a given
computational cost.  Let us assume that the cost for the coherent
analysis scales as a power of $T_{\rm obs}$: $C_{\rm coh} = AT_{\rm
  obs}^k$ for some constants $A$ and $k$. So if we assume a fixed
value of the computational cost $C_0$, then $T_{\rm obs} =
(C_0/A)^{1/k}$.  Thus, the sensitivity is $h_{0}^{\rm coh} \propto
T_{\rm obs}^{-1/2} = (C_0/A)^{-1/2k}$.  For a semi-coherent search on
the other hand, we have $C_0 \approx A N_{\rm stacks}T_{\rm coh}^k$
(this is true if the cost of combining the different stacks is
negligible).  In this case we get $T_{\rm coh} = (C_0/AN_{\rm
  stacks})^{1/k}$.  Using $h_0^{\rm semi-coh} \propto N^{-1/4}T_{\rm
  coh}^{-1/2}$ we get
\begin{equation}
  \label{eq:21}
  \frac{h_0^{\rm semi-coh}}{h_0^{\rm coh}} = N_{\rm stacks}^{(2-k)/4k}\,.
\end{equation}
This tells us that a semi-coherent search is not effective (i.e.
$h_0^{\rm semi-coh} > h_0^{\rm coh}$) for $k\leq 2$, and it gets more
and more effective for larger $k$ (this conclusion is robust: it is
not affected by the approximation of neglecting the cost of the
semi-coherent step).

In our present case, recall from equations
\eqref{eq:19}--\eqref{eq:25} that we have very different scalings in
the regimes $T_{\rm obs} \gg P_{\rm orb}$ and $T_{\rm obs} \ll P_{\rm
  orb}$.  So, for binary systems for which the parameters have been
sufficiently constrained astronomically and we can afford large
integration times, we shall assume that we are better off doing a
simple coherent search.  For others, a semi-coherent strategy has a
much larger impact.\footnote{Semi-coherent searches might start
  becoming more important as soon as we need to start including other
  parameters, especially frequency derivatives and $\dot{P}_{\rm
    orb}$.}  For each potential source, our strategy is to first
estimate the number of templates that are required for a coherent
search.  With reasonable estimates of available computational
resources, and assuming $T_{\rm obs}= 2$ years, this determines the
maximum coherent integration time $T_{\rm coh}$ that can be analysed.
If $T_{\rm coh}$ is not much smaller than $P_{\rm orb}$, then by the
previous argument we assume that the gain in using a semi-coherent
search is not significant, and we restrict ourselves to a pure
coherent search.  On the other hand, if this $T_{\rm coh}$ does turn
out to be much less than $P_{\rm orb}$, then we consider a
semi-coherent search and estimate $T_{\rm coh}$ and $N_{\rm stacks}$
(assuming here the cost of the semi-coherent step to be comparable to
the coherent step).  Equation (\eqref{eq:9}) then yields an estimate of
the search sensitivity $h_0^{\rm sens}$ for each potential source.  We
take the statistical factor $F_{\rm stat}$ in Equation (\eqref{eq:9}) using just
the number of templates in the coherent step. There is as yet no
systematic study of the semi-coherent metric for binary systems, and
this estimate should be updated as soon as these calculations are
available.  We expect that this approximation will not make a
qualitative difference to our final results because of the weak
dependence of $F_{\rm stat}$ on the number of templates, but this
needs to be verified.  In the previous section, we estimated the
\emph{estimated} GW amplitude $h_0^{\rm exp}$ from each source, and
the ratio $h_0^{\rm exp}/h_0^{\rm sens}$ can be a useful
``detectability'' ranking for each source.

In the cases when $P_{\rm orb}$ is essentially unknown, the ranges of
$T_{\rm asc}$ and $a_{\rm x}\sin i$ that we have chosen depend on $P_{\rm
  orb}$.  This means that the parameter space region is not
rectangular and thus the template counting equations of
Section \ref{subsec:templates} will over-estimate the number of templates
significantly, though not by orders of magnitude.

\subsection{Numbers of templates}
\label{uncertainty}

\begin{figure}
\begin{center}
\includegraphics[width=8cm, clip]{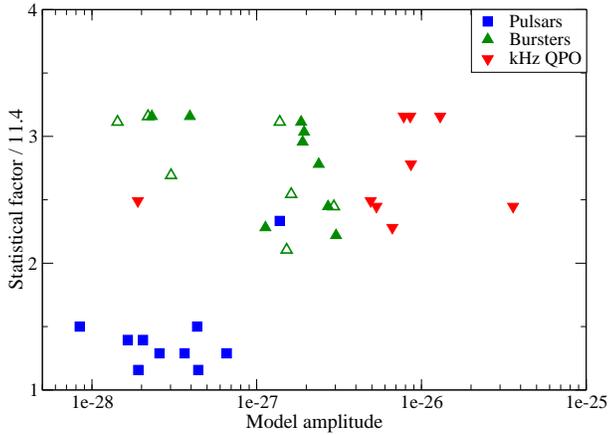}
\end{center}
\caption{The factor by which $F_{\rm stat}$ increases over the single
  template value ($F_{\rm stat}
= 11.4$) plotted against the predicted amplitude for the mountain
scenario (Equation \ref{mount_amp}).   The brightest and most
promising sources are the most poorly constrained, and this
is reflected in the effect on $F_{\rm stat}$. The difference in $F_{\rm stat}$ between the two strongest sources, Sco X-1 and GX 5-1, is primarily due to the fact that Sco X-1 has much better orbital constraints.}
\label{fstatscaled}
\end{figure}

We can now compute the effect on our searches of parameter uncertainty.  We will start by considering the effect on 
the statistical factor $F_{\rm stat}$, neglecting computational cost
issues.  In other words, we assume that it is feasible to do a coherent
fold.  In Tables \ref{pulsnohit}, \ref{burstnohit} and \ref{khznohit}
we summarize the parameter ranges assumed for the pulsars, bursters
and kHz QPO sources respectively.  The Tables show the associated
$N_{\lambda^i}$ for each parameter (for $T_{\rm obs} = 2$ years), the
resulting total number of templates that must be searched $N_{\rm temp}$, and the effect on
$F_{\rm stat}$.  Figure \ref{fstatscaled} shows the change
in the statistical factor for each source.  

In Figure \ref{mountpers}, were we showed results for a single template search, each source had $F_{\rm stat} = 11.4$.  This meant that we could plot a single detectability threshold curve for each detector.  When each source has a different $F_{\rm stat}$, however, the detectability threshold curves differ for each source.  One way of comparing detectability for sources with different noise threshold curves is to plot the ratio of emitted to detectable amplitude:  this is the approach that we took for the transients (Figure \ref{mounttrans}), and that we will adopt when we come to consider computational cost.  What we can also do, however, to see the effect, is to scale the predicted amplitude $h_0$ by $11.4/F_{\rm stat}$.  Predicted amplitude does not really change, of course, but it is a useful way of visualising the impact of the statistical factor.   

\begin{figure*}
\begin{center}
\includegraphics[width=16cm, clip]{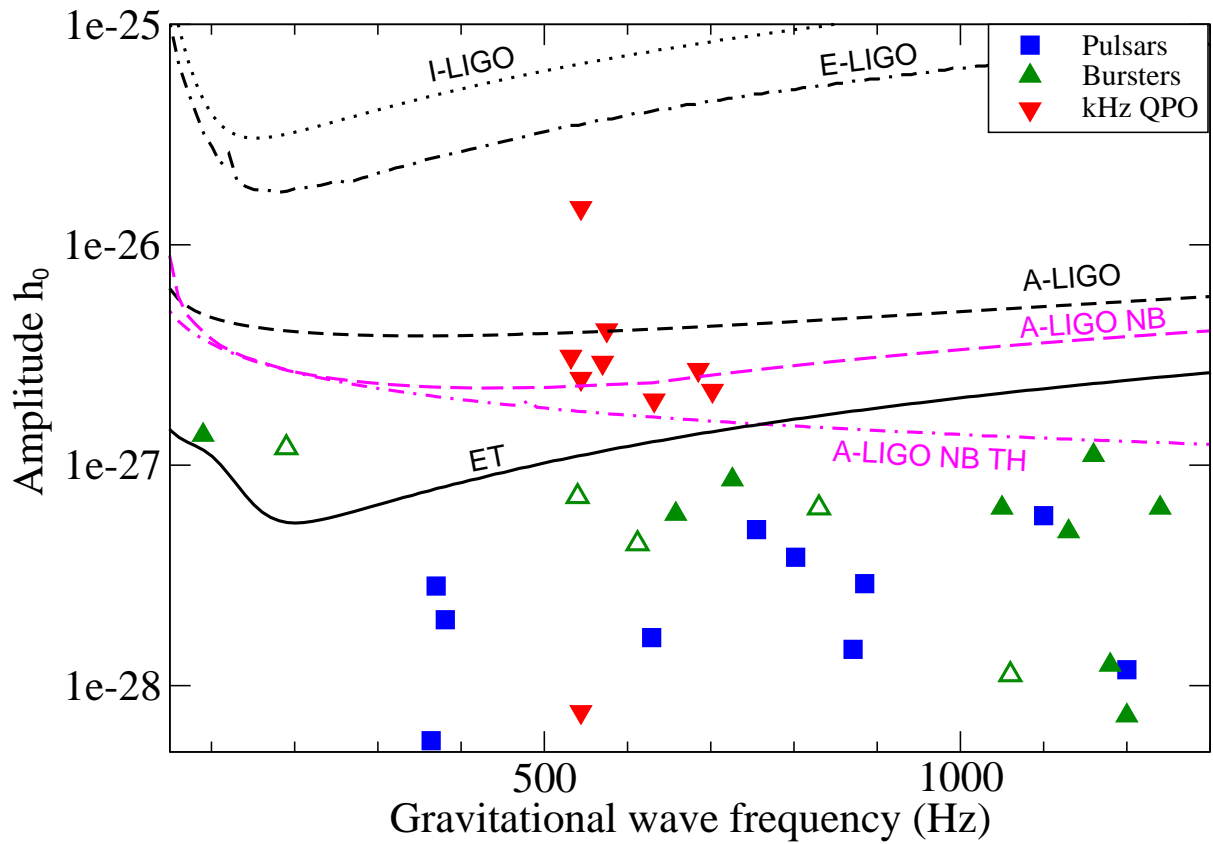}
\end{center}
\caption{Effect on detectability for the mountain scenario (assuming long-term average flux and a coherent fold with $T_{\rm
    obs} = 2$ years), taking into account the effect on $F_{\rm stat}$ associated with the fact that $N_{\rm temp} > 1$.  Compare to the best case detectability shown in Figure \ref{mountpers}.  As in Figure \ref{mountpers}, the noise curves are computed assuming $F_{\rm stat}$ = 11.4 (the single template value), but we have scaled the predicted amplitudes to reflect the fact that $F_{\rm stat}$ is larger.  Although this is not strictly ``correct'' (it is the thresholds that should move, not the predicted amplitudes) this is a useful way to visualise the impact.  See the text for more details.}
\label{mountpersnohit}
\end{figure*}

Figure \ref{mountpersnohit} shows the impact on the mountain scenario:  it should be compared to Figure \ref{mountpers}.  Only for the most tightly constrained
pulsars is detectability largely unchanged:  for the majority
of sources the detectable amplitude falls by a factor of 2-3 compared
to a single template search. Although this does not sound like a great
deal, it is sufficient to push all sources
except Sco X-1 below the detection threshold for Advanced
LIGO broad band\footnote{Sco5 X-1 would also be marginal for Enhanced LIGO, over a
  restricted frequency range.}.  The effect on the r-mode scenario,
summarized in Tables \ref{pulsnohit}-\ref{khznohit}, is
slightly less severe, leaving
Sco X-1, GX 5-1, GX 349+2 and 4U 1820-30 above the detection threshold for
Advanced LIGO broad band.   The situation is better for the narrow band configuration:  although none of the burst oscillation sources remain within range (for any of the emission scenarios considered), several of the kHz QPO sources are still viable.  The spins for these sources are poorly constrained.  However, the anticipated narrow band configurations (see top panel of Figure \ref{fig:sh}) have a reasonably broad bandwidth, leaving ample scope for searches.  

We can now take the final step and look at the impact of computational
constraints on searches involving multiple templates. The available computational power sets the length of data
$T_{\rm obs}$ that can feasibly be analysed within a given amount of
time.  In Table \ref{comphit} we summarize the impact in the situation
where we assume a maximum analysis time of 2 years, assuming that for
Advanced LIGO we have computational power 50 times greater than that
currently available within the LSC, while for the Einstein Telescope
we assume 100 times more computing than at present. These assumptions
are, of course, arbitrary, since the computing power that can be
applied to future searches depends not just on technology (e.g. Moore's
Law) but also on improvements in data analysis techniques, and of
course also on how much money is spent on computing by future
projects. We have taken numbers we feel are defensible, but they may
turn out to be conservative, especially for the Einstein Telescope.

\begin{figure*}
\begin{center}
\includegraphics[width=16cm, clip]{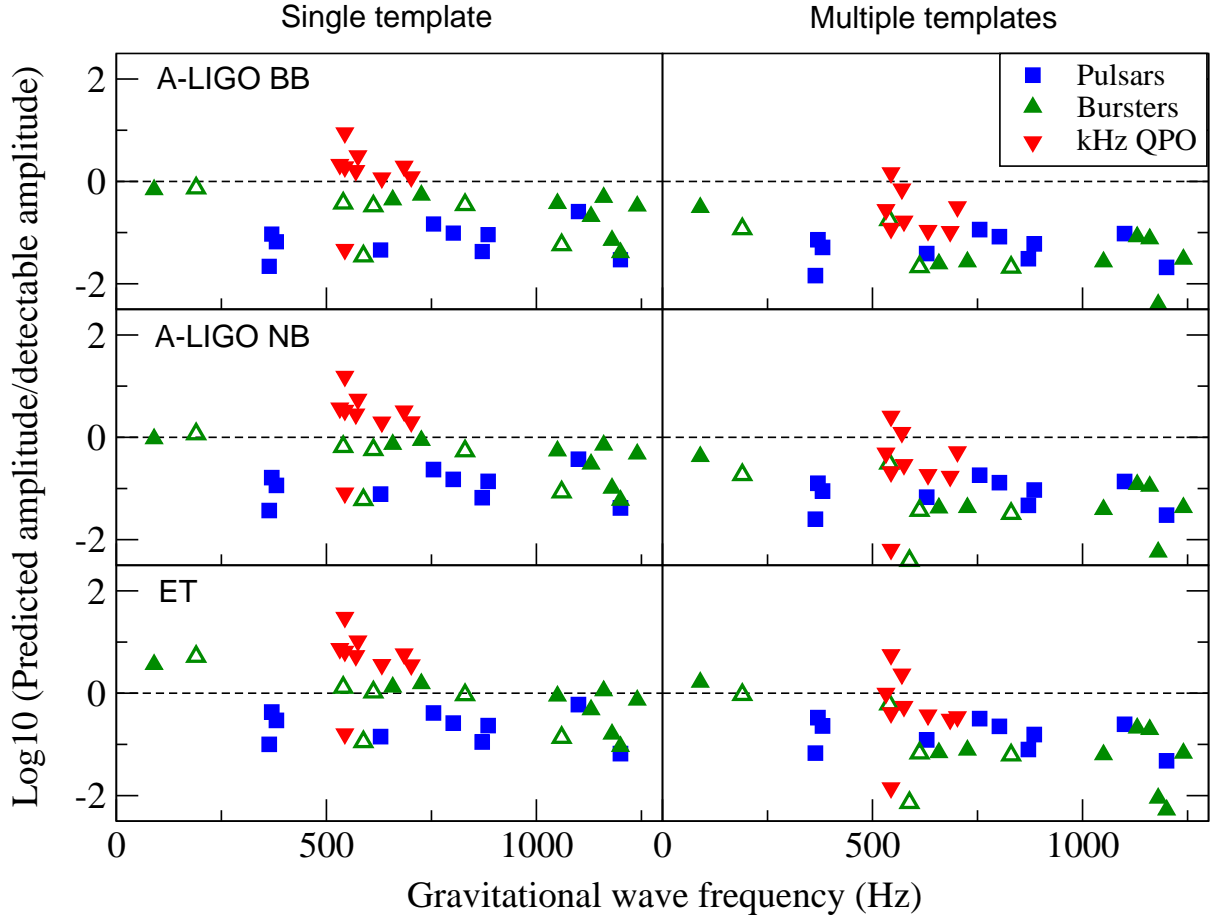}
\end{center}
\caption{Detectability for the mountain scenario (long-term average flux, maximum $T_{\rm obs} = 2$ years), taking into account computational limitations as well as the effect on $F_{\rm stat}$.  The computational cost of searching multiple templates affects the feasible integration time and the choice of coherent/semi-coherent search technique (Table \ref{comphit}).  In order to compare detectability for sources with different integration times we plot $\log_{10}$ of the ratio of predicted to detectable amplitude.  The three rows are for different detector configurations:  Top - Advanced LIGO broad band; Middle - Advanced LIGO narrow band envelope; Bottom - Einstein Telescope.  The left hand panels show detectability for the best case (single template) search, as in Figure \ref{mountpers}:  the right hand panels show the situation after including the statistical and computational limitation.  For Advanced LIGO we
  assume a 50-fold increase in
  computational efficiency as compared to the current situation;  for
  the Einstein Telescope we assume a 100-fold increase.  In both cases we assume a maximum analysis time of 2 years.}
\label{mountperscomphit}
\end{figure*}

\begin{figure*}
\begin{center}
\includegraphics[width=16cm, clip]{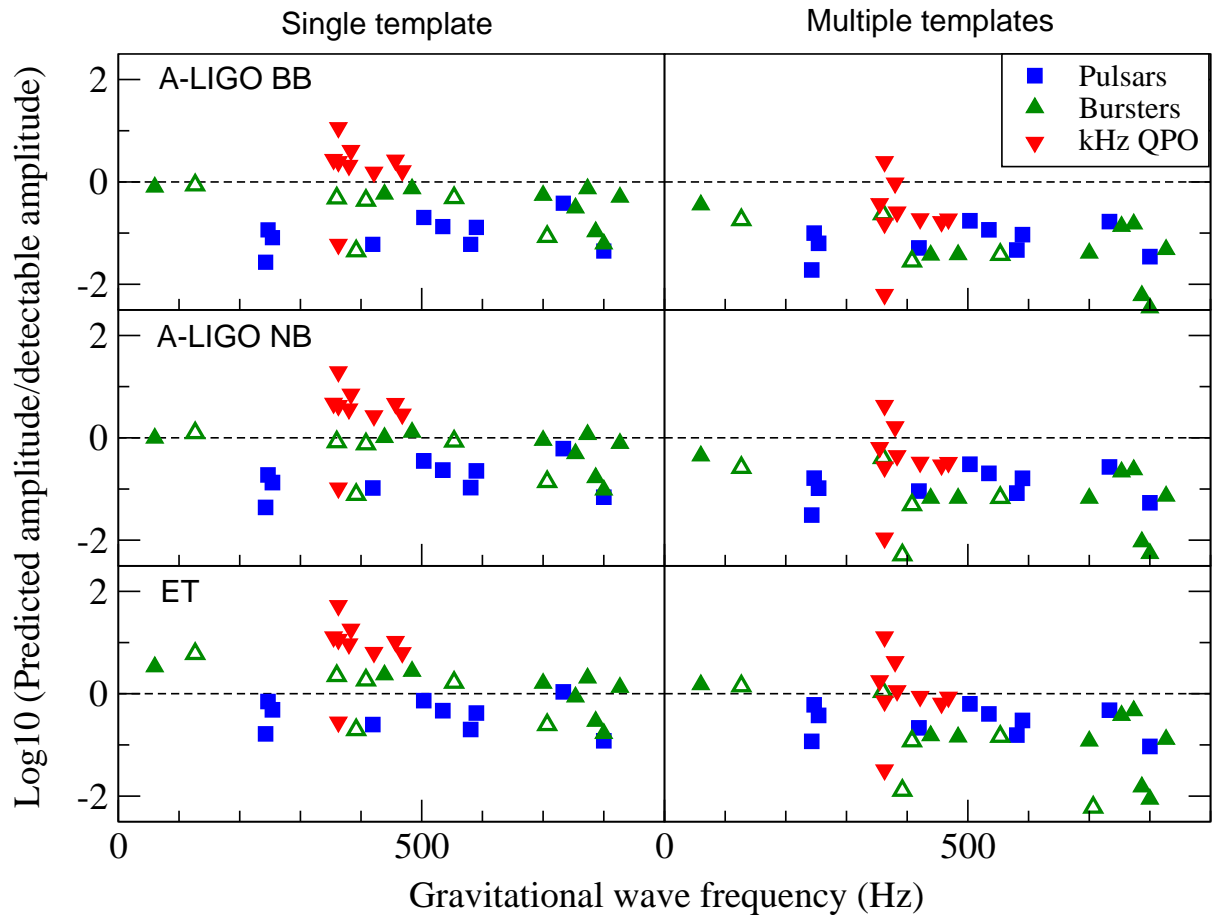}
\end{center}
\caption{As Figure \ref{mountperscomphit} but for the r-mode
  scenario. }
\label{rmodeperscomphit}
\end{figure*}

For the pulsars, computing constraints have little effect on the detectability
of the sources; however, these are not likely to be detectable in any case. But
for the bursters and the kHz QPO sources, computational constraints lead to a
major reduction in the $T_{\rm obs}$ that it is feasible to analyse. For many
sources we have to resort to a semi-coherent search, as a coherent fold is no
longer possible. Figures \ref{mountperscomphit} and \ref{rmodeperscomphit} show
the effects on detectability for the mountain and r-mode scenarios respectively,
taking into account both the change in $F_{\rm stat}$ and the reduced
observation times for our assumed computing power.

The additional impact of computational limitation is substantial. For Advanced LIGO's broad band configuration, only Sco X-1 and 4U 1820-30 (the latter in the r-mode scenario) remain above the detectability threshold.  These two sources are also the only two left above the narrow band envelope threshold, although there are other sources just below the envelope that might be detectable if the narrow band configuration were able to push closer to the thermal noise floor.  

Parameter uncertainty clearly poses a major problem, even for
stars where we have some indication of the spin rate\footnote{This
  problem is of course even more pronounced for the accreting neutron
  stars where we have no indication of spin rate.}. If the parameter
space volume can be reduced, however, then the statistical and computational restrictions will
have less impact. In the following section we will examine this in
more detail and consider how best to solve it.

\section{Discussion}
\label{disc}

\subsection{Current and future prospects}

It is clear from Section \ref{detect} that the detection of gravitational waves
from accreting neutron stars is a difficult task.  The X-ray bright kHz QPO
sources suffer from parameter uncertainty, forcing us to look
in addition at the much weaker but better constrained burst oscillation sources.
The best constrained sources, the accreting millisecond pulsars, are expected to
emit at such a low level that they are unlikely to be detectable by second
generation detectors.  This is particularly depressing since our
calculations have been carried out within the
context of a fairly optimistic modelling and analysis scenario.  We
have not as yet
considered any additional sources of spin-down, and have also neglected
physically reasonable complications such as spin variability which would
increase the number of templates still further. Spin variability would of course also lead to decoherence of the signal for sources where we cannot track the spin, reducing integration times to the decoherence time $T_{\rm decoh}$, defined by 

\begin{equation}
T_{\rm decoh}^2 \dot{\nu}_s = 1.
\end{equation}
One can get some idea of the worst case scenario by calculating the $\dot{\nu}_s$ that would result if the source were spinning up at the maximal rate due to the accretion torque (Equation 1).  The results are given in Table \ref{comphit}:  for the bright kHz sources this worst case decoherence time could be $\sim$ 1 week, and this is something that will need to be considered in future studies of detectability.    

In order to improve
prospects for those sources that are in principle 
detectable, we clearly need to find ways of improving source constraints and
removing computational limitations.  In the sections that follow we
detail the actions that would lead to the biggest improvements.

\subsection{Astronomical observations}

The major obstacle to detection by the current and imminent generation
of detectors is clearly parameter uncertainty, with spin uncertainty
having the largest impact (see Tables \ref{pulsnohit}-\ref{khznohit}).
Astronomical observations might help to constrain source properties.

The single
most valuable thing that could be done to improve the current
situation is to determine more precisely the NS spin in the bright kHz
QPO and burst sources.  A
substantial archive of high time resolution X-ray data exists for many
of these sources from RXTE and its predecessors.  However the only
deep search for pulsations in the literature is that carried out by
\citet{dib05} for 4U 1820-30.  Serious consideration should be given
to carrying out similar analysis for all of the kHz QPO and burst sources, most
particularly Sco X-1.  Thorough searches for intermittent pulsations, such as
those carried out by \citet{case07} and \citet{alt07}, would also be
worthwhile.

It may also be possible to find burst oscillations in the kHz QPO
sources:  five of the sources that we have analysed are known to
burst.   If a sufficient archive of bursts can be built up from these
sources burst oscillations may well be detected either in individual
bursts or by stacking power spectra from multiple bursts (the technique
used to find weak burst oscillations from EXO 0748-676 \citep{vil04}.  We also need to verify the burst oscillation 
frequencies for the seven sources with tentative or single burst detections, in
particular the four that are above the detection threshold for some of
the scenarios that we have examined:  XB 1254-690, 4U 1916-053, MXB
1730-335 and 4U 0614+09.

Identifying the orbital period can also make a substantial difference
to the number of templates searched.  Consider for example the
difference between the two kHz QPO sources Sco X-1 and GX 5-1.  Both
are expected to be strong emitters, but Sco X-1 suffers much less from
parameter uncertainty than GX 5-1 because the former has a well
constrained orbital period.  Three of the most promising kHz QPO
sources have no orbital constraints (GX 340+0, GX 5-1 and GX 17+2),
while a fourth (GX 349+2) has only a relatively weak constraint.
Several of the most promising burst oscillation sources also fall into
this category:  attention should focus on 4U 1608-522, KS 1731-260, 4U
1728-34, 4U 1702-429, and MXB 1730-335, sources that could be detectable
in some scenarios.  Immense progress has been made in recent years in
constraining orbital parameters for LMXBs using optical/IR
observations, particularly with the Bowen technique of
spectroscopy.  Identification of counterparts is often difficult, but
there would be a substantial payoff in terms of gravitational wave
detection prospects.

\subsection{Astrophysical modelling issues}

Both burst oscillation frequency and kHz QPO frequency are proxy
measures for the stellar spin rate.  The precise link to spin for each
measure is not clear because in neither case do we understand the
mechanism.  Astrophysical modelling to pin down the cause of the two
phenomena could therefore shed light on the reliability and accuracy
of the spin proxy.

Several models have been suggested for burst oscillations.  These include global
oscillations of the surface layers of the neutron star, large-scale drifting
vortices excited by the passage of the flame fronts, or hotspots linked in some
way to persistent pulsations.  All of the models have shortcomings, and progress
towards understanding this phenomenon has to some degree stalled.  The situation
for the kHz QPOs is however even worse.  There are several models, summarized in detail in \citet{vand06}. All involve either orbital 
motions of material within the disk or disk oscillations, and many also require
some mechanism to select preferred radii within the disk.  Developing testable
predictions that would distinguish between the different models, however, is
very difficult.   A substantial body of data exists within the RXTE archive to
test models if such predictions can be made.  The identification of a robust
link between kHz QPO frequency and stellar spin would have a major
impact in reducing parameter uncertainty for the most promising sources for
Advanced LIGO.

\subsection{Data analysis and detector issues}
\label{sec:disc-da}

In the absence of computational limitations, even taking into account the effect on statistics, there would be up to 8 sources potentially detectable by Advanced LIGO (Figure \ref{mountpersnohit}).  We therefore need to improve the data
analysis tools used for the searches.  We need significant
improvements in all relevant aspects of the data analysis pipelines: accelerating existing software through improvements to both 
software algorithms and computer hardware, and also developing other
data analysis techniques.  The software algorithms being used in the
LIGO data analysis software have improved continually over the years since the first LSC publication on periodic waves 
in 2003; we expect these improvements to continue.  It is possible that we may get an improvement of more than a order of magnitude 
in existing codes over the next several years.  The improvements in
computer hardware, even just following Moore's law, will yield an
additional factor of about 16 in the next 8 years, i.e. by the time we
expect Advanced LIGO to be operating.  Furthermore, computing
platforms like \texttt{Einstein@Home} allow us to increase the total
number of computers available to do the analysis. On top of this,
it might also be possible to use special purpose hardware for the periodic
wave searches.  Most of the analysis involves a large number of
relatively simple operations, and it might be quite feasible to design
chips which are efficient for these particular calculations. An
example of these are the GRAPE special-purpose computers which have
proven to be extremely useful in astrophysical N-body simulations
\citep{mak95} and in molecular dynamics calculations in condensed
matter physics.  It is hard to anticipate these developments, but
an improvement of 4 or 5 orders of magnitude in effective computing power
might be feasible.

Apart from these technical improvements, it is also possible to
develop new analysis methods.  A good example of this is the
cross-correlation method used in the Sco X-1 search \citep{abb07c}.
This was neither a matched filter nor a standard time-frequency
semi-coherent search. It was instead based on aperture synthesis, i.e.
the fact that we have multiple detectors in operation, and that they all
see essentially the same GW signal at any given time. This method has
so far mostly been used in the stochastic background searches, and
is now being adapted to the periodic wave searches \citep{dhu08}.
Apart from the computational efficiency, this method is also
relatively insensitive to the uncertainty in the signal model caused,
e.g. by pulsar glitches (though this leads to a correspondingly greater
computational cost in any follow-ups that must be done for verifying a
detection and for parameter estimation).  Further searches using this
method, in combination with the other methods discussed in this paper
seem to be quite promising, especially given that by the time of
Advanced LIGO there will be a third interferometer of comparable
sensitivity in Virgo.

In the near term while these improvements are in progress, it seems
reasonable to search for some of the brighter sources such as Sco X-1
not at the frequency implied by the kHz QPOs, but rather at lower frequencies
corresponding to the sweet spot of the detectors.  Not only are the
instruments more sensitive at these frequencies, but this is
computationally easier and is astrophysically well motivated if the
link between the kHz QPO and the spin frequency is not validated.

\begin{figure}
  \begin{center}
    \includegraphics[width=8cm, clip]{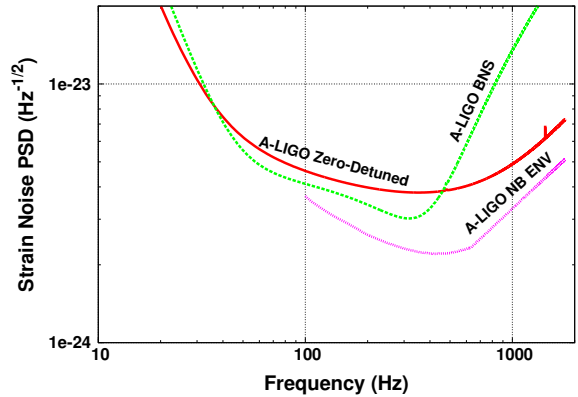}
    \caption{Some possible Advanced LIGO configurations. See text for
      discussion. }
    \label{fig:advligo}
 \end{center}
\end{figure}

We conclude this section with a brief discussion of possible Advanced
LIGO configurations and some of its implications for our purposes.
Figure \ref{fig:advligo} shows noise curves for three cases: an
example broad band configuration with some tuning of the signal
recycling cavity optimized for binary neutron star (BNS) inspirals,
the envelope of possible narrow band configurations, and the
zero-detuned broad band configuration. The zero-detuned curve and the
narrow band envelope are the same as in Fig.~\ref{fig:sh}.  The BNS
curve is relevant because binary neutron star inspirals will be among
the key targets for Advanced LIGO and the detector might possibly be
operated in this configuration for significant durations. Thus, while
BNS signals are not our concern in this paper, it is important to
consider the impact of this configuration for our purposes.  It is
obvious that the BNS curve affects us adversely above $\sim 500\,$Hz
which, as we see from Figure \ref{mountpers} is precisely where the
kHz QPO sources are expected to lie.  

These curves highlight the importance of spin measurements.  For all
potential sources with frequencies between, say, $150$ and $2000\,$Hz,
the narrow banding gains us no more than a factor of 2.  The best case
for narrow banding is then a compelling source whose frequency is very
well known and is within this range and is not detectable by the
broad-band curve.  The other possibility is to employ the zero-detuned
noise curve for detection and then use the narrow band curve as a
follow up to confirm detection or to increase SNR for extracting
better astrophysical information.  Given the present uncertainties in
the spin frequency, this latter possibility seems to be the best
option for detecting periodic waves from LMXBs.  This conclusion could
change if there turn out to be significant improvements in the narrow
band noise curves.

\subsection{Conclusions}

We have shown that gravitational wave observations of accreting neutron
stars will be challenging. It is therefore worthwhile considering what is to
be gained from such an effort: what new information gravitational wave
observations will bring to the study of these sources.

The input from electromagnetic observations that is required for these searches
is primarily on kinematical parameters like spin rates and orbital ephemerides.
The gravitational waves that are generated carry information on the orientation
and dynamics of the neutron star and its interaction with the accretion disk.
The elliptical polarization of the gravitational waves, which will
emerge from analysis of 
the detected signal, determines the orientation of the neutron-star
spin axis.  The degree of alignment of this with the inferred disk
orientation, especially for 
the slower-spinning (presumably younger) systems, will constrain models of
binary evolution and the formation of the NS.

In systems where observations and modelling constrain the NS spin to a relatively
narrow range, the GW frequency will determine whether the r-mode or mountain
scenario, or indeed some other scenario, is the appropriate one. The GW
amplitude will then determine the degree of mass asymmetry (mountains) or the
size of the velocity field (r-modes). In both cases this will open for the first
time a wealth of opportunities for studying the physics of NS
interiors.

For all detected systems, the measured GW amplitude and frequency will tell us
how much angular momentum is being carried away in GWs; by measuring or limiting
the rate of change of the GW frequency we can then infer the rate at which
angular momentum is being accreted from the disk. This will be an important
constraint on models of the disk and the NS magnetic field.

Clearly, if GW searches turn up signals in unexpected places in parameter space,
this will challenge prevailing models for these systems. For example, if the GW
frequency equals the X-ray pulsation frequency instead of twice or
four thirds its value, then this might imply that the pulsation
frequency is actually twice the true 
spin frequency.

Even in the worst case, where a sufficiently sensitive search fails to detect
GWs at the expected amplitude, the negative result could eliminate the GW option
for spin balance and demonstrate that this somehow resulted from a
propeller-type torque from the NS's magnetic field. Of course, one would have to
have confidence that the GWs were not coming out at a different frequency, so
the search would have to include that as a parameter.

The additional astronomical input that could make the difference between
detecting or not detecting GWs could come from a variety of observations.
Clearly, long-term X-ray timing is highly desirable, and here it is
disappointing that the future of RXTE is limited and no X-ray mission with
comparable capabilities seems to be planned for the period before Advanced LIGO
comes online. Thorough exploitation of the RXTE data archive is therefore
especially important. On the other hand, sensitive radio observations of X-ray
systems in quiescence may provide unexpected information, and this will be
easier with arrays like the Square Kilometre Array and its pathfinders.

\section{Acknowledgements}

We would like to thank Alan Levine for advice on ASM countrates, Jake
Hartman for advice on the effects of positional uncertainty on pulsar
timing, Reinhard Prix for help with the computational cost estimates,
Alberto Vecchio and Chris Messenger for discussions regarding the
parameter space metric, and Maria Alessandra Papa, Mike Landry and Graham
Woan for useful comments on a draft of this paper.  We are grateful to
Rana Adhikari, Peter Fritschel, Gregg Harry, Bangalore Sathyaprakash,
David Shoemaker, Kentaro Somiya and Ken Strain for pointing us to
Bench and for valuable discussions related to the Advanced LIGO noise
spectrum.  We thank Rana Adhikari also for providing us with the
Enhanced LIGO noise curve. ALW would like to
thank the Kavli Institute for Theoretical Physics for hospitality
during a KITP/MPA postdoctoral exchange visit, when this work was
started.  BK is grateful to the LSC continuous-waves
data analysis group for valuable discussions.  LB acknowledges support from the National Science Foundation
via grants PHY 05-51164 and AST 02-05956. BK and BFS acknowledge the
support of the DFG's special research centre SFB/Transregio 7
``Gravitational Wave Astronomy''.

\clearpage

\begin{deluxetable}{lccccc}
\tabletypesize{\small}
\tablewidth{0pt}
\tablecaption{Fluxes, distances and outburst properties for transients \label{sdata}}
\tablehead{\colhead{Source}  & \colhead{$\nu_s$} & \colhead{Distance} &
  \colhead{Long-term flux $F_{\rm av}$} &\colhead{Outburst duration} &
  \colhead{Outburst flux $F_{\rm ob}$}  \\ & (Hz) & (kpc) & ($\times
  10^{-8}$ ergs cm$^{-2}$ s$^{-1}$) & (days) & ($\times 10^{-8}$ ergs cm$^{-2}$ s$^{-1}$)
  \\}
\tablecolumns{6}
\startdata
\sidehead{Accreting millisecond pulsars}
IGR J00291+5934 & 598.892 $\pm$ 2e-08 & 5 $\pm$ 1  &  1.8e-03 & 13 & 1.6e-01  \\
Aql X-1 & 550.274 $\pm$ 9e-04 & 4.55 $\pm$ 1.35  &  1.2e-01 & 45 & 1.3  \\
SAX J1748.9-2021 & 442.361 $\pm$ 5e-08 & 8.1 $\pm$ 1.3 & 9.2e-03 & 61 & 4.0e-01  \\
XTE J1751-305 & 435.318 $\pm$ 4e-08 & 9 $\pm$ 3\tablenotemark{a} & 2.0e-03 & 10 & 2.9e-01  \\
SAX J1808.4-3658 & 400.975 $\pm$ 6e-09 & 3.5 $\pm$ 0.1 &  8.6e-03 & 20 & 3.5e-01  \\
HETE J1900.1-2455 & 377.296 $\pm$ 5e-09 & 4.7 $\pm$ 0.6 & 1.8e-02 & 730 & 9.0e-02  \\
XTE J1814-338 & 314.357 $\pm$ 1e-09 & 6.7 $\pm$ 2.9 & 1.3e-03 & 50 & 6.9e-02  \\
XTE J1807-294 & 190.624 $\pm$ 8e-08 & 8.35 $\pm$ 3.65\tablenotemark{a} & 1.4e-03 & 50 & 7.2e-02  \\
XTE J0929-314 & 185.105 $\pm$ 9e-09 & 7.8 $\pm$ 4.2\tablenotemark{a} & 2.7e-03 & 60 & 1.0e-01  \\
SWIFT J1756.9-2508 & 182.066 $\pm$ 7e-08 & 8 $\pm$ 4\tablenotemark{a}
& 1.4e-04 & 13 & 4.0e-02 \\
\sidehead{Burst oscillation sources}
XTE J1739-285 & 1122\tablenotemark{b} & 7.3 $\pm$ 3.3\tablenotemark{a} & 5.9e-03 & 196 & 8.2e-02  \\
4U 1608-522 & 620 & 4.1 $\pm$ 0.4  & 2.5e-01 & 100 & 2.0  \\
SAX J1750.8-2900 & 601 & 6.79 $\pm$ 0.14 & 3.5e-03 & 108 & 0.12  \\
GRS 1741.9-2853 & 589 & 7.2 $\pm$ 2.8 & 1e-02\tablenotemark{c}  & 30\tablenotemark{c}  & 1e-01\tablenotemark{c}  \\
4U 1636-536 & 581 & 6 $\pm$ 0.5  & 0.47 & \nodata & \nodata  \\
X 1658-298 & 567 & 12  $\pm$ 3 & 8.0e-02 & 872 & 0.77  \\
A 1744-361 & 530\tablenotemark{b} & 6 $\pm$ 3\tablenotemark{a}  & 5.3e-03 & 97 & 0.2  \\
KS 1731-260 & 524 & 7.2 $\pm$ 1.0 & 2.2e-01 & 1886 & 0.49  \\
4U 0614+09 & 415\tablenotemark{b}  & 2.5 $\pm$ 0.5\tablenotemark{a} & 1.2e-01 & \nodata & \nodata \\
4U 1728-34 & 363 & 5.2 $\pm$ 0.5 & 0.23 & \nodata & \nodata  \\
4U 1702-429 & 329 &  5.5 $\pm$ 0.2 & 0.13 & \nodata & \nodata  \\
MXB 1730-335 & 306\tablenotemark{b} & 9.25 $\pm$ 2.85 & 6.4e-02 & 24 & 0.59  \\
IGR J17191-2821 & 294\tablenotemark{b} & 7.5 $\pm$ 3.5\tablenotemark{a} & 6.6e-04 & 11 & 0.26  \\
4U 1916-053 & 270\tablenotemark{b} & 8.0 $\pm$ 2.2 & 6.9e-02 & \nodata & \nodata  \\
XB 1254-690 & 95\tablenotemark{b} & 13 $\pm$ 3 & 9.0e-02 & \nodata & \nodata  \\
EXO 0748-676 & 45 & 7.4 $\pm$ 0.9 & 4.6e-02 & \nodata & \nodata  \\
\sidehead{kHz QPO sources}
Cyg X-2 &  351 $\pm$ 34 & 10.55 $\pm$ 4.45 & 1.1 & \nodata & \nodata  \\
GX 340+0 & 343 $\pm$ 92  & 9.15 $\pm$ 5.15\tablenotemark{a} & 2.8 & \nodata & \nodata  \\
4U 1735-44 & 316 $\pm$ 32  & 8.5 $\pm$ 1.3 & 8.4e-01 & \nodata & \nodata  \\
GX 5-1 & 288 $\pm$ 69 &  7.85 $\pm$ 3.85\tablenotemark{a} & 5.3 & \nodata & \nodata  \\
4U 1820-30 & 285 $\pm$ 65 & 7.4 $\pm$ 0.6 & 1.4 & \nodata & \nodata  \\
Sco X-1 & 272 $\pm$ 40  & 2.8 $\pm$ 0.3 & 39 & \nodata & \nodata  \\
GX 17+2 & 272 $\pm$ 50 & 11.4 $\pm$ 2 &  1.8 & \nodata & \nodata  \\
XTE J2123-058 & 272 $\pm$ 50 & 9.6 $\pm$ 1.3 & 1.1e-03 & 53 & 8.7e-02  \\
GX 349+2 & 266 $\pm$ 13 & 4 $\pm$ 0.4 & 2.2 & \nodata & \nodata \\
\enddata
\tablenotetext{a}{Distance poorly constrained:  upper and/or lower
  limit set arbitrarily.}
\tablenotetext{b}{Burst oscillation frequency requires confirmation:  tentative
  detection, or only seen in one burst from this source.}
\tablenotetext{c}{No reliable measurement of flux possible
  for this Galactic Centre source due to field crowding.  The values
  given here are indicative estimates only.}
\end{deluxetable}

\clearpage

\begin{deluxetable}{lcccccccccc}
\rotate
\tabletypesize{\small}
\tablecaption{Parameter space and template requirements for the accreting
  millisecond pulsars. \label{pulsnohit}}
\tablewidth{0pt}
\tablehead{\colhead{Source}  & \colhead{$\nu_s \pm \Delta \nu_s$} & \colhead{$N_\nu$} &
  \colhead{$P_{\rm orb} \pm \Delta P_{\rm orb}$} & \colhead{$N_{P_{\rm orb}}$}
  &\colhead{$a_{\rm x} \sin i \pm \Delta a_{\rm x} \sin i$} & \colhead{$N_{a_p}$} &
  \colhead{$\Delta T_{\rm asc}$} & \colhead{$N_{T_{\rm asc}}$} &
  \colhead{$\log_{10}(N_{\rm temp})$}
  & \colhead{$F_{\rm stat}$} \\ & (Hz) & & (hours) & &
  (lt-s) & & (days) &  & & \\}
\startdata
IGR J00291+5934 & 598.89213053 & 18 & 2.4566922 & 23 & 6.4993e-02 & 1
& 4e-07  &  1 &  2.6   &  15.9 \\

& $\pm$ 2e-08 & (12) & $\pm$ 1.7e-06 & (15) & $\pm$ 2e-06 & (1) &  &
(1) & (2.2) & (14.7) \\
\\
Aql X-1 & 550.2745 & 7.6e+05 & 18.9479  & 1557 & 2.5 & 9692 & 3e-03  &
1049 & 15.6 &   26.6  \\

 &  $\pm$ 9e-04 & (5.1e+05) & $\pm$ 2.0e-04 & (1038) & $\pm$ 0.5 & (6462)
 & & (700) & (14.9)  & (26.0)  \\
\\
SAX J1748.9-2021 & 442.36108118  & 43  & 8.76525  & 139 & 0.38760 & 1  &
4e-06 & 1  & 3.8  &  17.1   \\

 & $\pm$ 5e-08 & (29) &  $\pm$ 3e-05 & (93) & $\pm$ 4e-05 & (1)
 &  & (1) & (3.4) & (15.9) \\
\\
XTE J1751-305 & 435.31799357  & 34 & 0.7070394 & 11 & 1.0125e-02  &
1 & 4e-7  & 1 & 2.5  &  15.9  \\

 & $\pm$ 4e-08 & (23)  & $\pm$ 6e-07 & (7) &  $\pm$ 5e-06 & (1)
 & & (1) & (2.2) & (14.7) \\
\\
SAX J1808.4-3658 & 400.975210221 & 6 & 2.013654711 & 1 & 6.028132e-02
& 1 & 1.0e-06  & 1 & 0.8   &  13.2  \\

 &  $\pm$ 6e-09 & (4) & $\pm$ 4e-09 & (1) &  $\pm$ 2.4e-07 & (1) & &
 (1)  & (0.6) & (13.2) \\
\\
HETE J1900.1-2455 & 377.296171971 & 5 & 1.3875717  & 11 & 1.841e-02  &
1 &  7e-06 & 1 & 1.7   &  14.7 \\

 & $\pm$ 5e-09 & (4)  & $\pm$ 1.4e-06 & (7) &  $\pm$ 1e-05 & (1) &   &
 (1) & (1.4) & (13.2)\\
\\
XTE J1814-338 & 314.35610879  & 9 & 4.274645250 & 1 & 0.390633  & 1 &
9e-07  & 1 & 1.0  &  13.2\\

 & $\pm$ 1e-08 & (7) &  $\pm$ 5.6e-08 & (1) &  $\pm$ 9e-06 & (1) &
 &(1) & (0.8) & (13.2) \\
\\
XTE J1807-294 & 190.62350694  & 68  & 0.6678935 & 1 & 4.819e-03  & 1 &
6e-06   & 1  & 1.8   &  14.7 \\

 &  $\pm$ 8e-08 & (46) &  $\pm$ 1e-07 & (1) & $\pm$ 4e-06 & (1) &  &
 (1)  & (1.7) & (14.7) \\
\\
XTE J0929-314 & 185.105254297  & 9 & 0.7263183  & 4 & 6.290e-03  & 1 &
1e-05 & 1 & 1.5  &   14.7 \\

 &  $\pm$ 9e-09 & (6) &  $\pm$ 8e-07 & (3) &  $\pm$ 9e-06 & (1) & & (1)  &(1.2) & (13.2)\\
\\
SWIFT J1756.9-2508 & 182.065804253  & 61 & 0.911696 & 62 & 5.942e-03
& 1 & 6e-05  & 1  & 3.6  &   17.1 \\

 & $\pm$ 7.2e-08 & (41) & $\pm$ 2.3e-05 & (42) &  $\pm$ 2.7e-05 & (1)
 & & (1) & (3.2) & (15.9)  \\

\enddata
\tablecomments{Parameter ranges searched for the pulsars, and number of
  templates resulting for the mountain (r-mode) scenario assuming a
  coherent fold with $T_{\rm obs} = 2$ years.  For all parameters
  apart from $T_{\rm asc}$ we give the absolute value as well as the
  range searched since both quantities affect the number of
  templates. Note that $\Delta T_{\rm asc}$ is the uncertainty on the
  measurement, so the searched range is twice this value.}
\end{deluxetable}

\clearpage

\begin{deluxetable}{lcccccccccc}
\rotate
\tabletypesize{\scriptsize}
\tablecaption{Parameter space and template requirements for the burst
  oscillation sources (details as for Table \ref{pulsnohit}). \label{burstnohit}}
\tablewidth{0pt}
\tablehead{\colhead{Source}  & \colhead{$\nu_s \pm \Delta \nu_s$} & \colhead{$N_\nu$} &
  \colhead{$P_{\rm orb} \pm \Delta P_{\rm orb}$} & \colhead{$N_{P_{\rm orb}}$}
  &\colhead{$a_{\rm x} \sin i \pm \Delta a_{\rm x} \sin i$} & \colhead{$N_{a_p}$} &
  \colhead{$\Delta T_{\rm asc}$} & \colhead{$N_{T_{\rm asc}}$} &
  \colhead{$\log_{10}(N_{\rm temp})$}
  & \colhead{$F_{\rm stat}$} \\ & (Hz) & & (hours) & &
  (lt-s) & & (days) & & &  \\}
\startdata

XTE J1739-285 & 1122  & 4.2e+09 & 120.1  & 2.2e+08 & 11.3
& 4.1e+05 & 2.5  & 1.3e+06 & 34.4 & 36.0 \\

 & $\pm$ 5 & (2.8e+09) & $\pm$ 119.9 & (1.5e+08) & $\pm$ 11.3 & (2.8e+05) & & (8.7e+05) & (33.7)  &
(36.0)  \\
\\
4U 1608-522 & 620 & 4.2e+09 & 67.5  & 1.0e+08 & 6.17  & 1.3e+05 &
1.4 & 3.9e+05 & 29.1 & 33.7 \\

 & $\pm$ 5 & (2.8e+09) & $\pm$ 57.5 & (6.7e+07) & $\pm$ 6.17 &
 (8.3e+04) & & (2.6+05) & (28.4) & (33.2) \\
\\
SAX J1750.8-2900 & 601 & 4.2e+09 & 120.1 & 1.2e+08 & 11.3  & 2.2e+05 & 2.5  & 7.0e+05 & 33.6 & 36.0 \\

 &  $\pm$ 5 & (2.8e+09) & $\pm$ 119.9 & (7.8e+07)  & $\pm$ 11.3 &
 (1.5e+05) &  & (4.7e+05)   & (32.9) & (35.5) \\
\\
GRS 1741.9-2853 & 589  & 4.2e+09 & 120.1  & 1.2e+08 & 11.3  &
2.2e+05 & 2.5 & 6.8e+05 & 33.6 & 36.0 \\

 & $\pm$ 5 & (2.8e+09) & $\pm$ 119.9 & (7.7e+07) &  $\pm$ 11.3 &
 (1.5e+05) &  & (4.6e+05) & (32.9)  & (35.5)  \\
\\
4U 1636-536 & 581  & 4.2e+09 & 3.7931263  & 238 & 0.735 & 1585&
2e-3  & 1112 & 17.8 & 27.9 \\

& $\pm$ 5 & (2.8e+09) &  $\pm$ 3.8e-06 & (159)  & $\pm$ 8.3e-02 & (1057) &  & (742 ) & (17.1) & (27.3) \\
\\
X 1658-298 & 567 & 4.2e+09 & 7.11610979  & 3 & 1.1 & 1.7e+04  &
9.3e-05  & 41 & 15.4 & 26.0 \\

 &  $\pm$ 5 & (2.8e+09) &  $\pm$ 8e-08 & (2) & $\pm$ 0.9 & (1.1e+04)
 &  &(27)   & (14.7)  & (26.0)  \\
\\
A 1744-361 & 530  & 4.2e+09 & 1.62  & 1.9e+07 & 0.125  & 1302 &
3.4e-02  & 6817 & 23.5 & 30.7 \\

 & $\pm$ 5 & (2.8e+09) & $\pm$ 0.37 & (1.3e+07) & $\pm$ 0.075 & (868) & & (4545)  &(22.8)  & (30.7 ) \\
\\
KS 1731-260 & 524  & 4.2e+09 & 121  & 1.0e+08 &  11.3  & 1.9e+05 &
2.5 & 6.1e+05 & 31.3 & 34.6 \\

 & $\pm$ 5 & (2.8e+09) & $\pm$ 119 & (6.7e+07)  &  $\pm$ 11.3 &
 (1.3e+05) &  & (4.1e+05)   & (30.6) & (34.6) \\
\\
4U 0614+09 & 415 & 4.2e+09 & 0.29 & 3.0e+06  & 7e-03 & 96 &  6.1e-03  &
300 &  20.2 & 29.0 \\

& $\pm$ 5 & (2.8e+09)  &  $\pm$ 0.04 & (2.0e+06)  & $\pm$ 7e-03
& (64)  & & (200)   & (19.5)  & (28.4) \\
\\

4U 1728-34 & 363  & 4.2e+09  & 5.1  & 1.5e+07 & 0.1  & 1200  & 0.1 & 3770
& 25.5  & 31.7 \\

 &  $\pm$ 5 & (2.8e+09) &  $\pm$ 4.9 & (9.7e+06) & $\pm$ 0.1 & (800)
 &  & (2513)    & (24.8)  & (31.7) \\
\\
4U 1702-429 & 329  & 4.2e+09 & 120.1  &  6.5e+07 & 11.3 & 1.2e+05  & 2.5
&  3.8e+05  & 32.9 & 35.5 \\

 & $\pm$ 5 & (2.8e+09) & $\pm$ 119.9 & (4.3e+07) & $\pm$ 11.3 &
 (8.2e+04)  &  & (2.6e+05)    &(32.2)  & (35.1) \\
\\
MXB 1730-335 & 306  & 4.2e+09 & 120.1  & 6.0e+07 & 11.3 & 1.1e+05 & 2.5
  & 3.6e+05 & 32.8 & 35.5 \\

 & $\pm$ 5 & (2.8e+09) & $\pm$ 119.9 & (4.0e+07) & $\pm$ 11.3 &
 (7.6e+04) &  & (2.4e+05)  & (32.1) & (35.1)  \\
\\
IGR J17191-2821 & 294  & 4.2e+09 & 120.1  & 5.8e+07 & 11.3  & 1.1e+05
& 2.5  & 3.4e+05 & 32.7  & 35.5  \\

 &  $\pm$ 5 & (2.8e+09) & $\pm$ 119.9 & (3.9e+07) &  $\pm$ 11.3 &
 (7.3e+04) & & (2.3e+05) &(32.0)  & (35.1) \\
\\
4U 1916-053 & 270  & 4.2e+09 & 0.83351411  & 3  & 1.45e-02 & 94 &
1.4e-04  & 4 & 12.2 & 24.0  \\

 & $\pm$ 5 & (2.8e+09)  &  $\pm$ 2.5e-07 &(2)  &  $\pm$ 1.05e-02 & (63) & &
(3)  & (11.5) & (23.2) \\
\\
XB 1254-690 & 95  &4.2e+09 & 3.9334  & 2606 & 0.98  & 390 & 3e-03
&365  & 17.7 & 27.9 \\

 & $\pm$ 5 & (2.8e+09) & $\pm$ 2e-04 &(1737)  &  $\pm$ 0.12 &(260)  & & (243)  & (17.0) & (27.3) \\
\\
EXO 0748-676 & 45  & 4.2e+09 & 3.8241072  & 9& 0.475  & 593 &
1.3e-03 & 40 & 14.5 & 25.3 \\

 & $\pm$ 5 & (2.8e+09) &  $\pm$ 2.4e-06 & (6) & $\pm$ 0.365 & (395)
 &  & (27)  & (13.8)  & (25.3)\\

\enddata
\end{deluxetable}

\clearpage

\begin{deluxetable}{lcccccccccc}
\rotate
\tabletypesize{\small}
\tablecaption{Parameter space and template requirements for the kHz QPO sources (details as for Table \ref{pulsnohit}).  \label{khznohit}}
\tablewidth{0pt}
\tablehead{\colhead{Source}  & \colhead{$\nu_s \pm \Delta \nu_s$} & \colhead{$N_\nu$} &
  \colhead{$P_{\rm orb} \pm \Delta P_{\rm orb}$} & \colhead{$N_{P_{\rm orb}}$}
  &\colhead{$a_{\rm x} \sin i \pm \Delta a_{\rm x} \sin i$} & \colhead{$N_{a_p}$} &
  \colhead{$\Delta T_{\rm asc}$} & \colhead{$N_{T_{\rm asc}}$ } &
  \colhead{$\log_{10}(N_{\rm temp})$}  & \colhead{$F_{\rm stat}$} \\ & (Hz) & & (hours) & &
  (lt-s) & & (days) &  & & \\}
\startdata

Cyg X-2 & 351 & 2.8e+10 & 236.26560000  & 1 & 13.5 & 2.0e+04 &
3.0e-02 & 3228   & 18.0 & 27.9 \\

 & $\pm$ 34 & (1.9e+10) & $\pm$ 8.3e-08 & (1) & $\pm$ 1.6 & (1.4e+04) &  &   (2152) & (17.4)  & (27.3)\\
\\
GX 340+0 & 343 & 7.6e+10 & 120.1 & 8.4e+07 & 11.3 & 1.6e+05 & 2.5 &
5.0e+05 & 34.5 & 36.0 \\

 & $\pm$ 92 & (5.1e+10)  & $\pm$ 119.9 & (5.6e+07) & $\pm$ 11.3 &
 (1.1e+05) &  & (3.3e+05) & (33.8) & (36.0)  \\
\\
4U 1735-44 & 316  & 2.7e+10 & 4.6520042  & 177 & 0.693 & 7825  &
3.0e-3 & 761 & 19.1 & 28.4 \\

& $\pm$ 32 & (1.8e+10)  & $\pm$ 7.7e-06 & (118)  & $\pm$ 0.693 &(5217)  & &(508)    & (18.4) & (27.9) \\
\\

GX 5-1 & 288 & 5.7e+10 & 120.1 & 6.9e+07 & 11.3  & 1.3e+05 &
 2.5 & 4.1e+05 & 34.1 & 36.0 \\

& $\pm$ 69 & (3.8e+10)  & $\pm$ 119.9 & (4.6e+07) & $\pm$ 11.3 &
(8.7e+04) & & (2.7e+05)   & (33.4) & (35.5) \\
\\
4U 1820-30 & 285 & 5.4e+10 & 0.182781083  & 9 & 1.35e-02 & 74 &
2.2e-04 & 28   & 14.5 & 26.0\\

& $\pm$ 65 & (3.6e+10)  & $\pm$ 2.8e-08 &(6)  & $\pm$ 6.5e-03 & (50) & & (19)  &(13.8)  &(25.3)  \\
\\
Sco X-1 & 272 & 3.3e+10 & 18.89551 & 58 & 1.44 & 1827 &  3.0e-03 &
350   & 17.6 & 27.9\\

& $\pm$ 40 & (2.2e+10) & $\pm$ 2e-05 & (39) & $\pm$ 0.18 & (1218) & & (234)  & (16.9) & (27.3) \\
\\
GX 17+2 & 272 & 4.2e+10 & 120.1 & 6.2e+07 & 11.3 & 1.2e+05  & 2.5 &
3.7e+05 & 33.8 & 36.0\\

& $\pm$ 50 & (2.8e+10) & $\pm$ 119.9 & (4.2e+07) & $\pm$ 11.3 &
(7.9e+04) & & (2.5e+05) & (33.1) & (35.5) \\
\\
XTE J2123-058 & 272 & 4.2e+10 & 5.956833 & 1657 & 1.6 & 3210 &
1.0e-03 &  422 & 19.5 & 28.4\\

& $\pm$ 50 & (2.8e+10)  & $\pm$ 5.6e-05 & (1105) & $\pm$ 0.3 & (2140) & & (281)    & (18.8) & (28.4) \\
\\
GX 349+2 & 266 & 1.1e+10 & 22.5 & 4.0e+05 & 3.5  & 3.2e+04 &
0.47 & 1.0e+05  & 24.8 & 31.7\\

& $\pm$ 13 &  (7.2e+09) & $\pm$ 0.1 & (2.7e+05) & $\pm$ 3.5 &
(2.1e+04) & & (6.6e+04)  & (24.1) & (31.2)\\

\enddata
\end{deluxetable}

\clearpage

\begin{deluxetable}{lcccccc}
\tabletypesize{\small}
\tablewidth{0pt}
\tablecaption{Computational cost effects for future performance scenarios. \label{comphit}}
\tablehead{\colhead{Source}   & \colhead{$\nu$}   & \multicolumn{2}{c}{$\times$ 50
    improvement} &  \multicolumn{2}{c}{$\times$ 100 improvement}  & \colhead{$T_{\rm decoh}$}\\
&  (Hz)  & \colhead{$N_{\rm stacks}$} & \colhead{$T_{\rm obs}$ (days)}  &  \colhead{$N_{\rm stacks}$} & \colhead{$T_{\rm obs}$ (days)} & \colhead{(years)} }
\tablecolumns{7}
\startdata
\sidehead{Accreting millisecond pulsars}
IGR J00291+5934 &  599  & 1 (1) & 730 (730) &  1 (1) & 730 (730) & 1.26 \\
Aql X-1 &  550 &  1 (1) & 512 (730) &  1 (1) & 645 (730) & 0.17 \\
SAX J1748.9-2021 & 442  & 1 (1) & 730 (730)  & 1 (1) & 730 (730) & 0.35 \\
XTE J1751-305 &  435 &  1 (1) &  730 (730)  & 1 (1) &  730 (730) & 0.67 \\
SAX J1808.4-3658  & 401  & 1 (1) & 730 (730) & 1 (1) & 730 (730) & 0.82 \\
HETE J1900.1-2455  & 377 & 1 (1) & 730 (730) & 1 (1) & 730 (730) &
0.43 \\

XTE J1814-338  & 314 & 1 (1) & 730 (730) & 1 (1) & 730 (730) & 1.12 \\

XTE J1807-294  & 191 & 1 (1) &  730 (730) & 1 (1) & 730 (730)& 0.86  \\
XTE J0929-314  & 185 & 1 (1) & 730 (730) & 1 (1) & 730 (730) & 0.66 \\
SWIFT J1756.9-2508 & 182 & 1 (1) & 730 (730) & 1 (1) & 730 (730) & 2.81 \\
\sidehead{Burst oscillation sources}
XTE J1739-285 & 1122 & 1.5e+05 (1.5e+05) & 0.1 (0.1) & 1.5e+05 (1.5e+05) & 0.1 (0.1) & 0.48 \\
4U 1608-522 & 620  & 8372 (7209) & 3.1 (3.6) & 7865 (6655) & 3.3 (3.9)  & 0.13 \\
SAX J1750.8-2900 & 601  & 1.5e+05 (1.5e+05) & 0.1 (0.1) & 1.5e+05 (7.3e+04) & 0.1 (0.2) & 0.67  \\
GRS 1741.9-2853 & 589 & 1.5e+05 (1.5e+05) & 0.1 (0.1) & 1.5e+05 (7.3e+04) & 0.1 (0.2) & 0.37  \\
4U 1636-536 & 581 & 1 (1) & 98 (167) & 1 (1) & 123 (211) & 0.07    \\
X 1658-298 & 567 & 1 (1) & 611 (730) & 1 (1) & 730 (730) &  0.08\\
A 1744-361 & 530 & 1 (1) & 1.2 (2.1) & 1 (1) & 1.5 (2.6) & 0.61 \\
KS 1731-260 & 524 & 2.4e+04 (2.1e+04) & 0.6 (0.7) & 2.4e+04 (2.1e+04) & 0.6
(0.7) & 0.08  \\
4U 0614+09 & 415 &  1 (1) & 14.9 (25.6) & 1 (1) & 18.8 (32.3) & 0.31  \\
4U 1728-34 & 363 &  9.6e+04 (8.2e+04) & 3.6 (4.2) & 9.1e+04 (7.7e+04) & 3.8 (4.5) & 0.11  \\
4U 1702-429 & 329 & 1.5e+05 (7.3e+04) & 0.1 (0.2) & 7.3e+04 (7.3e+04) & 0.2 (0.2) & 0.14  \\
MXB 1730-335 & 306 & 7.3e+04 (7.3e+04) & 0.2 (0.2) & 7.3e+04 (7.3e+04) & 0.2 (0.2) & 0.11  \\
IGR J17191-2821 & 294 & 7.3e+04 (7.3e+04) & 0.2 (0.2) & 7.3e+04
(7.3e+04) & 0.2 (0.2) & 1.39  \\
4U 1916-053 & 270 & 1 (1) & 730 (730) & 1 (1) & 730 (730) & 0.13  \\
XB 1254-690 & 95 & 1 (1) & 102 (175) & 1 (1) & 128 (220) & 0.07  \\
EXO 0748-676 & 45 & 1 (1) & 730 (730) & 1 (1) & 730 (730) & 0.17  \\
\sidehead{kHz QPO sources}
Cyg X-2 & 351 & 123 (1) & 60 (51) & 1 (1) & 40 (73) & 0.02 \\
GX 340+0 & 343 & 1.5e+05 (7.3e+04) & 0.1 (0.2) & 1.5e+05 (7.3e+04) & 0.1 (0.2) & 0.02 \\
4U 1735-44 & 316 & 1 (1) & 35 (61) & 1 (1) & 44 (76) & 0.03  \\
GX 5-1 & 288 & 1.5e+05 (7.3e+04) & 0.1 (0.2) & 1.5e+05 (7.3e+04) & 0.1 (0.2) & 0.01   \\
4U 1820-30 & 285 & 1 (1) & 730 (730) & 1 (1) & 730 (730) & 0.03  \\
Sco X-1 & 272 & 1 (1) & 110 (188) & 1 (1) & 138 (237) & 0.02 \\
GX 17+2 & 272 & 7.3e+04 (7.3e+04) & 0.2 (0.2) & 7.3e+04 (7.3e+04) & 0.2 (0.2) & 0.02  \\
XTE J2123-058 & 272 & 1 (1) & 26 (44) & 1 (1) & 33 (56) & 0.85  \\
GX 349+2 & 266 & 1908 (1625) & 41 (48) & 1781 (1514) & 44 (51) & 0.05  \\
\enddata
\tablecomments{Computational cost effects on data analysis, assuming a 50 or 100 times improvement in computational power
  over current capabilities, for the mountain (r-mode) scenario.  Where $N_{\rm stacks} = 1$, the fold is coherent and
  $T_{\rm obs}$ is the maximum feasible quantity of data that can be
  folded.  Where $N_{\rm stacks} > 1$, the fold is semi-coherent and
  $T_{\rm obs}$ is the length of each individual data segment.  The assumed spin
  frequency is given for each source to assist in cross-referencing this Table with
  the Figures.  We also give the decoherence time $T_{\rm decoh}$, as defined in Section \ref{disc}.}
\end{deluxetable}

\end{document}